\documentclass[twocolumn,pra,amsmath,amssymb]{revtex4}
\usepackage{graphicx}
\usepackage{dcolumn}
\usepackage{bm}
\usepackage{epstopdf}
\usepackage{mathrsfs}
\usepackage{amsmath}
\usepackage{amssymb}

\begin{document}
\title{Quantum-to-Classical Correspondence and Hubbard-Stratonovich Dynamical Systems, a Lie-Algebraic Approach}

\author{Victor Galitski}
\affiliation{
Joint Quantum Institute and Department of Physics,
University of Maryland, College Park, MD 20742-4111}

\begin{abstract}
We propose a Lie-algebraic duality approach to analyze non-equilibrium evolution of closed dynamical systems and thermodynamics of interacting quantum lattice models (formulated in terms of Hubbard-Stratonovich dynamical systems). 
The first part of the paper utilizes a geometric Hilbert-space-invariant formulation of  unitary time-evolution, where a quantum Hamiltonian is viewed as a trajectory in an abstract Lie algebra, while the sought-after evolution operator is a trajectory in a dynamic group,  generated by the algebra via exponentiation. The evolution operator is uniquely determined by the time-dependent dual generators that satisfy a system of differential equations, dubbed here  dual Schr{\"o}dinger-Bloch equations, which represent a viable alternative to the conventional Schr{\"o}dinger formulation. These dual Schr{\"o}dinger-Bloch equations are derived and analyzed on a number of specific examples. It is shown that deterministic dynamics of a closed classical  dynamical system  occurs as action of a symmetry group on a classical manifold and is driven by the same dual generators  as in the corresponding quantum problem. This represents quantum-to-classical correspondence. In the second part of the paper, we further extend the Lie algebraic approach to a wide class of interacting many-particle lattice models. A generalized Hubbard-Stratonovich transform is proposed and it is used to show  that the thermodynamic partition function of a generic many-body quantum lattice model can be expressed in terms of traces of single-particle evolution operators governed by the dynamic Hubbard-Stratonovich fields. The corresponding Hubbard-Stratonovich dynamical systems are generally non-unitary, which yields a number of notable complications, including  break-down of the global exponential representation. Finally,   we derive Hubbard-Stratonovich dynamical systems for the Bose-Hubbard model and a quantum spin model and use the Lie-algebraic approach to obtain new non-perturbative dual descriptions of these theories. \end{abstract}

\maketitle

\section{Introduction}
\label{sec:Intro}
The fundamental problem of quantum-to-classical correspondence is an old question that has existed since the birth of the quantum
theory, and which appears frequently in essentially all fields of physics under various covers, in particular, in the context of the path integral approach to quantum mechanics of Feynman~\cite{Feynman_PI}, Glauber's coherent states~\cite{Glauber} and their various generalizations~\cite{Klauder_CS,Perelomov_CS,Gilmore_CS}, integrable systems~\cite{Thacker_RMP}, classical and quantum chaos~\cite{Gutzwiller_book}, AdS/CFT correspondence~\cite{AdSCFT}, and more general Langlands duality~\cite{Langlands}, etc. One particularly simple variation of this general problem is to connect classical motion of a particle or an ensemble of classical particles under the influence of external  time-dependent fields to  unitary evolution of the corresponding externally-driven quantum system. In many cases, including the text-book harmonic oscillator problem, the one-to-one correspondence between  classical and quantum dynamics has been long known and well-established. However, the situation remains less clear in some other cases, notably  in dynamical systems where classical motion occurs on a manifold with constraints, which on the quantum side usually manifest themselves through  topological terms in a quantum action or non-trivial commutation relations in the Hamiltonian formalism~\cite{altland_condensed_2010}. The simplest such examples are an externally driven spin~\cite{TLS_laser,TLS_Nakamura,TLS_Girard,TLS_Kou,TLS_Seifert,TLS_Mooji,Awschalom_spin} or a two-photon system with time-varying parameters~\cite{Gilmore_RMP}. Various quasiclassical approaches~\cite{solari_semiclassical_1987,kochetov_quasiclassical_1998,stone_semiclassical_2000,Semiclas_JC} have been proposed to analyze such systems, and the results of these analyses are often suggestive  of classical and quantum unitary dynamics representing two sides of the same coin, but no simple general summary of such correspondence appears to exist to encompass all relevant cases on a systematic basis. 

Apart from the pure fundamental interest to the problems of unitary quantum evolution and quantum-to-classical correspondence, the recent experimental advances in the fields of quantum optics, atomic and molecular physics~\cite{noneq_RMP,noneq1,noneq2,noneq3,noneq4,noneq5,noneq6,noneq7,noneq8,noneq9,noneq10,noneq11,NLGRVG,ARCGGR,AdSSC,TG_exp,HO_exp,Rigol_etal,Integrable_relax}, and quantum control~\cite{Qcontr1,Qcontr1.5,Qcontr2-,Qcontr2,Qcontr3,Qcontr3.5,Qcontr4} make these fundamental questions  also practically important and experimentally relevant. Models that had existed before only in theorist's imagination can now be realized and studied in the  laboratory, where unitary relaxationless dynamics can be directly detected and analyzed. On the theory side however, the analytical tools to study such non-equilibrium dynamics are still being actively developed~\cite{TP_Review,Keldysh_Rev,DAlessandro_book,Grit1,Grit2}.

Motivated by these recent advances and our recent theoretical results~\cite{VG_Qfl,Fermionization,BWZW,Our_TLS}, we use the latter as well as results from the old~\cite{Magnus,Wilcox,WeiNorm1,WeiNorm2} and more recent~\cite{Alhassid1,Alhassid2,Vourdas,ChemPhysRev,MagnusRev} mathematical physics literature and quantum control theory~\cite{DAlessandro_book} to develop a general Lie-algebraic framework that describes both unitary quantum dynamics and classical non-equilibrium evolution on an equal footing. We utilize the approach pioneered by Magnus~\cite{Magnus,MagnusRev} and derive a set of differential equations, referred to below as dual Schr{\"o}dinger-Bloch equations (DSBE), that make quantum-to-classical correspondence manifest. The major point emphasized throughout the paper is that the unitary quantum evolution is largely decoupled from the Hilbert space and can be analyzed completely independently of it based on the minimal Lie algebraic structure provided by the underlying Hamiltonian. More specifically, we first consider a Hamiltonian expressed in terms of operators that close under commutation into a finite-dimensional Lie algebra, ${\cal A}$, which is specified by its structure constants, $\left[\check{J}_a,\, \check{J}_b \right] = i \sum\limits_{c=1}^d f_{ab}^{\phantom{ab}c} \check{J}_c$. The unitary quantum evolution operator, $\check{U}(t)$ lies in the Lie group, $G_{\cal A} = \exp{(i {\cal A})}$, which is generated by the algebra via exponentiation and it is uniquely determined by a trajectory, ${\Phi}^a(t)$, in the dynamical group as follows $\check{U}(t) = \exp\left[ -i \sum\limits_a \Phi^a(t)  \check{J}_a \right]$. The classical equations of motion for the quantum-mechanical averages, $M_a(t) = \langle \Psi(t) | \hat{J}_a |\Psi(t) \rangle$ follow from the structure constants of the algebra, $\dot{M}_b(t) = \sum\limits_{a,c = 1}^d f_{ba}^{\phantom{ab}c}\, b^a(t) M_c(t)$. Classical motion generally takes place on a classical manifold with constraints, ${\cal M}$, via action of a symmetry group, $G({\cal M})$, in the $d$-dimensional adjoint  representation, ${\bf M}(t) = e^{-\sum_a \Phi^a(t) \hat{f}_a} {\bf M}(0)$, with $\hat{f}_a$ representing the corresponding generators. It is demonstrated that this group action is determined by the very same dual generators, ${\bm \Phi}(t)$, that appear in the quantum problem. This manifests quantum-to-classical correspondence. The dual generators, $\Phi_a(t)$, satisfy a system of non-linear differential equations, the DSBE. New results derived here include a specific form of these dual  Schr{\"o}dinger-Bloch equations for a number of dynamical systems.  

The second part of the paper focuses on a class of more complicated Hamiltonians, bilinear in algebra generators, that encompasses a wide variety of interacting many-body quantum systems of interest to both condensed matter physics and cold atoms, which include various quantum lattice spin models and Hubbard-type models. We utilize the path integral formalism to formulate a generalized Hubbard-Stratonovich decomposition~\cite{VG_Qfl} and to prove that thermodynamic properties of such many-body systems can be represented as a linear combination of traces of  quantum evolution operators  corresponding to all possible realizations of the Hubbard-Stratonovich fields. This approach gives rise to Hubbard-Stratonovich quantum dynamical systems, which  generally involve non-unitary evolution and require a complex extension of the underlying Lie algebra. It is argued and demonstrated on an explicit example of the familiar Bose-Hubbard model that in treating interacting many-body Hamiltonian, it is useful to Hubbard-Stratonovich decouple the terms associated with single-particle Lie algebra generators instead of focusing only on the interacting terms quartic in creation/annihilation operators in the Fock basis; i.e., the hopping terms, $b^\dagger(i) b(j)$, and interaction terms $b^\dagger(i) b(i)b^\dagger(j) b(j)$ represent the same level of complexity from this point of view and are to be decoupled into linear combinations of $b^\dagger(i)$, $b(i)$, and $b^\dagger(i) b(i)$, which together with the identity span the solvable four-dimensional harmonic oscillator algebra, $\mathfrak{h}_4$. We solve the corresponding Hubbard-Stratonovich dynamical system exactly and obtain a new non-perturbative representation for the partition function of the Bose-Hubbard model. We also analyze a frustrated quantum spin model and, using the dual Schr{\"o}dinger-Bloch equation, derive a dual representation of the theory in terms of group generators. It is argued that such dual descriptions may be helpful in circumventing the sign problem in certain cases.

Our paper is structured as follows: In Sec.~\ref{sec:InvariantForm}, we, starting from the standard Schr{\"o}dinger equation in a Hilbert space, derive a Hilbert-space-invariant form of the Schr{\"o}dinger equation for quantum evolution that relies only on the primitive Lie-algebraic structure of the dynamical system. In Sec.~\ref{sec:DualSBE}, we utilize the global exponentiation conjecture (that quantum evolution can be {\em globally} exponentiated from the algebra, $\check{U}(t) = \exp\left[ -i {\bm \Phi}(t) \cdot \check{\bm J} \right]$) and derive a general form of the dual Schr{\"o}dinger-Bloch equation for the Hilbert-space-invariant generators, ${\bm \Phi}(t)$.  Sec.~\ref{sec:DSBEExamples} provides a number of examples of the usage of the dual Schr{\"o}dinger-Bloch equations: Sec.~\ref{sec:DSBEh4} derives and solves DSBE for the harmonic oscillator algebra, Sec.~\ref{sec:DSBEsu2} derives DSBE for the $\mathfrak{su}(2)$  spin algebra, Sec.~\ref{sec:contra} establishes a relation between the latter two dynamical systems, Sec.~\ref{sec:DSBEh6} discusses the two-photon algebra, and Sec.~\ref{sec:DSBEsu(N)} formulates a generic five-step procedure of reverse-engineering exact solutions using DSBE in the higher-rank $\mathfrak{su}(N)$ dynamical systems. Sec.~\ref{sec:QCC} establishes  quantum-to-classical correspondence and Sec.~\ref{sec:JC}  illustrates the meaning of this correspondence on the simple example of  linearly-driven Jaynes-Cummings model that is solved exactly. Sec.~\ref{sec:many-body} extends the Lie-alegraic ideas to interacting many-particle systems: a generic Lie-algebraic lattice model is introduced in Sec.~\ref{sec:Lielattice} and Sec.~\ref{sec:GenHS} derives a generalized  Hubbard-Stratonovich transform and formulates a generic Hubbard-Stratonovich dynamical system. Sec.~\ref{sec:ExamplesLielatt} includes two examples of this Lie-algebraic approach
to interacting quantum models: In Sec.~\ref{sec:BH}, we solve exactly the Hubbard-Stratonovich dynamical system for the harmonic oscillator algebra to derive an exact, previously unknown, representation of the partition function of the Bose-Hubbard model. Sec.~\ref{sec:DualHS} formulates, on the example of a frustrated quantum spin model, a dual path-integral approach to interacting spin systems where the integration over Hubbard-Stratonovich fields is replaced with path-integration over the dual generators in the dynamical group. Sec.~\ref{sec:Summary}  provides a summary and discussion of many open questions. Appendices~\ref{sec:GenProp} and \ref{app:clasM} are logically connected to the first  ``single-particle'' part of the paper, namely they relate to Secs.~\ref{sec:DualSBE} and \ref{sec:DSBEExamples} and Sec.~\ref{sec:QCC}, correspondingly: Appendix~\ref{sec:GenProp} summarizes general algebraic properties and classification of  possible unitary quantum dynamics and appendix~\ref{app:clasM} reviews the construction of the classical manifold that plays a central role in the discussion of quantum-to-classical correspondence.

\section{From Schr{\"o}dinger Picture to a Hilbert-Space-Invariant Formulation}
\label{sec:InvariantForm}

\subsection{Dynamical Lie algebra}
\label{ss:dynLA}

Consider the non-equilibrium  Schr{\"o}dinger equation 
\begin{equation}
\label{SE1}
\left\{
\begin{array}{ll}
i \partial_t |\Psi(t) \rangle = \hat{\cal H}_L(t) |\Psi(t) \rangle \\
|\Psi(0) \rangle = |\psi_0 \rangle,
\end{array}
\right.
\end{equation}
where $\hat{\cal H}_L(t)$ is an $L \times L$ Hermitian Hamiltonian matrix acting on the wave-functions, $|\Psi(t) \rangle$,
in an $L$-dimensional complex Hilbert space ${\cal H}il(L)$ spanned by the linear combinations, $|\psi \rangle = \sum\limits_{k=1}^L c_k | k \rangle$, where $|k\rangle \in {\cal H}il(L)$ are orthonormal basis vectors
in the Hilbert space and $c_k \in \mathbb{C}$. $|\psi_0 \rangle = \sum\limits_{k=1}^L c_k^{(0)} | k \rangle \in {\cal H}il(L)$ is a normalized initial state. Note that the normalized wave-functions  lie on the hyper-surface of the $(2L-1)$-dimensional sphere, $S^{2L-1}$. In this Section, we assume that $L < \infty$ but the key results will be generalized to a specific class of 
infinite-dimensional representations in appendix~\ref{SSsec:Infinite_d}.

Any such dynamic Hamiltonian can be written in the form, $\hat{\cal H}_L(t) = \sum\limits_{i = 1}^m \beta^i(t) \hat{\cal O}_i,$ 
where $\hat{\cal O}_i$ are Hermitian linearly-independent matrices and $\beta^i(t)$ are time-dependent parameters. Also, assume, without loss of generality, that ${\rm Tr}\, \hat{\cal O}_i = 0$.
 If this is not so from the outset, we can always redefine our operators as follows $\hat{\cal O}_i \to \left( \hat{\cal O}_i - {\hat{I}_L \over L} {\rm Tr}\, \hat{\cal O}_i \right)$, where $\hat{I}_L$ is the $L \times L$ identity matrix, and trivially solve the Abelian part of the Schr{\"o}dinger equation that will give rise to a pure phase dynamics of the wave-function as follows, $|\Psi(t) \rangle \to \exp\left[ - {i \over L} \sum\limits_{i = 1}^m {\rm Tr}\, \hat{\cal O}_i \int\limits_0^t \beta^i(s) ds \right] |\Psi(t) \rangle$. All operators in the remaining part of the Hamiltonian will be traceless and therefore our assumption does not reduce the practical generality of the analysis. This implies that these traceless operators, $\hat{\cal O}_i$, certainly belong to an $L$-dimensional defining representation of the special linear algebra $\mathfrak{sl}(L,\mathbb{C})$ and the unitarity constraint further requires the operators to belong to the $L$-dimensional unitary representation of $\mathfrak{su}(L)$. 

However, the quantum dynamical system (\ref{SE1}) may generate a much smaller Lie algebra than the $(L^2 - 1)$-dimensional $\mathfrak{su}(L)$ (recall that $L$ is the dimensionality of the Hilbert space).  In fact, we argue below that this algebra, ${\cal A}$, often has little-to-no relation to a particular Hilbert space. Consider now all possible matrix commutators  $\hat{\cal O}_{ij} =-i \left[ \hat{\cal O}_i,\, \hat{\cal O}_j \right]$ and multiple commutators, $\hat{\cal O}_{i,jk} = -i \left[ \hat{\cal O}_i,\, \hat{\cal O}_{jk} \right]$, {\em etc.} that are guaranteed to eventually close into a finite-dimensional Lie algebra, ${\cal A} \subset \mathfrak{su}(L)$ (or more precisely into its $L$-dimensional representation, $T_L\left[{\cal A}\right]$). Note that  $m \leq d={\rm dim}\, {\cal A} \leq (L^2 - 1)$. 
 
Choose  a basis in $T_L\left[{\cal A}\right]$, i.e., a set of $d$ linearly-independent $(L \times L)$ Hermitian traceless matrices, $\hat{J}_a \in T_L\left[{\cal A}\right]$, which themselves are linear combinations of matrices $\hat{\cal O}_i$ in the original Hamiltonian and their  multiple commutators. These generators will satisfy a specific set of commutation relations, 
\begin{equation}
\label{sc}
-i \left[ \hat{J}_a,\, \hat{J}_b \right] = \sum\limits_{c = 1}^d f_{ab}^{\phantom{ab}c}  \hat{J}_c,
\end{equation} 
where the structure constants are guaranteed to be real, $f_{ab}^{\phantom{ab}c} \in\mathbb{R}$. Finally, we can re-write the matrix Hamiltonian in terms of the generators as follows
\begin{equation}
\label{HLsum}
\hat{\cal H}_L(t) = \sum\limits_{a=1}^d b^a(t) \hat{J}_a.
\end{equation}
Eqs.~~(\ref{sc}) and (\ref{HLsum}) summarize the fact that an {\em arbitrary} time-dependent Schr{\"o}dinger equation formulated in a {\em finite-dimensional Hilbert space} can be reduced to the Lie-algebraic form (see also, appendix~\ref{SSsec:Infinite_d} for a discussion of  certain theories in infinite-dimensional Hilbert spaces) .

\subsection{Exponential image and dynamical group}
Lie algebras usually appear as a means to study Lie groups, and are often considered by-products
of the latter. In the context of non-equilibrium quantum mechanics however, a reverse view is much
more useful~\cite{DAlessandro_book}. Below, we treat the Lie algebra, ${\cal A}$, arising from the underlying Hamiltonian as a primary structure, 
while the Lie group it generates, $G_{\cal A} = \exp (i {\cal A})$, as secondary. A particular Hilbert space will be a tertiary
structure of minor or no importance. To proceed with this program, we use structure constants in Eq.~(\ref{sc}) to define an 
{\em abstract Lie algebra}, ${\cal A}$, spanned by the generators, $\check{J}_a$,
\begin{equation}
\label{comm}
\left[ \check{J}_a,\, \check{J}_b \right] \equiv {\rm ad}_{\check{J}_a} \check{J}_b = i f_{ab}^{\phantom{ab}c} \check{J}_c,
\end{equation}
where here and below a sum over repeating indices is assumed, unless stated otherwise. Here we put inverse hats on top of the {\em abstract generators}, $\check{J}_a$, of the {\em abstract algebra}, ${\cal A} = {\rm span}\, \left\{ \check{J}_1,\ldots, \check{J}_d \right\}$, to distinguish them from their specific matrix representation arising from the conventional Schr{\"o}dinger formulation in the Hilbert space, ${\cal H}il(L)$. We note that the word, ``abstract,'' here implies that $\check{J}$'s are not matrix objects, nor are they operators acting on a wave-function, but rather vectors in a linear space equipped with the commutator that makes it into a Lie algebra. It also implies that the products, $\check{J}_a \check{J}_b$, $\check{J}_a^3$, {\em etc.} do not make sense in the abstract Lie algebra, but only commutators do. The products are of course well-defined given a representation but normally have little relation to the underlying dynamical system. E.g., the square of an $\mathfrak{su}(2)$-generator from its two-dimensional representation (i.e., half-integer spin described by the Pauli matrices) is $\hat{\sigma}_{x,y,z}^2/4 = \hat{I}_2/4$ and is proportional to an identity matrix that commutes with any other observable. On the other hand, the square of a generator from any higher-dimensional representation of $\mathfrak{su}(2)$ (e.g., spin-$1$ particle described by $3 \times 3$ matrices) is non-trivial and as such may in principle have non-trivial dynamics. In what follows, we  mostly concentrate on Lie-algebraic invariants, which are defined as objects or statements, whose definition or proof relies only on the commutators of the generators in the dynamic Lie algebra and their linear combinations and does not involve any other algebraic structures.  

However, a particular faithful representation can be used as a tool to prove certain Lie-algebraic-invariant statements and identities and demonstrate 
that they are in fact independent of the representation itself. The Hadamard lemma (think, unitary rotation of an operator into the Heisenberg or interaction picture),
\begin{equation}
\label{Hadamard}
e^{-i \check{X}}  \check{Y} e^{+i \check{X}} = e^{-i {\rm ad}_{\check{X}}} \check{Y} \in {\cal A},
\end{equation}
and  the Baker-Campbell-Hausdorff (BCH) relation for $e^{i\check{X}} e^{i \check{Y}} = e^{i \check{Z}}$ (think, terms in a time-ordered exponential),
\begin{equation}
\label{BCH}
\check{Z} =  \check{X} + \int\limits_0^1 ds \psi \left( e^{i {\rm ad}_{\check{X}}} 
e^{is\, {\rm ad}_{\check{Y}}} \right)  \check{Y},
\end{equation}
stand out as perhaps the most important among such invariant identities [in Eqs.~(\ref{Hadamard}) and (\ref{BCH}), we  assumed that $\check{X},\check{Y} \in \cal{A}$ and used the logarithmic derivative of the Gamma-function, defined via its series, e.g., $\psi(e^z) = \sum\limits_{n=0}^\infty {B_n \over n!}  z^n$, with $B_n$ being the Bernoulli numbers]. 

 While the usefulness of Eq.~(\ref{BCH}) for practical calculations appears to be limited, it does provide a nice 
 illustration of the invariance of the BCH-product. Let us recall however that in the case of an arbitrary finite-dimensional Lie algebra, the BCH identity is convergent in a close vicinity of the origin, $\check{0} \in {\cal A}$, but Eq.~(\ref{BCH})  does not always converge globally~\cite{Suzuki}, with the algebra as simple as $\mathfrak{sl}(2,\mathbb{C})$ being the classic counter-example. Another manifestation of this problem is that there is generally no guarantee that in a particular matrix representation a product of two matrix exponentials of algebra elements can be written as a single  matrix exponential of an algebra element. This ``complication'' is also closely tied to the fact that a Lie algebra is guaranteed to exponentiate  into its Lie group only locally, but does not always do so globally. 

One can argue however that in the context of quantum dynamical system (\ref{SE1}), the dynamic algebra does  exponentiate onto a Lie group.  Here we recall that  the covering problem of global exponentiation of  an arbitrary finite-dimensional Lie algebra arises due to non-compact generators that do not necessarily form a subgroup~\cite{Gilmore_book}. However, such generators are prohibited in the quantum dynamical system described
by Hermitian finite-dimensional matrices. The very existence of a finite-dimensional Hilbert space guarantees that  the set of matrix exponentials and their products, form a subset of $SU(L)$ group, which is indeed compact. This suggests that the entire dynamic group is a compact manifold that can be globally generated by an exponential map. An argument here is that any two points of this compact Riemannian manifold are connected by a minimizing geodesics and the exponentials from the algebra generate such geodesics. Therefore, it seems that we can safely define a compact abstract group, generated by the abstract dynamic algebra via exponentiation that we denote as
\begin{equation}
\label{GA}
G_{\cal A} = e^{i {\cal A}}.
\end{equation}
We emphasize that while this dynamical symmetry group may be a subgroup of $SU(L)$, it generally has little to do with a particular Hilbert space, which we have used here only to argue in favor of the existence of a global exponential map onto the Lie group, $G_{\cal A}$. We will refer to the assumption about the existence of this map as to ``global exponentiation conjecture'' (we call it a ``conjecture,'' because the mathematical rigor of the above arguments is not entirely satisfactory and the background of the author is insufficient to be able to formulate a general statement in the form of a rigorous theorem). Finally, let us stress that this conjecture does {\em not}  assume that the specific form of BCH formula (\ref{BCH}) and/or Magnus series~\cite{Magnus,MagnusRev} are globally convergent, but it does imply, in particular, the following: Given a faithful matrix representation, the matrix product of two exponentials of algebra elements gives rise to an exponential of another algebra element, i.e., if $\hat{X},\, \hat{Y} \in T_L[{\cal A}]$, then, $-i \ln ( e^{i \hat{X}} e^{i \hat{Y}} ) \in T_L[{\cal A}]$.

\subsection{Geometric Alternative to Unitary Evolution in a Hilbert space}
\label{ss:geom}

Returning to our quantum dynamical system~(\ref{SE1}), we observe however, that since the familiar Schr{\"o}dinger equation is written in terms of a wave-function, it is manifestly representation-specific. Let us now  cast it into a representation-invariant form. In the framework of the Schr{\"o}dinger formulation~(\ref{SE1}), our goal is to find a norm-preserving solution governed by a unitary $L \times L$ evolution matrix as follows
\begin{equation}
\label{solSE}
|\Psi(t) \rangle = \hat{U}_L(t) |\psi_0\rangle = \sum\limits_{k,p=1}^L U_k^{\phantom{k}p}(t) c_p^{(0)} |k\rangle.
\end{equation}
We can now substitute this solution in the Schr{\"o}dinger equation to arrive to
\begin{equation}
\label{SE2}
\left\{
\begin{array}{ll}
i \partial_t \hat{U}_L(t) = \hat{\cal H}_L(t) \hat{U}_L(t) \\
\hat{U}_L(0) = \hat{I}_L.
\end{array}
\right.
\end{equation}

\begin{figure}
\begin{center}
\includegraphics[width=0.5\textwidth]{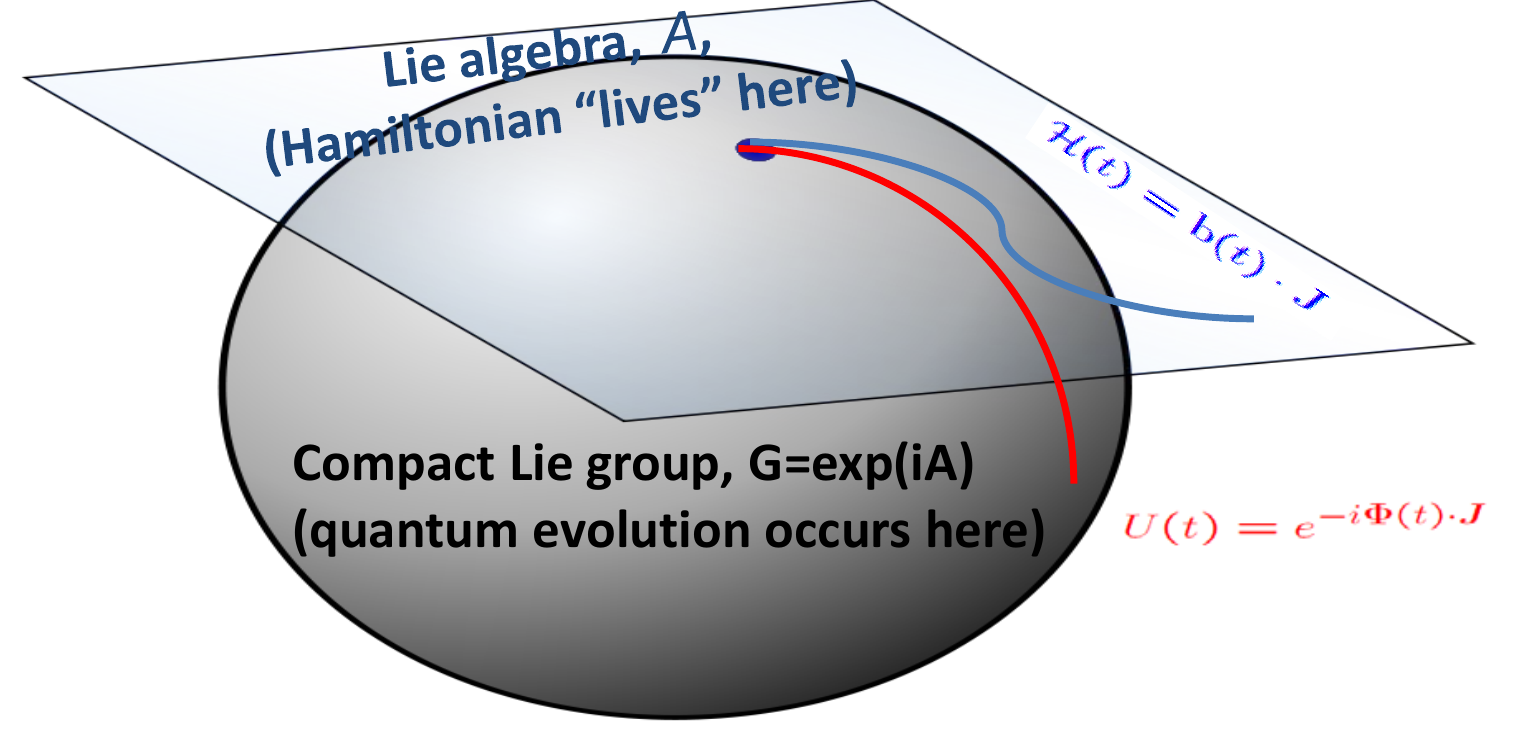}\\
\caption{Cartoon illustration of the geometrical view of quantum dynamics: The Hamiltonian is a trajectory in a Lie algebra, ${\cal A}$, which 
is a finite-dimensional vector space. This trajectory gives rise to unitary evolution, which is a trajectory in the Lie group, $G_{\cal A} = e^{i{\cal A}}$, generated by the algebra via exponentiation. Topologically, this group is a  compact connected manifold. As emphasized in the text, neither ${\cal A}$ nor $G_{\cal A}$ are associated with any particular Hilbert space.}
\label{fig:A->G}
\end{center}
\end{figure}

As evident from (\ref{SE2}), the quantum evolution operator is decoupled from the initial condition, which is well-known. What is also known~\cite{Gilmore_book}, but perhaps less widely appreciated, is that the unitary evolution is also decoupled from the Hilbert space and is representation-invariant. To see this, multiply (\ref{SE2}) by $\hat{U}_L^{-1}(t) \equiv \hat{U}_L^\dagger(t)$ from the right to find $i \partial_t \hat{U}_L(t) \hat{U}_L^\dagger(t) = \hat{\cal H}_L(t)$. Based on the global exponentiation conjecture of the previous section, we conclude that since at $t=0$ the identity, $\hat{U}_L(0) = \hat{I}_L$, certainly belongs an $L$-dimensional representation of the group,  $T_L\left[G_{\cal A}\right]$, it will stay there at all times.  The same conjecture also implies that the solution can be sought in the form~\cite{Magnus}, $\hat{U}_L(t) = \exp \left[ - i { \Phi}^a(t) \hat{J}_a\right]$, where ${ \Phi}^a(t)$ are real functions of time [we will also use the vector notation, ${\bm \Phi} = \left(\Phi^1,\Phi^2,\ldots,\Phi^d \right)$, $\hat{\bm J} = \left(\hat{J}_1,\hat{J}_2,\ldots,\hat{J}_d \right)^{\rm\bm T}$ and a dot-product
 ${\bm \Phi}(t) \cdot \hat{\bf J} \equiv { \Phi}^a(t) \hat{J}_a$]. Now we observe that the product, $i \partial_t \hat{U}_L(t) \hat{U}_L^\dagger(t)$, is representation-invariant, because it can be calculated using the BCH-identity only. Therefore, one can drop the representation indices and write the Schr{\"o}dinger equation in an ``abstract form:''
\begin{equation}
\label{SE3}
\left\{
\begin{array}{ll}
i \partial_t \check{U}(t) \check{U}^{-1}(t) = i \left[\partial_t  e^{-i {\bm \Phi}(t) \cdot \check{\bm J} } \right]  e^{i {\bm \Phi}(t) \cdot \check{\bm J} }  = b^a(t) \check{J}_a \\
\\
\check{U}(t) \in G_{\cal A}\,\, \mbox{and}\,\, \check{U}(0)=\check{1} \equiv e^{i {0}}.
\end{array}
\right.
\end{equation}
Note that eventhough, we have ``derived'' this equation from the textbook Schr{\"o}dinger equation (\ref{SE1}), Eq.~(\ref{SE3}) has no ``recollection'' about the Hilbert space and can be analyzed completely independently of it. In fact, we might have as well started
from Eq.~(\ref{SE3}), which does not involve the notion of a wave-function at all.  In this approach, the Hamiltonian
\begin{equation}
\label{HinA}
\check{H}(t) = {\bf b}(t) \cdot \check{\bm J}
\end{equation}
 is a trajectory in the abstract Lie algebra, ${\cal A}$, while the evolution operator is a trajectory in the abstract
Lie group, $\check{U}(t) \subset G_{\cal A} = \exp\left[ i {\cal A} \right]$. To solve the quantum-dynamical system (\ref{SE3}) implies to find the latter from the former. Fig.~\ref{fig:A->G} illustrates this geometric view of quantum dynamics.

\section{Dual Schr{\"o}dinger-Bloch Equation}
\label{sec:DualSBE}
The evolution operator in Eq.~(\ref{SE3}) is a trajectory in the Lie-group, generated by the Lie-algebra of the driving Hamiltonian, and therefore has the form
\begin{equation}
\label{S&S+}
\check{U}(t) = \exp \left[ - i { \Phi}^a(t)  \check{J}_a \right]\,\, \mbox{and}\,\,
\check{U}^{-1}(t) = \exp \left[ + i { \Phi}^a(t)  \check{J}_a \right].
\end{equation}
We note that since the vector, ${\bm \Phi}(t)$, defines the group, there appears a problem of its proper parameterization that we will discuss at a later stage (see, Sec.~\ref{ssec:param}).  However, we shall not dwell on this subtlety for the moment and treat the corresponding differential equation (\ref{SE3}) rather freely, keeping in mind that there exists a finite-dimensional matrix representation that always can be used to justify the corresponding algebraic manipulations. 

 Eqs.~(\ref{SE3}) and (\ref{S&S+}) lead to the following structure, $\check{H}(t)  = i\lim\limits_{\epsilon \to 0} \left\{ {1 \over \epsilon} \left( e^{-i \left[\Phi^a(t) + \epsilon \dot{\Phi}^a(t) \right]\check{J}_a} e^{i \Phi^a(t)\check{J}_a} - e^{i 0} \right) \right\}$,
which had been considered before by many, including notably Feynman (see, e.g., Ref.~[\onlinecite{Wilcox}] for a related discussion). Using the Hadamard lemma (\ref{Hadamard}) and the related operator/matrix identity, ${\partial \over \partial \epsilon} e^{\check{X} + \epsilon{Y}} \Bigr|_{\epsilon \to 0} = \int_0^1 ds e^{(1-s)\check{X}} \check{Y} e^{s\check{X}}$, we obtain
\begin{equation}
\label{dSE'}
\left[ \dot{\Phi}^a(t)  \int\limits_0^1 ds e^{ - is \Phi^b(t) {\rm ad}_{\check{J}_b} } - b^a(t) \right] \check{J}_a = \check{0},
\end{equation}
which is to be supplemented by the initial condition, ${\bm \Phi}(t) = {\bf 0}$. We see that Eq.~(\ref{dSE'}) involves nothing but commutators and therefore is representation invariant in the sense that the vector
function, ${\bm \Phi}(t)$, determines evolution operators in all Hilbert spaces.
 
The operators, $\left( {\rm ad}_{{\bm \Phi} \cdot \check{\bm J}} \right)$, naturally give rise to a $d$-dimensional matrix representation of ${\cal A}$, which is known as the adjoint representation of the algebra, and it is  uniquely determined by the structure constants. The representation is constructed in a standard way by assigning to each vector ${\bm \Phi}$, the following $(d \times d)$-matrix, $F_b^{\phantom{b}c}[{\bm \Phi}]= {\bm \Phi} \cdot {\bm f}_b^{\phantom{b}c}$,
\begin{equation}
\label{regrep}
-i\left( {\rm ad}_{{\bm \Phi} \cdot \check{\bm J}} \right) \check{J}_b = {\bm \Phi} \cdot {\bm f}_b^{\phantom{b}c} \check{J}_c. 
\end{equation}
The matrices, $\hat{f}_1,\ldots \hat{f}_d$  correspond to the adjoint representation of algebra generators $\check{J}_1,\ldots,\check{J}_d$. Notationwise, here and below, the hat on top of the $f$-matrices signifies the fact that they are indeed
matrices (rather than abstract objects) and the bold fonts correspond to $d$-dimensional vectors, e.g.,  $\hat{\bm f} = \left(\hat{f}_1,\hat{f}_2,\ldots,\hat{f}_d \right)^{\rm \bm T}$. Note also that if we go back to the matrix elements, $ f_{ab}^{\phantom{ab}c}$ we  recover the structure constants of the algebra by construction. Therefore, the adjoint representation is basically a set of $d \times d$ matrices built directly from the structure constants. Using these matrix notations, we can write Eq.~(\ref{dSE'}) in the following compact form
\begin{equation}
\label{dSE}
\int\limits_0^1 ds e^{s {\bm \Phi}(t) \cdot \hat{\bm f} }\, \dot{\bm \Phi}(t) = {\bf b}(t),\,\,\,\, {\bm \Phi}(0) = {\bm 0}
\end{equation}
This is the main equation used throughout the paper. We will refer to it as {\em the dual Schr{\"o}dinger-Bloch equation (DSBE)}, and use it for actual calculations of quantum and classical dynamics. Let us emphasize two important properties of the DSBE, which will be elaborated upon in the following sections: (i)~Eq.~(\ref{dSE}) has no ``recollection'' of the original Hilbert space and in its explicit form does not contain the Planck constant; (ii)~DSBE is purely real, i.e., $\hat{\bm f}$ are real-valued matrices and the sought-after functions ${\bm \Phi}(t)$ are real as well. 

A word of caution is in order: Since any matrix, including $({\bm \Phi} \cdot \hat{\bm f})$ in Eq.~(\ref{dSE}), commutes with itself, it is tempting to calculate the integral over $s$ in (\ref{dSE}) and rewrite it terms of the inverse matrix and its exponential, c.f., Refs.~[\onlinecite{Alhassid1},~\onlinecite{Alhassid2}]. However, one should avoid doing so and exercise care at this point, because  the adjoint-representation matrix does not necessarily have an inverse, i.e., $\hat{F}^{-1}[{\bm \Phi}]=\infty$. We emphasize that this is not a minor point, but an essential complication, which should be kept in mind when analyzing the DSBE. A correct approach is to calculate the matrix exponential, $e^{s {\bm \Phi}(t) \cdot \hat{\bm f} }$, first, and  evaluate the integral over $s$ in Eq.~(\ref{dSE}) only afterwards (reversing the order of these operations may lead to errors in actual calculations).

Concluding this section, let us also recall that the adjoint representation defines the Cartan-Killing inner product in the algebra,
\begin{equation}
\label{CK}
\left( {\bm \Phi},\, {\bm \Theta} \right) = {\rm Tr}\, \left\{ \left( {\bm \Phi} \cdot \hat{\bm f} \right) \, 
\left( {\bm \Theta} \cdot \hat{\bm f} \right) \right\},
\end{equation}
which is used as a tool to classify the algebras~\cite{Gilmore_book}. This Cartan-Killing product is non-singular for semisimple algebras and in this case the Cartan-Killing product of basis vectors, can be identified with the metric that gives rise to geometric structures.

\section{Examples of the dual Schr{\"o}dinger-Bloch equation}
\label{sec:DSBEExamples}

As argued in appendix~\ref{sec:GenProp}, the variety of {\em all possible} quantum dynamical systems formulated in a finite-dimensional Hilbert space and certain theories formulated in an infinite-dimensional Hilbert space (see appendix~\ref{SSsec:Infinite_d}), reduce to an analysis of dynamical systems in universal simple algebras (for practical purposes, understanding driven $\mathfrak{su}(L)$ systems should be sufficient) and  dynamics in non-universal solvable algebras (that are argued to be comparatively trivial and always exactly solvable via factorization). In this section, we demonstrate the application of the DSBE for both solvable (harmonic oscillator) and simple [$\mathfrak{su}(L)$] algebras.

\subsection{DSBE for the Weyl-Heisenberg algebra, $\mathfrak{h}_4$}
\label{sec:DSBEh4}
Consider the four-dimensional Weyl-Heisenberg algebra $\mathfrak{h}_4 = {\rm span}\, \left\{ \check{1}, \check{a}, \check{a}^\dagger, \check{a}^\dagger \check{a} \right\}
= {\rm span}\, \left\{ \check{1}, \check{x}, \check{p}, \check{a}^\dagger \check{a} \right\}$, associated with a driven harmonic oscillator with the Hamiltonian
\begin{equation}
\label{HO}
\check{\cal H}_{\mathfrak{h}_4}(t) = \omega(t) \check{a}^\dagger \check{a} +  \alpha^*(t) \check{a}^\dagger + \alpha(t) \check{a} + b^0(t)\check{1},
\end{equation}
where the operators $\check{a} \equiv \check{a}^-$ and $\check{a}^\dagger \equiv \check{a}^+$  
satisfy the canonic commutation relations, $\left[ \check{a},\,\check{a}^\dagger\right] = \check{1}$ 
and $\left[ \check{a}^\dagger\check{a},\,\check{a}^{\pm}\right] = \pm \check{a}^\pm$ and $\check{x} = (\check{a} + \check{a}^\dagger)/\sqrt{2}$
and $\check{p} = i( \check{a}^\dagger- \check{a} )/\sqrt{2}$ satisfy $\left[\check{x},\, \check{p}\right] = i \check{1}$. Want to find the solution to Eq.~(\ref{SE3}) corresponding to (\ref{HO}). To illustrate  an application of the general factorization scheme outlined in appendix~\ref{sec:Fact}, let us decompose the solvable algebra $\mathfrak{h}_4$ into an Abelian and nilpotent components as follows
\begin{equation}
\label{h4}
\mathfrak{h}_4 = {\rm span}\left\{\check{a}^\dagger \check{a}\right\} + {\rm span}\left\{\check{1}, \check{x}, \check{p} \right\} 
= \mathfrak{u}(1) + \mathfrak{h}_3,
\end{equation}
where the latter, $\mathfrak{h}_3$, is a nilpotent subalgebra. Solve the other ${\mathfrak{u}(1)}$-component by introducing the evolution matrix
$\check{U}_{\mathfrak{u}(1)}(t) = \exp { \left[ - i \check{a}^\dagger \check{a} \int\limits_0^t \omega(t') dt' \right]}$. The full evolution operator is given by the product 
$\check{U}_{\mathfrak{h}_4} = \check{U}_{\mathfrak{u}(1)}(t) \check{U}_{\mathfrak{h}_3}(t)$, where the last factor is sought in the form
\begin{equation}
\label{Sh3}
\check{U}_{\mathfrak{h}_3} (t) = \exp\left\{ - i \left[ \Phi^0(t) \check{1} + \Phi^1(t) \check{x} + \Phi^2(t) \check{p} \right] \right\}
\end{equation}
and is to satisfy the Schr{\"o}dinger equation with the rotated Hamiltonian,
\begin{equation}
\label{Hh3}
\check{\cal H}_{\mathfrak{h}_3}(t) = b^0(t) \check{1} + b^1(t) \check{x} + b^2(t) \check{p},
\end{equation} 
where $b^1(t) + i b^2(t) = {1 \over \sqrt{2}} \alpha(t) e^{i \int\limits_0^t \omega(t') dt'}$ (recall that $\exp\left[ - i \Omega {\rm ad}_{\check{a}^\dagger \check{a}} \right] \check{a}^{\pm} = e^{\mp i \Omega} \check{a}^\pm$). Using the regular representation of $\mathfrak{h}_3$,  we can present the operator (\ref{regrep}) in the DSBE (\ref{dSE}) as follows:
\begin{equation}
\label{Fh3}
\hat{F}_{\mathfrak{h}_4}[{\bm \Phi}(t)] \equiv {\bm \Phi}(t) \cdot \hat{\bm f} = \left(
\begin{array}{ccc}
0 & 0 & 0 \\
\Phi^2(t) & 0 & 0\\
- \Phi^1(t)  &0  & 0\\
\end{array}
\right),
\end{equation} 
 Due to the nilpotent structure of the algebra, the matrix (\ref{Fh3}) exponentiates easily, leaving behind only two non-vanishing terms in the series, and yields $\int\limits_0^1 ds e^{s{\bm \Phi}(t) \cdot \hat{\bm f}} = {\small \left(
\begin{array}{ccc}
1 & 0 & 0 \\
\Phi^2(t)/2 & 1 & 0\\
-\Phi^1(t)/2  &0  & 1\\
\end{array}
\right)}$. The dual Schr{\"o}dinger-Bloch equation therefore reduces to 
\begin{equation}
\label{dSEh312}
\dot{\Phi}^{1,2}(t) = b^{1,2}(t)
\end{equation}
and 
\begin{equation}
\label{dSEh30}
\dot{\Phi}^{0}(t) + {1 \over 2} \left[ \dot{\Phi}^{1}(t){\Phi}^{2}(t) - \dot{\Phi}^{2}(t){\Phi}^{1}(t)\right] = b^{0}(t).
\end{equation}
Hence, we find the exact quantum evolution as follows:
\begin{equation}
\label{Phi12h3}
\Phi^{1}(t) + i \Phi^{2}(t) = {1 \over \sqrt{2}} \int\limits_0^t dt' \alpha(t')  e^{i \int\limits_0^{t'} \omega(t'') dt''}
\end{equation}
and 
\begin{eqnarray}
\nonumber
\Phi^{0}(t) = &&\!\!\!\!\! \int\limits_0^t dt' b^0(t') - {1 \over 2} \int\limits_0^t dt' \int\limits_0^{t'} dt'' 
\left| \alpha(t') \right| \left| \alpha(t'') \right|\\
&& \!\!\!\!\! \times
\sin \left\{ \int\limits_{t''}^{t'} d\tau\left[ \omega(\tau)  + \phi(t') - \phi(t'') \right] \right\},
\label{Phi0h3}
\end{eqnarray}
where $\alpha(t) = \left|\alpha(t) \right| e^{i \phi(t)}$. 

Finally, note that the phase space associated with the harmonic oscillator algebra is not compact, which seems to be in conflict 
with the arguments of Secs.~\ref{ss:dynLA} and \ref{ss:geom} (see also Fig.~\ref{fig:A->G}), which argued that the dynamical group must be compact. The resolution of the paradox is in that the harmonic oscillator algebra does not have a finite-dimensional unitary representation and hence no finite-dimensional Hilbert space exists, which was a key assumption of Sec.~\ref{ss:dynLA}. On the other hand, the exponential representation still works perfectly well (see also, Sec.~\ref{sec:contra} and appendix~\ref{SSsec:Infinite_d}).

\subsection{DSBE for the $\mathfrak{su}(2)$-algebra}
\label{sec:DSBEsu2}
Now consider the simple algebra, $\mathfrak{su}(2)$, defined via the commutation relations
\begin{equation}
\label{su(2)comm}
\left[ \check{J}_\alpha, \, \check{J}_\beta \right] = i \varepsilon_{\alpha \beta \gamma} \check{J}_\gamma,
\end{equation}
where the Greek indices run over $1,2,3 \equiv x,y,z$ and here and below in this section we make no distinction between
the covariant and contravariant indices. We are interested in spin dynamics governed by the Hamiltonian, 
\begin{equation}
\label{Hsu(2)}
\check{\cal H}_{\mathfrak{su}(2)} (t) = {\bf b}(t) \cdot \check{\bm J}.
\end{equation}
We follow the general prescription outlined above and also in Ref.~[\onlinecite{Our_TLS}], where the $\mathfrak{su}(2)$ case was considered in the context of two-level system dynamics, and seek the solution in the form $\check{U}_{\mathfrak{su}(2)} = \exp \left[ - i {\bm \Phi}(t) \cdot 
\check{\bm J} \right]$. The adjoint operator in the DSBE (\ref{dSE}), $\left( - i {\rm ad}_{{\bm \Phi}(t) \cdot \check{\bf J}} \right)$, has a regular matrix representation as follows
\begin{equation}
\label{adsu(2)}
\hat{F}_{\mathfrak{su}(2)}[{\bm \Phi}(t)] \equiv {\bm \Phi} \cdot \hat{\bm f} =
\left(
\begin{array}{ccc}
0 & -\Phi_z &\Phi_y\\
\Phi_z & 0 & -\Phi_x\\
-\Phi_y & \Phi_x & 0\\
\end{array}
\right),
\end{equation}
and describes a generator of $SO(3)$-rotations. The exponential of this operator is therefore a finite rotation around the axis 
${\bf n} = {\bm \Phi}/\Phi$, and the exponential from Eq.~(\ref{dSE}) reads
\begin{equation}
\label{expadsu(2)}
e^{ s {\bm \Phi} \cdot \hat{\bm f} } = \delta_{\alpha \beta} \cos{ s\Phi } +
n_\alpha n_\beta \left[1 - \cos{ s \Phi} \right] - \varepsilon_{\alpha \beta \gamma} n_\gamma \sin{s\Phi }.
\end{equation}

We now present the time-derivative as follows $\dot{\bm \Phi} = \dot{\Phi} {\bf n} + {\Phi} \dot{\bf n}$, and keeping in mind that $\dot{\bf n} \cdot {\bf n} = 0$, obtain the DSBE for the $\mathfrak{su}(2)$ algebra
\begin{equation}
\label{DSEsu(2)}
\dot{\Phi}\, {\bf n} + \sin{\Phi}\, \dot{\bf n} + \left(1 - \cos{\Phi} \right) \left[ {\bf n} \times \dot{\bf n} \right] = {\bf b}(t).
\end{equation}
This equation is not formally solvable for a generic time-dependence of the driving field. However, it could be used to prove certain exact relations [such as, e.g., $\Phi(t) = \int_0^t d\tau {\bf b}(\tau) \cdot {\bm n}(\tau)$] and generate exactly solvable
models for two-level-system dynamics~\cite{TLS_laser,TLS_Nakamura,TLS_Girard,TLS_Kou,TLS_Seifert,TLS_Mooji,Awschalom_spin}  by inverting the trajectories in the SU(2) group back onto the algebra.

\subsection{Contracting the DSBE from the $\mathfrak{u}(2)$ algebra into Weyl-Heisenberg}
\label{sec:contra}
The harmonic oscillator algebra and spin algebra as well as their dual Schr{\"o}dinger-Bloch equations discussed above appear to be completely different: driven dynamics in $\mathfrak{h}_4$  is solvable, while that in $\mathfrak{su}(2)$ is not exactly-solvable; the former algebra is associated with the non-compact Euclidean classical phase space $(x,p) \in \mathbb{R}^2$, while the latter with the magnetization dynamics on the Bloch sphere, ${\bm M} \in S^2$. However, despite these drastic distinctions, the two algebras are known to be closely connected and this connection beautifully manifests itself via Lie-algebraic contraction. We recap the main idea of this particular contraction~\cite{Gilmore_book}, because such procedures generally provide an effective means to reduce Schr{\"o}dinger-Bloch duals of an algebra to the corresponding  equations in more primitive algebras that
descend from it via contraction. 

Consider the four-dimensional $\mathfrak{u}(2)$ algebra that is obtained from $\mathfrak{su}(2)$ by adding to it a one-dimensional Abelian component,
\begin{equation}
\label{u(2)}
\mathfrak{u}(2) = {\rm span}\, \left\{ \check{J}_0, \check{\bm J} \right\} = \mathfrak{u}(1) + \mathfrak{su}(2),
\end{equation}
where $\left[ \check{J}_0,\, \check{J}_\alpha \right] = {0}$. Define a parameter-dependent linear transform of the $\mathfrak{u}(2)$ generators: $\check{\cal J}_0 = \check{J}_0$, $\check{\cal J}_{1,2} (\epsilon) = \epsilon \check{J}_{y,x}$, and $\check{\cal J}_3 (\epsilon) = \check{J}_z + \epsilon^{-2} \check{J}_0$. 
The non-trivial commutation relations in terms of the new operators are $\left[ \check{\cal J}_3,\, \check{\cal J}_{1,2} \right] = \mp i \check{\cal J}_{2,1}$ and $\left[ \check{\cal J}_1,\, \check{\cal J}_{2} \right] = - i \epsilon^2 \check{\cal J}_3 + i \check{\cal J}_0$. In the limit of $\epsilon \to 0$, these commutation relations become identical to those in the Weyl-Heisenberg algebra, $\mathfrak{h}_4$. I.e.,
\begin{eqnarray}
\nonumber
\lim\limits_{\epsilon \to 0}\, \mathfrak{u}_\epsilon(2) &\equiv& \lim\limits_{\epsilon \to 0}\, {\rm span}\, \left\{ \check{\cal J}_0, \check{\cal J}_1(\epsilon), \check{\cal J}_2(\epsilon), \check{\cal J}_3(\epsilon) \right\}\\
&=& {\rm span}\, \left\{ \check{1}, \check{x}, \check{p}, \check{a}^\dagger \check{a} \right\} = \mathfrak{h}_4.
\label{contraction}
\end{eqnarray}

Note that the DSBE for $\mathfrak{u}(2)$ follow from Eq.~(\ref{DSEsu(2)}) by adding to it a trivial equation for the Abelian $\mathfrak{u}(1)$ component. Let us however not solve it but instead re-write the ``trajectory'' in the algebra in terms of the 
new operators $\Phi^a(t) \check{J}_a = \Theta^a(t) \check{\cal J}_a$, which leads to $\Theta^0 = \Phi^0 - \epsilon^{-2} \Phi^z$, $\Theta^1 = \Phi^y/\epsilon$, $\Theta^1 = \Phi^y/\epsilon$, and $\Theta^3 = \Phi^3$. The corresponding substitutions lead to the evolution operator, $S_{\mathfrak{u}_\epsilon(2)} = \exp \left[ - i \sum\limits_{a=0}^3\Theta^a(t) \check{\cal J}_a \right]$ and new set of equations for $\Theta^a(t)$ that in the $\epsilon \to 0$ limit crossover to the DSBE for $\mathfrak{h}_4$. 

\subsection{Six-dimensional two-photon algebra, $\mathfrak{h}_6$}
\label{sec:DSBEh6}
The two-photon operator algebra, $\mathfrak{h}_6 = {\rm span}\left\{ \check{1}, \check{a}, \check{a}^\dagger, \check{a}^\dagger \check{a}, \check{a}^2, (\check{a}^\dagger)^2 \right\}$, is discussed in detail in the review by Gimlore {\em et al.}~\cite{Gilmore_RMP} Hence, we just reiterate the main results in a slightly different form relevant to the preceding discussion. $\mathfrak{h}_6$ is not semisimple, but in accordance with the general rule, it can be decomposed into a simple component and a nilpotent component as follows
\begin{eqnarray}
\nonumber
\mathfrak{h}_6 = && {\rm span} \left\{  {1 \over 4} \left[(\check{a}^\dagger)^2 + \check{a}^2 \right],  {i \over 4} \left[(\check{a}^\dagger)^2 
- \check{a}^2 \right], {1 \over 2}\check{a}^\dagger \check{a} \right\} \\
&+& {\rm span} \left\{ \check{1}, \check{x}, \check{p} \right\} = \mathfrak{su}(2) + \mathfrak{h}_3.
\label{h6}
\end{eqnarray}
Using the canonic commutation relations, $\left[\check{a},\, \check{a}^\dagger \right] = \check{1}$, we can verify that the first three generators in Eq.~(\ref{h6}) give way to the  commutation relations identical to those of the ``usual'' spin  $\check{J}_{x,y,z}$ operators~(\ref{su(2)comm}). We can also explicitly verify that $\left[\mathfrak{su}(2),\, \mathfrak{h}_3\right] \subset \mathfrak{h}_3$ in (\ref{h6}). Hence, we can utilize the DSBE Eq.~(\ref{DSEsu(2)}) to determine the dynamics of the first component and the factorization trick (\ref{Stot}) or/and the DSBE for the nilpotent component (\ref{dSEh312}, \ref{dSEh30}) to determine the total evolution matrix,
\begin{equation}
\label{Sh6}
\check{U}_{\mathfrak{h}_6}(t) = \check{U}_{\mathfrak{su}(2)}(t)\, \check{U}_{\mathfrak{h}_3}(t).
\end{equation}

\subsection{Reverse-engineering exact dynamics in higher-rank groups; $SU(N)$ examples}
\label{sec:DSBEsu(N)}
\subsubsection{Dynamic ``eightfold way''}

The next level of complexity after $\mathfrak{su}(2)$ is provided by rank-$2$ algebras, such as $\mathfrak{su}(3)$. We first focus on an $\mathfrak{su}(3)$ dynamical system~\cite{BlochSU(3)} associated with the latter $8$-dimensional algebra, which is well-studied in the context of the standard model. To calculate directly the exponentials of  $8$-dimensional matrices in the DSBE (\ref{dSE}) is a cumbersome exercise and we proceed differently in a way generalizable to other higher-rank algebras. 

First, recall that there exists a three-dimensional (defining) representation of $\mathfrak{su}(3)$ that can be constructed as follows. Define, $\hat{\tau}_{\nu;\alpha} \equiv \hat{\tau}_{(l,m);\alpha}$ with $1 \leq l <m \leq 3$ and $\alpha = x,y,z \equiv 1,2,3$ to be three-dimensional matrices such that the only possibly non-zero elements are $(l,l)$, $(l,m)$, $(m,l)$, and $(m,m)$ [here the notation implies, $\left( \mbox{row},\, \mbox{column} \right)$] and the corresponding $2\times 2$ matrices are  the familiar Pauli matrices, $\hat{\sigma}_{\alpha}$. The standard  Gell-Mann matrices, introduced in relation to the famous ``eight-fold way,''
are expressed via $\tau$'s as follows: $\hat{\lambda}_1 = \hat{\tau}_{(1,2);x}$, $\hat{\lambda}_2 = \hat{\tau}_{(1,2);y}$, $\hat{\lambda}_3 = \hat{\tau}_{(1,2);z}$, $\hat{\lambda}_4 = \hat{\tau}_{(1,3);x}$, $\hat{\lambda}_5 = \hat{\tau}_{(1,3);y}$, $\hat{\lambda}_6 = \hat{\tau}_{(2,3);x}$, $\hat{\lambda}_7 = \hat{\tau}_{(2,3);y}$, $\hat{\lambda}_8 = {1 \over \sqrt{3}} \left[ \hat{\tau}_{(2,3);z} +  \hat{\tau}_{(1,3);z} \right]$. Note that only two among these Gell-Mann matrices are diagonal, $\hat{\lambda}_3$ and $\hat{\lambda}_8$, and they correspond to the three-dimensional representation of mutually commuting Cartan operators. We are interested in studying $\mathfrak{su}(3)$ dynamics driven by the time-dependent Hamiltonian,
\begin{equation}
\label{Hsu(3)}
\hat{T}_3 \left[ \check{\cal H}_{\mathfrak{su}(3)} (t) \right] = \sum\limits_{a=1}^{8} b^a(t) \hat{\lambda}_a,
\end{equation}
where $\hat{\lambda}_{1,2,\ldots,8}$ are the eight three-dimensional Gell-Mann matrices. 
Now consider the operator in the exponential of the DSBE (\ref{dSE}) in this representation, i.e., the operator $\left( {\Phi}^a \hat{\lambda}_a \right)$. There exists a unitary time-dependent $SU(3)$ rotation that diagonalizes this matrix into a combination of Cartan generators
\begin{equation}
\label{U3diag}
{\Phi}^a(t) \hat{U}_3(t)  \hat{\lambda}_a  \hat{U}_3^\dagger(t) = \omega_1(t) \hat{\lambda}_3 + \omega_2(t) \hat{\lambda}_8.
\end{equation}
Now, we can use the Bulgac-Kusnezov construction~\cite{BulKuz} or alternatively seek the unitary rotation as a product of three $SU(2)$ rotations
\begin{equation}
\label{U3}
\hat{U}_3(t) = \!\! \prod\limits_{\nu=(i,j>i)}\!\!\! \exp \left[ - i {\bm \chi}^{\nu} (t) \cdot \hat{\bm \tau}_{\nu} \right],
\end{equation}
where we can take ${\bm \chi}^{\nu} (t)$ to be ``$xy$-vectors,'' i.e., ${\bm \chi}^{\nu} (t) \cdot \hat{\bm \tau}_{\nu} \equiv  { \chi}^{\nu;x} (t)  \hat{\bm \tau}_{\nu;x} + { \chi}^{\nu;y} (t)  \hat{\bm \tau}_{\nu;y}$. Note that per elementary properties of the Pauli matrices, we can write 
\begin{equation}
\label{U3'}
\hat{U}_3(t) = \!\!\! \prod\limits_{\nu=(i,j>i)}\!\!\!  \Bigl[ \hat{\tau}_{\nu;0} \cos \left| {\bm \chi}^{\nu} (t) \right| - i \left( {\bm n}_\nu(t) \cdot \hat{\bm \tau}_{\nu} \right) \sin \left| {\bm \chi}^{\nu} (t) \right| \Bigr],
\end{equation}
where ${\bm n}_\nu(t) =  {\bm \chi}^{\nu} (t) /\left| {\bm \chi}^{\nu} (t) \right|$ is a two-dimensional unit vector. We emphasize that unlike 
Eq.~(\ref{U3}), Eq.~(\ref{U3'}) is {\em not} representation-invariant. However it is useful for actual calculations and can be utilized to relate $\Phi^a(t)$ and $\omega_{1,2}(t)$ in Eq.~(\ref{U3diag}), and these relations are indeed representation-invariant. 

Now, present the unitary rotation in the eight-dimensional adjoint representation using the invariant form (\ref{U3}) and write the DSBE as follows
\begin{equation}
\label{DSBEsu(3)'}
\hat{U}_8^\dagger (t)\hat{U}_8(t) \int\limits_0^1 ds e^{s \Phi^a(t) \hat{f}_a} \hat{U}_8^\dagger (t)\hat{U}_8(t) 
\dot{\bm \Phi}(t) = {\bf b}(t),
\end{equation} 
where ${f}_a$ are the known $\mathfrak{su}(3)$-generators in the eight-dimensional representation. This rotation brings the form $\Phi^a(t) \hat{f}_a$ to a diagonal form and we have
\begin{equation}
\label{DSBEsu(3)''}
\hat{U}_8^\dagger (t) \int\limits_0^1 ds e^{s \left[ \omega_1(t) \hat{f}_3 + \omega_2(t) \hat{f}_8 \right]} \hat{U}_8(t) \dot{\bm \Phi}(t) =  {\bf b}(t).
\end{equation} 
Recall that $\hat{f}_{a=3} = {\rm diag}\, \left\{ 2,-2,0,-1,1,1,-1,0 \right\}$ and $\hat{f}_{a=8} = \sqrt{3} {\rm diag}\, \left\{ 0,0,0,1,-1,1,-1,0 \right\}$. Therefore, the matrix exponentiation of the diagonal matrices in Eq.~(\ref{DSBEsu(3)''}) is simple and yields
for $\hat{\cal D}(t)  = \int\limits_0^1 ds e^{-is \left[ \omega_1(t) \hat{f}_3 + \omega_2(t) \hat{f}_8 \right]}$ the following explicit expression
\begin{eqnarray}
\nonumber
\hat{\cal D}(t)  
= {\rm  diag}\, \{ && \!\!\!\!\!\! w [2 \omega_1], w^* [2 \omega_1], 1, w [\sqrt{3} \omega_2 - \omega_1], \\
\nonumber
&& \!\!\!\!\!\!  w^* [\sqrt{3} \omega_2 - \omega_1], w [\sqrt{3} \omega_2 + \omega_1],\\
&& \!\!\!\!\!\! w^* [\sqrt{3} \omega_2 + \omega_1], 1 \},
\label{D}
\end{eqnarray}
where 
\begin{equation}
\label{w}
w(2z) \equiv \int_0^1 ds e^{-2isz} = e^{-iz} {\sin{z} \over z}.
\end{equation}
Note that $\lim\limits_{z \to 0} w(z) = 1$. The third and eight matrix elements that are identically equal to one in the diagonal matrix function $\hat{\cal D}(t)$ in Eq.~(\ref{D}) correspond to the zero eigenvalues of the generic matrix, $\left( \Phi^a  \hat{f}_a \right)$, which therefore does not have an inverse. It is a generic feature of the adjoint representation, which descends from the simple fact that all elements in a Lie algebra commute with themselves. Finally, the Schr{\"o}dinger-Bloch dual for $\mathfrak{su}(3)$ can be written as 
\begin{equation}
\label{dSBEsu(3)f}
\hat{U}_8^\dagger(t) \hat{\cal D}(t)  \hat{U}_8(t) \dot{\bm \Phi}(t) = {\bf b}(t),
\end{equation}
where $\hat{U}_8(t)$ is the eight-dimensional representation of the unitary $SU(3)$ rotation (\ref{U3}) that diagonalizes the rotating three-dimensional ``Hamiltonian,''  $\left( \Phi^a(t) \hat{\lambda}_a \right)$, which has three real time-dependent eigenvalues, $\left[\omega_2(t)/\sqrt{3} \pm \omega_1(t) \right]$ and $\left[ - 2 \omega_2(t)/\sqrt{3} \right]$ [see, Eq.~(\ref{U3diag})], and the diagonal matrix $\hat{D}(t)$ is given by Eqs.~(\ref{D}) and (\ref{w}). Note that the main difficulty in deriving the dual dynamical system (\ref{dSBEsu(3)f}) is to find the frequencies, $\omega_{1,2}(t)$, and the rotation matrix, $\hat{U}_3(t)$, with the latter to be expressed in terms of the representation-invariant generators. Technically, deriving these quantities reduces to solving a generic cubic algebraic equation, which is cumbersome but should always be practically doable. This observation also suggests that an explicit derivation of the DSBE for an arbitrary simple algebra, ${\cal A}$, is possible if the dimensionality of the minimal faithful representation is $L_{\rm min} \leq 4$ and connects this problem nicely to the Galois theory.

However, even if an explicit representation invariant dynamic $\mathfrak{su}(3)$ system is written down, to solve it [i.e., to find eight dynamic generators, $\Phi^a(t)$] may be a hopeless goal, as it generally amounts to the gargantuan task of solving a system of eight coupled differential equations. On the other hand, just like in the $\mathfrak{su}(2)$ case, the inverse problem of determining driving fields that give rise to certain non-linear evolution is much simpler. Eq.~(\ref{dSBEsu(3)f}) should then be viewed as a result for the eight-dimensional ``$\mathfrak{su}(3)$ Zeeman magnetic field,'' ${\bf b}(t)$.

\subsubsection{Generalizing the results to $\mathfrak{su}(N)$}
The above discussion is not specific to $\mathfrak{su}(3)$, but generalizes to an arbitrary simple algebra and in particular to $\mathfrak{su}(N)$. 
 $\mathfrak{su}(N)$ is the most symmetric among all algebras and can be viewed as $N(N-1)/2$ $\mathfrak{su}(2)$'s tied together. 
 The $N$-dimensional defining representation of  $\mathfrak{su}(N)$ can be constructed in much the same way as in $\mathfrak{su}(3)$, by introducing Pauli matrices, $\hat{\tau}_{\nu;\alpha}$ with $\nu \equiv (i,j)$, such that $1 \leq i < j \leq N$ and $\alpha = x,y,z$. However, only $(N-1)$ linearly-independent diagonal generators (Cartan operators, $\check{h}_i$) exist, which is exactly the rank of the algebra, $\mathfrak{Ran}\mbox{\bf k}\, \mathfrak{su}(N) = N-1 = r$. Therefore, the entire algebra is spanned by $N(N-1)/2$ Pauli matrices of the $x$ and $y$-type (or equivalently, by raising/lowering operators, $\hat{\tau}_{\nu;\pm}$) and the $(N-1)$ Cartan operators, which of course correctly reproduces the dimension of the algebra ${\rm dim} \left[\mathfrak{su}(N)\right] = 2 N(N-1)/2 + (N-1) = (N^2 - 1) = d$. 

In order to reverse engineer an exact quantum dynamics in $\mathfrak{su}(N)$, one can follow the following five-step prescription: ({\bf i})~First, choose $r = N-1$ arbitrary frequency functions for the mutually commuting Cartan operators, $\omega_j(t)$, with $j = 1,2, \ldots, r$. ({\bf ii})~Second, choose $N(N-1)/2$ arbitrary two-dimensional vector fields, ${\bm \chi}_{\nu}(t)$, and use them to build an $SU(N)$ rotation as a product of $N(N-1)/2$  non-commuting $SU(2)$ rotations, $\hat{U}_N(t) = \prod_\nu \left[ \hat{\tau}_{\nu;0} \cos \chi_\nu(t) - i \left( {\bm n}_\nu(t) \cdot \hat{\bm \tau}_\nu \right) \sin \chi_\nu(t) \right]$, c.f.  Eq.~(\ref{U3}). ({\bf iii})~Determine the generators of the exact solution from ${\Phi}^a(t) \hat{T}_N [\check{J}_a] =  \sum\limits_{j=1}^r \omega_j(t) \hat{U}_N^\dagger(t) \hat{T}_N [\check{h}_j] \hat{U}_N(t)$. Note that the latter is a doable straightforward calculation in the $N$-dimensional representation described above. 
({\bf iv})~Determine the unitary rotation and the diagonal operator $\hat{\cal D}(t) = \int\limits_0^1 ds \exp \left\{ -is \sum\limits_{j=1}^r \omega_j(t)  \hat{T}_{N^2-1} [\check{h}_j] \right\}$, which are expressed in terms of the simple universal function $w(z)$ in Eq.~(\ref{w}) but involving some non-universal combinations of $\omega$'s. ({\bf v})~Finally, evaluate ${\bf b}(t) = \hat{U}_{N^2-1}^\dagger(t) \hat{\cal D}(t) \hat{U}_{N^2-1}(t) \dot{\bm \Phi}(t)$, which therefore provides a precise driving field needed to perform the desired evolution.
 
 Let us mention here that the $\mathfrak{su}(4)$ case may be of interest to quantum computing. A particular realization of this $15$-dimensional algebra is a pair of coupled half-integer spins with the most generic Hamiltonian of the form
\begin{equation}
\label{Hsu(4)}
\hat{\cal H}_{\mathfrak{su}(4)}(t) = {\bf b}_1(t) \cdot \hat{\bm \sigma}_1 + {\bf b}_2(t) \cdot \hat{\bm \sigma}_2 +  
\sum\limits_{\alpha,\beta} J^{\alpha \beta}(t) \hat{ \sigma}_{1,\,\alpha} \hat{ \sigma}_{2,\,\beta},
\end{equation}
where $\hat{\sigma}_{\alpha}(i)$ are the standard Pauli matrices and we allow all the 15 parameters to be functions of time. The full Hilbert space of the problem is spanned by the wave-functions of the form, $| \psi \rangle =  c_1 |\!\!\uparrow\uparrow\rangle + c_2 |\!\!\uparrow\downarrow\rangle +c_3 |\!\!\downarrow\uparrow\rangle +c_4 |\!\!\downarrow\downarrow\rangle$. The normalized states reside on the seven-dimensional sphere and include, in particular, entangled states. Using the proposed five-step procedures, one may engineer time pulses that ``untie'' the entangled states back into pure states and vice versa thereby untying the corresponding second Hopf fibration, $S^7/S^3 = S^4$, which was proposed to describe two-spin entanglement~\cite{2Hopf_entangl,3Hopf_entangl}.

\subsection{Two remarks on the global exponential representation}

\subsubsection{Square-root and fractional powers of the evolution matrix}
If the dynamic symmetry group can be represented globally as an exponential of algebra elements, this representation leads to a very curious ``natural'' definition of matrix powers, as follows. Assume that $G_{\cal A}$ is a dynamic group and consider its element $\check{U} = \exp \left[ - i {\bm \Phi} \cdot \check{\bm J} \right]$.
Define an arbitrary real ($\alpha \in\mathbb{R}$) power of $\check{U}$ as follows
\begin{equation}
\label{U^}
\check{U}^\alpha = \exp \left[ - i \alpha {\bm \Phi} \cdot \check{\bm J} \right] \in G_{\cal A}.
\end{equation}
For example a square root of an $SU(2)$ matrix (in two-dimensional representation), $\hat{U} = e^{{i \over 2} {\bm \Phi} \cdot \hat{\bm \sigma}}$, would be simply $\sqrt{\hat{U}} = e^{{i \over 4} {\bm \Phi} \cdot \hat{\bm \sigma}}$. 

It appears that using such fractional powers one can construct mixed Schr{\"o}dinger-Heisenberg representations of quantum evolution (which  may conceivably be used for derivation of generalized ``rotating-wave approximations''). Consider for example an average of operator products ({\em not} a Lie-algebraic invariant), $C_{ab}(t) = \langle \Psi(t) | \hat{J}_a \hat{J}_b | \Psi(t) \rangle$ with $| \Psi(t) \rangle= \hat{U}_t | \psi_0 \rangle \equiv \exp \left[ - i  {\bm \Phi} \cdot \hat{\bm J} \right]  | \psi_0 \rangle$. We can write it as follows
\begin{equation}
\label{Cabmix}
C_{ab}(t) = \langle \psi(0) | \hat{U}^{-{ 1\over 2}}_t \hat{U}^{-{ 1\over 2}}_t \hat{J}_a \sqrt{\hat{U}_t} \hat{U}^{-{ 1\over 2}}_t \hat{J}_b \sqrt{\hat{U}_t} \sqrt{\hat{U}_t} | \psi_0 \rangle
\end{equation}
or equivalently as
\begin{equation}
\label{Cabmix2}
C_{ab}(t) = \langle \Psi^{(1/2)}(t) | \hat{J}_a^{(1/2)}(t) \hat{J}_b^{(1/2)}(t) | \Psi^{(1/2)}(t) \rangle,
\end{equation}
where $| \Psi^{(1/2)}(t) \rangle = \exp \left[ - {i \over 2}  {\bm \Phi} \cdot \hat{\bm J} \right] | \psi_0 \rangle$ and 
$\hat{J}_a^{(1/2)}(t) = e^{ {i \over 2} {\rm ad}_{{\bm \Phi} \cdot \hat{\bm J}}} \hat{J}_a$ are the wave-function and operator in such a mixed $1/2$-Heisenberg-$1/2$-Schr{\"o}dinger representation. 

While it is admittedly unclear whether the exotic mixed representations and our ability to raise group elements to an arbitrary power are useful for practical calculations, the existence of this operation brings up a more serious question about its multivaluedness  and most importantly about the ``analytical properties'' of the global exponential map. This leads to the following section. 

\subsubsection{Parameterization problem}
\label{ssec:param}
Here we point out an important problem of parametrization of the dynamic symmetry group that is present in both the DSBE ``single-exponential approach''~\cite{Magnus} advocated here and product representation methods routinely used in the mathematical and quantum control literature~\cite{WeiNorm1,WeiNorm2,MagnusRev}, i.e., where $\hat{U}(t) = \prod\limits_{a=1}^d e^{-i \chi_a(t) \hat{J}_a}$ (no summation). Namely, the problem is to determine a domain where the dual generators, ${\bm \Phi}(t)$ [or ${\bm \chi}(t)$], are actually defined. A na{\"\i}ve view is that they can be taken arbitrary $d$-dimensional vector fields in a one-to-one correspondence with the driving field, ${\bm b}(t)$, which at any given time, $t$, is a vector of the $d$-dimensional vector space, $\mathbb{R}^d$,  associated with the underlying $d$-dimensional algebra, ${\cal A}$. This na{\"\i}ve  approach however can not be entirely correct, because the generators are supposed to parameterize group elements, and the group $G_{\cal A} = e^{i {\cal A}}$ is in most cases quite different from and in some intuitive sense ``smaller than'' the Euclidean space, $\mathbb{R}^d \sim {\cal A}^*$. For example, the $\mathfrak{su}(2)$ algebra is just $\mathbb{R}^3$ as a vector space, while its covering group, $SU(2)$, is topologically a three-dimensional sphere. This entire $SU(2)$ group is reproduced in the two-dimensional defining representation, via the elementary identity: $\hat{U}[{\bm \Phi}] =  e^{-{i \over 2} {\bm \Phi} \cdot \hat{\bm \sigma}} = \hat{I}_2 \cos\left[ {\Phi \over 2} \right] - i \left( {\bf n} \cdot \hat{\bm \sigma} \right) \sin\left[ {\Phi \over 2} \right]$. Clearly, if $\Phi_1 = \Phi_2 + 4\pi k$ and ${\bm n}_1 = {\bm n}_2$ or if
$\Phi_1 =  4\pi k - \Phi_2$ and ${\bm n}_1 = -{\bm n}_2$ (with $k \in \mathbb{Z}^+$ being positive integer), both ${\bm \Phi}_1$ and ${\bm \Phi}_2$ give rise to the same actual group element.  Interestingly, these twisted conditions are much reminiscent to the Hopf fibration, $S^2 = S^3/S^1$, with the modulus of dual generator, $\Phi = |{\bm \Phi}|$ to be associated with $S^1$ and its ``direction,'' ${\bm n}$ with $S^2$. This example suggests that in fact proper parameterization of the  dynamic groups may be far from trivial.

Let us now recall that what seems to be a standard approach in the mathematical literature to relating the Lie group and its Lie algebra is to define the elements of the latter as equivalence classes of curves passing through the identity in the former. While the author is not qualified to provide any valuable insight into the mathematical aspects of this old mathematical construction, we would like to speculate that in the context of quantum dynamical systems, a reverse procedure, if at all possible, would have been of great interest in that the algebra rather than the dynamic group is suggested as a primary structure. An interesting mathematical  question is whether one can define group elements of $G_{\cal A} = e^{i {\cal A}}$ as equivalence classes of the algebra elements instead along the lines of the Hadamard construction in~[\onlinecite{comment}]. Let us conclude this somewhat obscure mathematical discussion, by noting that the formal question of parametrization of dual generators may in fact be of importance to physics, where proper parametrization of the dual generators may circumvent the sign problem in certain interacting theories, see, e.g., Sec.~\ref{sec:DualHS}, where a dual Hubbard-Strtonovich approach is presented for a quantum spin model.

\section{Quantum-to-Classical Correspondence}
\label{sec:QCC}
We have been calling Eq.~(\ref{dSE}) ``the dual Schr{\"o}dinger-Bloch equation,'' but did not explain why we have chosen to associate Bloch's name with the equation that seemingly focuses on quantum dynamical systems only. We show below that the equation is in fact equally applicable to describe the corresponding classical evolution. Note that we have been using the units where the Planck constant is set to one, $\hbar = 1$. Had we not chosen this convention for the units, we nevertheless would still have found that in most sensible physically-motivated cases, the DSBE does not actually include the Planck constant anyway and therefore is unchanged in the semiclassical limit, $\hbar \to 0$. Also, it has been emphasized that the Lie algebraic approach is by construction representation-invariant, and hence all results apply equally well to large-dimensional representations, including the limit $L \to \infty$, usually associated with the quasiclassical approximation. These arguments suggest, and the simple discussion below explicitly shows, that the DSBE equation~(\ref{dSE}) is indeed as applicable to describe classical dynamical systems as it is to describe the corresponding  unitary quantum evolution.

\subsection{Generalized Bloch equations}
In a typical non-equilibrium quantum-mechanical problem with a well-defined  Hilbert space, ${\cal H}il(L)$, we are interested in calculating observables such as averages 
\begin{equation}
{\bf M}(t) = \langle \psi_0 | \hat{\bm J}(t) | \psi_0 \rangle
\label{<i>}
\end{equation}
and correlation functions. E.g., a general two-point correlator  can be presented as
\begin{equation}
C(t_1,t_2) = \sum\limits_{k,p} c^{ab}(t_1,t_2) \langle \psi_0 | \hat{J}_a(t_1)  \hat{J}_b(t_2) | \psi_0 \rangle
\label{C},
\end{equation}
where $| \Psi_0 \rangle \in {\cal H}il(L)$ is a wave-function that describes an initial state of the actual physical system
  and $\hat{\bm J}(t) = \hat{U}^\dagger (t) \hat{\bm J} \hat{U}(t)$ are Heisenberg operators acting in ${\cal H}il(L)$. 
 Note that these Heisenberg operators belong to the algebra, because
\begin{equation}
\label{Rcl}
\check{J}_a(t) = \exp \left[ -i  {\rm ad}_{{\bm \Phi}(t) \cdot \check{\bm J}}\right]\,  \check{J}_a = R_a^{\phantom{a}b}(t) \check{J}_b,
\end{equation}
which involves nothing but commutators. On the other hand, the observables such as the matrix elements (\ref{<i>}) depend on the Hilbert space and the initial conditions and are not algebraic invariants. However, the amount of information we ``need from the Hilbert space'' is very small for the purpose of calculating (\ref{<i>}) and reduces to the initial conditions for the averages, ${\bf M}(0) =  \langle \Psi_0 | \hat{\bm J} | \Psi_0 \rangle$. In fact, one can specify arbitrary initial conditions without reference to any Hilbert space (or quantum mechanics all together), with the caveat that an arbitrary classical choice may fail to satisfy  quantization requirements, which automatically arise in the quantum-mechanical formulation. Given a set of initial conditions, ${\bf M}(0)$,  the full time-dependence follows
\begin{equation}
\label{M(t)}
{\bf M}(t) =\hat{R}[{\bm \Phi}(t)]  {\bf M}(0),
\end{equation}
where $\hat{R}[{\bm \Phi}(t)] = e^{-\hat{F}\left[{\bm \Phi}(t)\right]}$ are the matrices uniquely determined by the solutions to the DSBE and therefore they are {\em Lie-algebraic invariants}. 
As to correlations functions, only averages of the commutators are expressed in terms of Lie-algebraic invariants in the same way
as ${\bf M}(t)$. E.g., a correlator involving a commutator of two operators $C_{a_1,a_2} (t_1,t_2) = -i \langle \psi_0 |  \left[\hat{J}_{a_1}(t_1),\, \hat{J}_{a_2}(t_2) \right] | \psi_0 \rangle$ evolves into 
\begin{equation}
\label{Heis}
C_{a_1,a_2} (t_1,t_2) = R_{a_1}^{\phantom{a}b_1}(t_1)\, R_{a_2}^{\phantom{a}b_2}(t_2)\, f^{\phantom{b_1 b}c}_{b_1 b_2}\, M_c(0), 
\end{equation}
where $f_{b_1 b_2}^{\phantom{b b}c}$ are the structure constants (\ref{comm}) of the underlying algebra, ${\cal A}$, and the tensor 
$\hat{R}(t)$ is a Lie-algebraic invariant defined via (\ref{M(t)}). Therefore to find the correlator, $C_{a_1 a_2}$, the only required piece of information that is not representation-invariant  (and thus may depend on the Hilbert space) is the initial conditions, ${\bf M}(0)$. Let us note here that this not generally true for an arbitrary correlation function of type (\ref{C}), which involves operator products apart from ``pure commutators,'' or more generally for any observable that is not a member of the algebra. This can be seen by assuming the contrary and finding a counter-example, e.g.,  in, perhaps, the simplest case of the two-dimensional representation of $\mathfrak{su}(2)$ with the ``correlator''  $\langle \uparrow | \hat{\sigma}_z^2(t)  | \!\!\uparrow \rangle \equiv 1$, which is identically equal to one independently of evolution, while for all higher-dimensional representations,  $\hat{T}_{L>2}^2[\check{J}_z]$ is not proportional to the identity matrix and may exhibit a non-trivial time-dependence.

Let us now confine ourselves to considering  Lie-algebraic-invariant  dynamics in the sense defined above, i.e., averages, correlators, Berry phase, etc. The averages can be described in terms of generalized Bloch equations that follow from  the Schr{\"o}dinger equation (\ref{SE1}), or equivalently, from the familiar Heisenberg equations of motion
$$
\dot{\check{{\bf J}}}(t) = i \left[ \check{\cal H}(t),\, \check{\bm J}(t) \right],
$$
where the Hamiltonian, $\check{\cal H}(t) = {\bf b}(t) \cdot \check{\bm J}$ and the structure constants (\ref{comm}) of ${\cal A}$ yield
\begin{equation}
\label{Bloch}
\dot{M}_b(t) = f_{ba}^{\phantom{ba}c}\, b^a(t) M_c(t),
\end{equation}
which have to be supplemented by the initial conditions, ${\bf M}(0)$. The generalized Bloch equations (\ref{Bloch}) are the analogue to the {\em classical equations of motion} for an arbitrary Lie algebra. For some solvable algebras, such as the harmonic oscillator algebra, they reduce to the Newton equations of motion in a Euclidean space (or more precisely to Hamilton's equations of motion).  For the simple algebra, $\mathfrak{su}(2)$, these are the usual Bloch equation on the two-dimensional sphere, $\dot{\bf M} = {\bf b}(t) \times {\bf M}$. In general, equations (\ref{Bloch}) represent classical dynamics on a non-trivial manifold with constraints, ${\cal M}$,  embedded into the $d$-dimensional Euclidean space. 

We have shown that the solutions to both quantum and classical dynamics are uniquely determined by the generators, $\Phi^a(t)$. The classical solution is governed  by the action
of the operator, $\hat{R}$, in Eqs.~(\ref{M(t)}) or equivalently  below
\begin{equation}
\label{q(t)2}
{\bf M}(t) =e^{-{\bm \Phi}(t) \cdot \hat{\bm f} } \,{\bf M}(0).
\end{equation}
Recall that the classical averages form a $d$-dimensional vector that is acted upon by the $(d \times d)$-dimensional matrix,
$\hat{F}\left[{\bm \Phi}(t)\right] = {\bm \Phi}(t) \cdot \hat{\bm f}$  from the adjoint representation of the underlying algebra.
The Casimir invariants constrain the vector of classical averages to lie within a classical manifold, ${\cal M}$. 
In most cases of physical significance, this manifold is a symmetric Riemannian space with the metric that is uniquely 
determined by the Cartan-Killing  product~(\ref{CK}). Therefore, Eq.~(\ref{q(t)2}) represents action of a symmetry group, $G({\cal M})$, on the classical manifold, ${\cal M}$. Topological construction of the classical manifold as a coset representative of the dynamic symmetry group is discussed, e.g., in Ref.~[\onlinecite{Gilmore_RMP}]. For completeness, we also review it in appendix~\ref{app:clasM}, where it is argued that in the context of dynamical system at hand, the Perelomov's construction, which is slightly different from that of Gilmore in that it allows an arbitrary reference states to be used, is preferable and it is natural to associate the reference state with the initial condition, $|\psi_0\rangle$, so that the classical manifold is ${\cal M} = G_{\cal A}/H(|\psi_0\rangle)$, where $H(|\psi_0\rangle)$ is the maximum stability subgroup of the initial state. 

Note also, that the classical symmetry group is generally expected to be smaller than the covering group generated by the algebra, {\em i.e.}, $G({\cal M}) \subset G_{\cal A} =\exp(i {\cal A} )$. E.g., in the previously considered case of the $\mathfrak{su}(2)$ algebra~\cite{Our_TLS}, $G(S^2) = SO(3)$ is the group of three-dimensional rotations to which $SU(2) = G_{\mathfrak{su}(2)}$ is a double-cover, $SU(2)/SO(3)=\mathbb{Z}_2$.  

Concluding this section, we reiterate its main conclusion that quantum-to-classical correspondence manifests itself in that the dynamics of classical averages~(\ref{q(t)2}) and correlation functions~(\ref{Heis}) are governed by the same generators, ${\bm \Phi}(t)$, that determine evolution of the quantum wave-function, see Eqs.~(\ref{solSE}) and (\ref{S&S+}), and satisfy the dual Schr{\"o}dinger-Bloch equations~(\ref{dSE}).

\subsection{Example of quantum-to-classical correspondence for a linearly-driven Jaynes-Cummings model}
\label{sec:JC}
Here we provide an explicit simple example that illustrates  quantum-to-classical correspondence and demonstrates the 
 Lie-algebraic approach in action. Consider the driven Jaynes-Cummings Hamiltonian:
\begin{equation}
\label{JC1}
\hat{\cal H}_{\rm JC}(t) = \Lambda_b(t) \hat{b}^\dagger \hat{b} + \Lambda_s(t) {\hat{\sigma}_z \over 2}
+ {g(t) \over 2} \left( \hat{b}^\dagger \hat{\sigma}_- + \hat{b} \hat{\sigma}_+ \right).
\end{equation}
Here $\hat{b}^\dagger$/$\hat{b}$ are canonical boson creation/annihilation operators, $\hat{\sigma}_{z}$ and 
$\hat{\sigma}_\pm = \left( \hat{\sigma}_x \pm i \hat{\sigma}_y \right)$ are the standard Pauli matrices, and $\Lambda_b(t)$,
$\lambda_s(t)$, and $g(t)$ are arbitrary functions of time. As the title of this section indicates, we will be specifically 
interested in the linearly-driven Landau-Zener problem with $  \Lambda_s(t) = -\Lambda_b(t) = \lambda t$ and $g(t)\equiv g = {\rm const}$.

First note that the operator $\hat{N} = \hat{b}^\dagger \hat{b} +  {\hat{\sigma}_z \over 2}$ commutes with the Hamiltonian and any other
operator present in Eq.~(\ref{JC1}). I.e., 
\begin{equation}
\label{[N,A]=0}
\left[\hat{N},\,\hat{b}^\dagger \hat{b} \right] = \left[\hat{N},\,\hat{\sigma}_z  \right] =  
\left[\hat{N},\,\hat{b}^\dagger \hat{\sigma}_-  \right] =  \left[\hat{N},\,\hat{b} \hat{\sigma}_+ \right] = 0.
\end{equation}
Therefore, we can treat the operator $\hat{N}$ as a $c$-number. 

Second, let us also define the following operators
\begin{equation}
\label{Sigma+}
\hat{\Sigma}_+ = {1 \over 2 \sqrt{2N + 1}} \hat{b} \hat{\sigma}_+,
\end{equation}
\begin{equation}
\label{Sigma-}
\hat{\Sigma}_- = {1 \over 2 \sqrt{2N + 1}} \hat{b}^\dagger \hat{\sigma}_-,
\end{equation}
and
\begin{equation}
\label{Sigma3}
\hat{\Sigma}_3 = \hat{b}^\dagger \hat{b} - \hat{N} \equiv  {\hat{\sigma}_z \over 2}.
\end{equation}
These operators span the $\mathfrak{su}(2)$ algebra, i.e.
\begin{equation}
\label{Su(2)}
\left[\hat{\Sigma}_+,\, \hat{\Sigma}_- \right] = 2\hat{\Sigma}_3 \mbox{  and  }
\left[\hat{\Sigma}_3,\, \hat{\Sigma}_\pm \right] = \pm  \hat{\Sigma}_\pm.
\end{equation}
The Hamiltonian (\ref{JC1}) becomes
\begin{equation}
\label{JC2}
\hat{\cal H}_{\rm JC}(t) = \Lambda_b(t) \hat{N} + \Lambda_3(t) \hat{\Sigma}_3
+ \tilde{g}(t)  \left( \hat{\Sigma}_+ + \hat{\Sigma}_- \right),
\end{equation}
with $\Lambda_3(t) = \left[\Lambda_s(t)- \Lambda_b(t)\right]$ and $\tilde{g}(t) = \sqrt{2N + 1} g(t)$.
The first Abelian term in Eq.~(\ref{JC2}) gives rise to a trivial phase dynamics and can be factorized away. 
In the specific case of linear drive, we  end up with the Landau-Zener Hamiltonian for $\hat{\cal H}_\Sigma = \hat{\cal H}_{\rm JC}(t) - \Lambda_b(t) \hat{N}$:
\begin{equation}
\label{JCLZ}
\hat{\cal H}_\Sigma =2 \lambda t \hat{\Sigma}_3
+ \tilde{g}  \left( \hat{\Sigma}_+ + \hat{\Sigma}_- \right).
\end{equation}

{\em The evolution operator --} Since the algebra of $\Sigma$-operators is isomorphic to $\mathfrak{su}(2)$, we can look
for solution in any faithful representation, including the two-dimensional spin-$1/2$ representation. This maps the original problem onto
the exactly-solvable Landau-Zener problem in its canonic form~\cite{LZ}:
\begin{equation}
\label{LZ1/2}
i\partial_t \Psi(t) = \left[ \lambda t \hat{\sigma}_z + \tilde{g} \hat{\sigma}_x \right] \Psi(t),
\end{equation}
where $\Psi(t)$ is a two-component spinor. Note that the symmetry of Eq.~(\ref{LZ1/2}) with respect to
time-reversal demands that if $\Psi_1(t) ={u_t \choose v_t}$ is a particular solution, then $\Psi_2(t) = {v^*_{-t} \choose u^*_{-t}}$
 is also a solution. If an arbitrary particular solution is known, then a full $S$-matrix 
can be constructed as follows~\cite{VG_Qfl}
\begin{equation}
\label{ULZ}
\hat{U}_{\rm LZ}(t) = {1 \over |u_0|^2 - |v_0|^2} \Biggl(
\begin{array}{cc}
u_0^* u_t - v_0 v^*_{-t} & u_0 v^*_{-t} - v_0^* u_t \\
u_0^* v_t - v_0 u^*_{-t} & u_0 u^*_{-t} - v_0^* v_t
\end{array}
\Biggr),
\end{equation}
where $\Psi_1(0)= {u_0 \choose v_0}$ is an initial condition that the known solution satisfies. 

A particular exact solution was derived, e.g., by Lim and Berry~\cite{Lim_Berry} as follows:
\begin{equation}
\label{Psi1}
\Psi_1(t) = e^{- {\pi \over 8} \gamma} { -i \pi \nu^{1/2} D_{\nu - 1} \left[ \eta(t) \right] 
\choose
 D_{\nu } \left[ \eta(t) \right]},
\end{equation} 
where $\gamma = \tilde{g}^2/\lambda$, $\nu = i \gamma/2$, and $D_\nu(z)$ are known parabolic cylinder functions with 
well-studied asymptotes. They give rise, in particular, to the standard Landau-Zener exponential transition rate as well as
to non-analytic corrections to it, as discussed in the Appendix of Ref.~[\onlinecite{Lim_Berry}]

From Eqs.~(\ref{ULZ}), (\ref{Psi1}), and the identity, $\hat{U}_{\rm LZ}(t) = e^{-{i \over 2} {\bm \Phi}_{\rm LZ}(t) \cdot \hat{\bm \sigma} }=
\hat{I}_2 \cos \left( {\Phi_{\rm LZ} \over 2} \right) - i\left( {\bm n}_{\rm LZ} \cdot \hat{\bm \sigma} \right) \sin \left( {\Phi_{\rm LZ} \over 2} \right)$, we can find the representation-invariant and Hilbert-space independent dual generators. 
The evolution operator for the original Jaynes-Cummings problem therefore reads (here we restored the trivial phase dynamics):
\begin{equation}
\label{UJZ}
\hat{U}_{\rm JC} = \exp\left\{ - i  \left[ { \lambda \over 2} t^2 \hat{N} + {\bm \Phi}_{\rm LZ}(t) \cdot \hat{\bm \Sigma} \right] \right\},
\end{equation}
Note that short of exotic initial conditions with entangled spin-boson states, simple initial states (e.g., ``spin-up'' at $t=0$) will evolve {\em in exactly the same way} as they would in a pure spin-$1/2$ problem with the same driving field and the presence of the bosonic bath will manifest itself only in a rescaling of the Landau-Zener parameter. Another way to make the same statement is to focus on evolution of the classical averages, which occurs via action of the $SO(3)$ rotation group
\begin{equation}
\label{JC_clas}
\Biggl(
\begin{array}{c}
[2N + 1]^{-{1 \over 2}} \langle  \hat{b}^\dagger \hat{\sigma}_- + \hat{b} \hat{\sigma}_+ \rangle\\
i[2N + 1]^{-{1 \over 2}} \langle   \hat{b} \hat{\sigma}_+ - \hat{b}^\dagger \hat{\sigma}_-  \rangle\\
\langle  \hat{\sigma}_z/2 \rangle
\end{array}
\Biggr)
\equiv {\bm M}(t) = e^{{\bm \Phi}_{\rm LZ}(t) \cdot \hat{\bm f}} {\bm M}(0),
\end{equation}
where the $(3 \times 3)$ finite-rotation matrix, $\hat{F}_{\mathfrak{su}(2)}[{\bm \Phi}_{\rm LZ}(t)]$, is given by Eq.~(\ref{adsu(2)}). Note that the classical theory is formulated here in terms of the averages of  products of the bosonic and spin operators, as opposed to a formulation in terms of  $\langle \hat{b}(t) \rangle$ and $\langle \hat{\bm \sigma}(t) \rangle$. The latter representation was used~\cite{Altland_etal} in the context of the Dicke
model (which in effect is a non-trivial generalization of the Jaynes-Cummings model for a higher-dimensional representation of spin). The ``semi-classical approximation''  used there  involved replacing the average of operator products with a product of averages. For certain driven dynamics, this 
approximation can be shown to become reliable  when the dimensionality of the representation (or equivalently the number of spins in the Dicke model) becomes large, but it is the least reliable in the opposite limit of the Jaynes-Cummings model.


%
\section{Lie-algebraic Approach to Interacting Many-Particle Quantum Models}
\label{sec:many-body}

\subsection{Lie-algebraic formulation of a quantum lattice model}
\label{sec:Lielattice}
So far we have focused on single-particle quantum dynamical problems for a Hamiltonian linear in algebra generators. We now address a more general Hamiltonian associated with an interacting many-body quantum system.  The general system we have in mind is a lattice model with pair-wise interactions between particles associated with the sites. In the Lie-algebraic sense, it is equivalent to having a representation of a finite-dimensional algebra associated with each site (e.g., $\mathfrak{h}_4$ for bosons, $\mathfrak{u}(2)$ for spins, $\mathfrak{u}(3)$ for ``quarks,'' etc.). 

Define a discrete set, ${\cal S}$, e.g. corresponding to a lattice in physical space. Assign to each site $i \in {\cal S}$, a finite dimensional Lie algebra, ${\cal A}(i) = {\rm span}\, \left\{ \check{J}_0(i), \ldots, \check{J}_{d_i}(i) \right\}$ and its $L_i$-dimensional representation. Define an interacting Hamiltonian as a form bilinear in generators of the single-particle algebras. In what follows, we shall focus on the case where the single-particle entities on all sites are identical to each other, i.e.,  $\forall i \in {\cal S}$, ${\cal A}(i)  \sim {\cal A}$ and the corresponding representations are the same $L(i) \equiv L$. This is not the most general case that can be considered (e.g., the text-book Dicke model discussed in the previous section does not belong to this category as it involves two different algebras, $\mathfrak{h}_4$ and $\mathfrak{u}(2)$, on two ``sites'').  Let us also require  that the algebra ${\cal A}$ includes an Abelian $\mathfrak{u}(1)$ component, $\check{J}_0$ (i.e., $\left[ \check{J}_0,\, \check{J}_a \right] = 0$), which always can be incorporated into the theory. In what follows,  we identify it with the unity operator, $\hat{J}_a(i)  \hat{J}_0(j) = \hat{J}_a(i)$. Note that for the interacting problem, it is important to specify the physical dimension of the single-particle Hilbert space as the many-particle algebra generated by the interacting Hamiltonian will generally include powers, $\check{J}^n$, which acquire a precise meaning only given a representation.  With this, let us write down the interacting Hamiltonian in the following form 
\begin{equation}
\label{Hint}
\hat{\cal H} = \sum\limits_{i \in {\cal S}} \sum\limits_{a=1}^d \mu^a(i) \hat{J}_a(i) + {1 \over 2 {\Omega}_{\cal S}} \sum\limits_{i,j \in {\cal S}} \sum\limits_{a=1}^d \lambda^a(i,j) \hat{J}_a(i)  \hat{J}_a(j),
\end{equation}
where ${\Omega}_{\cal S}$ is the number of sites in the discrete set, ${\cal S}$, $\mu^a$ are real parameters that may correspond to external fields or a chemical potential, and $\lambda^a(i,j) = \lambda^a(j,i)$ is a symmetric interaction matrix. Note that in Eq.~(\ref{Hint}), we have excluded from consideration interactions between different elements in the algebra, but even with this simplifying assumption the model (\ref{Hint}) still remains very general and  includes a huge number of condensed matter Hamiltonians  (excluding certain topological lattice models). For example, the hopping terms of the Bose-Hubbard model can be written as, $\hat{b}_1^\dagger \hat{b}_2  + \hat{b}_2^\dagger \hat{b}_1 = \hat{x}_1 \hat{x}_2 + \hat{p}_1 \hat{p}_2$, where $\hat{x} = (\hat{b} + \hat{b}^\dagger )/\sqrt{2}$ and $\hat{p} = i( \hat{b}^\dagger - \hat{b} )/\sqrt{2}$ on any site. Note that the usual density-density interactions, $\hat{b}_1^\dagger \hat{b}_1 \hat{b}_2^\dagger \hat{b}_2$, also are bi-linear in terms of generators $\hat{b}^\dagger \hat{b}$ from $\mathfrak{h}_4$. In this Lie-algebraic sense, there are not considered any more complicated than the hopping terms, eventhough writing them down involves an expression quartic in terms of boson creation/annihilation operators. 

There are two ways one can proceed along in considering the Hamiltonian (\ref{Hint}). A first approach is to define new composite operators $\hat{X}^{ab}_{ij} = \hat{J}_a(i) \hat{J}_b(i)$, and take all non-zero terms in the Hamiltonian or their special linear combinations and commute them with one another, $\left( {\rm ad}^n_{\hat{X}} \hat{X'} \right)$, continuing doing so until the algebra closes. For any finite-dimensional representation of single-particle algebra, ${\cal A}$, it is certainly guaranteed to happen because any matrix satisfies its characteristic equation and so the number of linear-independent powers of $\hat{X}^n$ is limited (note however that for infinite-dimensional representations, e.g., for the boson case with no restriction on their number on a site, it is not generally so). In any case, the many-particle Hamiltonian (\ref{Hint}) then can be viewed as a linear form in and a member of the new, possibly huge, many-particle algebra, ${\cal A}_{\rm int}$, that gets generated this way. Such an approach  may be a sensible thing to pursue if the many-particle algebra formulated in terms of some physically-relevant generators closes into a tractable form or factorizes into simpler components.  This would not be unexpected in integrable cases, or when the rank of the many-particle algebra (or equivalently the number of conserved quantities or Casimir operators) grows steadily with the ``system size.''  According to the preceding ``single-particle'' section, as long as the many-particle algebra  remains finite-dimensional, the classical and quantum dynamics (formulated in terms of the new generators of ${\cal A}_{\rm int}$) are related to each other, and therefore classical integrability and quantum integrability are in a one-to-one correspondence as well, in the sense that the dynamics of the averages in the many-particle algebra, $\left\langle \hat{X}(t) \right\rangle$, determines the dynamics of the operators and their commutators. Again, the existence of such correspondence may be established via the dual Schr{\"o}dinger-Bloch equations (\ref{dSE}) for ${\cal A}_{\rm int}$, although its practical application may be challenging. 

\subsection{Hubbard-Stratonovich Dynamical Systems}
\label{sec:GenHS}
However, a general case,  relevant to most ``real-life Hamiltonians,'' is that the many-particle algebra ``explodes''  fast and often becomes infinite-dimensional. Also the integrability status of such generic models is rarely known.  Having this sort of a case in mind, we proceed via a different route, generalizing a construction of Ref.~[\onlinecite{VG_Qfl}]. We concentrate on the thermodynamic problem first, i.e., on a time-independent interacting Hamiltonian (\ref{Hint}). We assume the applicability of a path-integral representation for the following quantity 
\begin{equation}
\label{Z}
{\cal Z}(u) = {\rm Tr}\, e^{-\hat{\cal H} u}, \,\, u = \beta + it,
\end{equation}
where the trace is over the entire $L^{{\Omega}_{\cal S}}$-dimensional Hilbert space of the many-particle problem. Notice that we allow the parameter $u$ in Eq.~(\ref{Z}) to be a complex number such that $u=\beta =1/T$ corresponds to the partition function, while ${\cal Z}(u = it)$ reflects the property of a unitary evolution~\cite{TSVGTLS} ({\em e.g.}, describing a free evolution after a quench in system parameters).  Now assume that the usual path-integral representation of the partition function, ${\cal Z}$, applies, i.e.,
\begin{eqnarray}
\nonumber
{\cal Z}(u) = \int\limits_{{\bf M}_k(0) = {\bf M}_k(u)}&&\!\!\!\!\!\! \prod_{k \in {\cal S}} {\cal D} [{\bf M}_k(\tau)] 
e^{-i\sum\limits_{k \in {\cal S}} S_{\rm Berry}[{\bf M}_k]}\\
&&\!\!\!\!\!\!\!\!\!\!\!\! \times \exp\left\{- \int_0^u d\tau H[{\bf M}_k(\tau)] \right\},
\label{Pathint}
\end{eqnarray}
where $S_{\rm Berry}[{\bf M}_k]$ is a topological term on a site $k$, enforcing proper commutation relations for single-particle generators, and $H[{\bf M}_k(\tau)]$ is a Hamiltonian (\ref{Hint}) with all operators replaced with the corresponding coherent-state vector fields. The integration over $\tau$ in Eq.~(\ref{Pathint}) is along a straight line in the complex plane, $\tau = \tau' + i\tau''$, that goes from $0$ to $u = \beta + it$ and the coherent-state trajectories are required to form loops, ${\bf M}(0) = {\bf M}(u)$. Note that the existence and validity of such a path-integral representation is a ``big if'' from a formal mathematical standpoint. One problem related to the previous discussion is the choice of a classical manifold, ${\cal M}$, where the loop-trajectories reside. The single-particle construction does not necessarily apply here, because it relied on single-particle pure states, while here entangled states are possible and generally unavoidable.  This is however not a concern if group action covers the entire Hilbert space (e.g., spin-$1/2$ systems) and also some version of the path integral construction is more likely to remain reliable in cases where the classical manifold is representation-independent (e.g., the lattice bosonic models associated with the solvable single-particle algebra, $\mathfrak{h}_4$). At this point, we just assume that this is so and that classical trajectories remain in a singly-connected classical manifold ${\bf M}(\tau) \subset {\cal M}$. Now let us focus on the interaction term in the action
\begin{equation}
\label{Sint}
e^{-S_{\rm int}} = \exp\left[- {1 \over 2 N_{\cal S}} \sum\limits_{i,j;a}  \lambda^a(i,j)  \int\limits_0^\beta d\tau  {M}_{i,a}(\tau) {M}_{j,a}(\tau) \right].
\end{equation}
We can decouple the interaction terms for each non-zero $\lambda^a(i,j) \ne 0$, using the following identity
\begin{eqnarray}
\nonumber
 e^{-\hat{z}_u {\lambda} M_1 M_2} = &&\!\!\!\!\!\!
\int\limits_{-\infty}^{\infty}  {d A \over \sqrt{2 \pi |{\lambda}|}} e^{-{A^2 \over 2 |{\lambda}|} + {1 \over 2} \hat{z}_u{\lambda} \left(M_1^2 + M_2^2\right)}
\\
&&\!\!\!\!\!\!\!\!\!\!\!\!\!\!\! \times \exp \left\{ \sqrt{-\hat{z}_u {\rm sgn}\, {\lambda}} A \left(M_1 + M_2\right) \right\}, 
\label{HSgen}
\end{eqnarray}
where ${\lambda} \in \mathbb{R}$ and $\hat{z}_u = u/|u| = (\beta + it)/\sqrt{\beta^2 + t^2}$ is a unit vector in the complex plane associated with the choice of the time-integration contour in Eqs.~(\ref{Z})  and (\ref{Pathint}) (e.g., for thermodynamics, it is just equal to one, $\hat{z}_\beta = 1$). If we implement this transform (\ref{HSgen}) in the path-integral (\ref{Sint}),  it renormalizes the on-site interactions as follows, 
\begin{equation}
\label{tilambda}
\lambda^a(i,i) \to \tilde{\lambda}^a(i,i)=
\lambda^a(i,i) - {1 \over 2} \sum\limits_{j \in l(i)} \lambda^a(i,j)
\end{equation}
[where the summation is over all interacting links containing a given site, $l(i)$], but it does not generate any other terms. We now can introduce another set of Hubbard-Stratonovich fields on sites, $B_i(\tau)$, to decouple the
remaining on-site interactions to arrive to a representation in which for each particular realization of the Hubbard-Stratonovich fields, $A_l(\tau)$ and $B_i(\tau)$, the remaining path integrals over the coherent states decouple on different sites and correspond to {\em a single-particle driven dynamics}, where the time-evolution in the algebra is governed  by the corresponding Hubbard-Stratonovich fields. Therefore, we can ``roll back'' the path-integral construction into the Hamiltonian formulation to find the following main general result
\begin{eqnarray}
\nonumber
&&\!\!\!\!\!\! {\cal Z}(u)  = 
\int\limits_{-\infty}^{\infty}  \prod\limits_{a;l \in {\rm links}} \!\!\!\!\left[{\cal D}  A_l^a(\tau) \right]
\prod\limits_{a;i \in {\rm sites}} \!\!\!\! \left[{\cal D} B_i^a(\tau) \right] \prod\limits_i z_i\left[ {\bf A}, {\bf B} \right]
\\
&&\!\!\!\!\!\!\!\! \!\! \times \exp \left\{- {N_S \over 2} \int\limits_0^u d\tau \left[
 \sum\limits_{a;l} {\left| A^a_l(\tau) \right|^2 \over \left| {\lambda}^a_l \right|}  + \sum\limits_{a;i} {\left| B^a_i(\tau) \right|^2 \over |\tilde{\lambda}^a_i|}
\right]
 \right\}, 
\label{HSgenmain}
\end{eqnarray}
where $l$ denotes all lattice links $l = (i,j)$ for which the interaction is non-zero, $i$ is a site index, and 
$$
z_i= {\rm Tr}\, \hat{\rho}_i (u)
$$
 is the trace of the time-evolution matrix that governs the behavior of the single-particle quantum-dynamical system, where the driving field depends on a particular Hubbard-Stratonovich realization 
\begin{equation}
\label{nonunSE}
 \left\{
\begin{array}{ll}
- \partial_\tau \hat{\rho}(\tau) \hat{\rho}^{-1}(\tau)= \hat{H}_i(\tau)  \\
\hat{\rho}(0) = \hat{1}
\end{array}
\right.
\end{equation}
We will call this system of differential equations, the Hubbard-Stratonovich dynamical system. The corresponding Hamiltonian is generally {\em  non-Hermitian} and has the form
\begin{equation}
\label{nonHerH}
\hat{H}_i(\tau) = \sum\limits_{a=1}^d \left\{ {\mu}^a(i) - {F}_i^a(\tau) - \sqrt{-\hat{z}_u \tilde{\lambda}^a_i} B^a_i(\tau) \right\} \hat{J}_a,
\end{equation}
with $ {F}_i^a(\tau)  = \sum\limits_{l=(i,j)} \sqrt{-\hat{z}_u \lambda^a_l} A^a_l(\tau)$ and $\tilde{\lambda}$ defined in Eq.~(\ref{tilambda}).
Note that for a generic interacting Lie-algebraic model of type (\ref{Hint}), the time-evolution, $\hat{\rho}$, is not necessarily unitary, nor it is associated with any real form. From the Lie-algebraic point of view, the main  difference between the Schr{\"o}dinger equation Eq.~(\ref{SE2}) and Eq.~(\ref{nonunSE}) is that the latter is generally to be implemented in a {\em complex extension} of the underlying single-particle algebra. This leads to a number of catastrophic complications such as a break-down of the global exponentiation conjecture, which relied on the compactness of the underlying dynamic group (however, as noted in Appendix~\ref{SSsec:Infinite_d}, certain non-compact dynamical groups still allow global exponential map, see also next section~\ref{sec:BH}).  In the general case, we are forced to look for solution in a product form, see, e.g., Wei and Norman~[\onlinecite{WeiNorm1,WeiNorm2}],  or in a mixed form of type, $\hat{\rho}(\tau) = e^{-{\bf R}(\tau) \cdot \hat{\bm J}} e^{-i{\bm \Phi}(\tau) \cdot \hat{\bm J}}$, which may be ``admissible'' in certain cases (e.g., any $SL(2,\mathbb{C})$ matrix can be represented in this form). 

\section{Examples of the Lie-algebraic approach to many-particle lattice models}
\label{sec:ExamplesLielatt}
\subsection{Non-unitary quantum evolution on the example of the Bose-Hubbard model}
\label{sec:BH}

Consider the Bose-Hubbard model~\cite{MPABH,BH_Zoller,BHexp} on a two-dimensional simple square lattice in equilibrium,
\begin{equation}
\label{HBH}
\hat{\cal H}_{\rm BH} = - t\sum\limits_{l=\langle i,\, j \rangle} \hat{b}_i^\dagger \hat{b}_j  - \mu_0 \sum\limits_{i} \hat{n}_i
+ U \sum\limits_i \hat{n}_i \left( \hat{n}_i - 1 \right),
\end{equation}
where $t$ is the hopping amplitude, $\mu_0$ is the bare chemical potential, $U$ is the on-site repulsion, the indices $i$ and $j$ label sites
of the square lattice, and $l=\langle i,j \rangle$ label its links. The bosonic creation/annihilation operators satisfy the familiar commutation relations, $\left[ \hat{b}_i,\,\hat{b}_j^\dagger \right] = \delta_{ij} \hat{1}$,
$\hat{n}_i = \hat{b}^\dagger_i \hat{b}_i$  is the density operator. On every site, $\hat{n}_i$, $\hat{b}^\dagger$, $\hat{b}_i$, and $\hat{1}$ span the harmonic oscillator algebra. 

The partition function of the model is
\begin{equation}
\label{ZBH}
{\cal Z}_{\rm BH} = \int \prod\limits_i  \left[ {\cal D}^2b_i(\tau)\right] e^{-S[\bar{b},b]},
\end{equation}
where the action is 
\begin{eqnarray}
\label{SBH}
\nonumber
{S}_{\rm BH} = - \int\limits_0^\beta d\tau \Bigl\{ &&\!\!\!\!\!\!\!\! 
\sum\limits_i \left[ \bar{b}_i(\tau) \left( \partial_\tau - \mu \right) b_i(\tau) + U \left| b_i(\tau) \right|^4 \right]\\
&&\!\!\!\!\!\!   - t \sum\limits_{\langle i,j \rangle} \bar{b}_i(\tau) b_j(\tau) \Bigr\}.
\end{eqnarray}
Here we used the prescription for operator-ordering of Ref.~[\onlinecite{JW_VG}], which was suggested in order to reproduce the correct result in the single-site toy model. Therefore, $\mu = \mu_0 + U$ in Eq.~(\ref{SBH}).
 
The conventional approach is to use Bogoliubov mean-field or Gross-Pitaevskii analysis that lead to a reliable mean-field description of the superfluid phase and classical fluctuations deep in the phase. This description however becomes unreliable on the Mott insulating side and in the vicinity of the superfluid-to-Mott-insulator transition as well as in the classical high-temperature liquid. To build an unbiased approach, we notice that from the Lie-algebraic point of view, the interaction terms are not any more complicated than the hopping terms, as both these terms represent bilinear forms in terms of  elements of the harmonic operator algebra.  In accordance with the general ``prescription'' from the previous section, we decouple the hopping terms first using the following algebraic identity
\begin{eqnarray}
\label{HSid} 
\nonumber
e^{\varepsilon_\tau t ( \bar{b}_i b_j + \bar{b}_j b_i )} = \int\! {d^2 a_l \over \varepsilon_\tau \pi / t } 
e^{- \varepsilon_\tau \left[ {1 \over t} \left|a_l \right|^2 +
\left( \bar{a}_l B_l + \bar{B}_l a_l \right) - t \left(n_i + n_j \right) \right]},
\end{eqnarray}
where $l = \langle i, j \rangle$ denotes a nearest-neighbor link (the links themselves form a square lattice), $\varepsilon_\tau$ corresponds to a single imaginary time-step in the truncated path integral, and 
\begin{equation}
\label{Bl}
B_l(\tau) \equiv b_i(\tau) - b_j(\tau),
\end{equation}
where we adopt the convention that for the horizontal links, $l = (i,i+\hat{\bf x})$, the plus sign in Eq.~(\ref{Bl}) corresponds to the $i$-th site on the left, and for the vertical links, $l = (i,i+\hat{\bf y})$, the plus sign corresponds to the $i$-th site in the bottom. This yields
\begin{eqnarray}
\label{ZBH2}
\nonumber 
{\cal Z}_{\rm BH} =&& \int \prod\limits_i  \left[ {\cal D}^2b_i(\tau)\right]  \prod\limits_l  \left[ {\cal D}^2 a_l(\tau)\right]
e^{-{1 \over t} \sum\limits_l \int\limits_0^\beta d\tau \left| a_l(\tau) \right|^2 }\\
&&\times \prod\limits_i \exp \Biggl\{ - \int\limits_0^\beta d\tau \Bigl[
 \bar{b}_i(\tau) \left( \partial_\tau - \tilde{\mu}\right) b_i(\tau)
 \nonumber \\
 && \phantom{aaa} + \left\{ \bar{\cal F}_i(\tau) b_i(\tau) + {\rm c.~c.} \right\} + U \left| b_i(\tau) \right|^4  \Bigr] \Biggr\},
\end{eqnarray}
where  $\tilde{\mu} = \mu_0 + U - 4t$ and
\begin{equation}
\label{curl}
{\cal F}_i(\tau) \equiv {\rm curl}_i a(\tau) = a_{(i,i+\hat{\bf x})} - a_{(i-  \hat{\bf x},i)} + a_{(i,i+\hat{\bf y})} - a_{(i-  \hat{\bf y},i)}
\end{equation}
is a curl of the Hubbard-Stratonovich field on the site $i$, which is a plaquette of the link lattice. 

\subsubsection{Hubbard-Stratonovich dynamical system for the harmonic oscillator algebra, $\mathfrak{h}_4$}

Let us now follow the general prescription of Sec.~\ref{sec:GenHS} and  decouple the on-site interaction as well with another set of real Hubbard-Stratonovich-fields to arrive to the following general expression
\begin{eqnarray}
\label{ZBH4}
{\cal Z}_{\rm BH} = &&\!\!\!\!\!\! \int   \prod\limits_l  \left[ {\cal D}^2 a_l(\tau)\right] \prod\limits_i  \left[ {\cal D}V_i(\tau)\right]
e^{-{1 \over t} \sum\limits_l \int\limits_0^\beta d\tau  \left| a_l(\tau) \right|^2 }\nonumber \\
&&\!\!\!\!\!\! \times e^{-{1 \over 4U} \sum\limits_i \int\limits_0^\beta d\tau V_i^2(\tau) } \prod\limits_i {\rm Tr}\, \hat{{\rho}}_i(\beta),
\end{eqnarray}
where the imaginary-time evolution operator, $\hat{\rho}_i(\tau)$, satisfies the following non-Hermitian Schr{\"o}dinger equation,
\begin{equation}
\label{nHSE}
-\partial_\tau \hat{{\rho}}_i(\tau) = \left\{ \left[ i V_i(\tau) - \tilde{\mu}\right] \hat{b}^\dagger \hat{b} +
 \left[ \bar{\cal F}_i(\tau) \hat{b} + {\cal F}_i(\tau) \hat{b}^\dagger \right] \right\} \hat{{\rho}}_i(\tau),
 \end{equation}
 with the initial condition, $\hat{{\rho}}_i(0) = \hat{1}$. Note that there is known to exist an ambiguity in selecting a channel in the Hubbard-Stratonovich transformation as discussed in Refs.~[\onlinecite{Kleinert1}], [\onlinecite{Kleinert2}], and [\onlinecite{Kleinert3}]. I  our case this subtlety manifests itself in that we can decompose the on-site term into linear combinations of either $\hat{b}^\dagger \hat{b}^\dagger$
 and $\hat{b}\hat{b}$ or $\hat{b}^\dagger\hat{b}$, which gives way to different single-particle algebras. 
 
We choose the latter decoupling, which leads to the  action in Eq.~(\ref{ZBH4}) that is not only local, but also corresponds to the (non-unitary) dynamics in the $\mathfrak{h}_4$ {\em solvable algebra.} Using the method introduced in Sec.~\ref{sec:DSBEh4}, we find the exact solution to Eq.~(\ref{nHSE}) as follows
 \begin{equation}
 \label{rhoh4}
 \hat{\rho}_i(\tau) = \exp \left[ - W_i(\tau) \hat{b}^\dagger \hat{b} \right]  e^{-\eta_{+,i}(\tau) \hat{b}^\dagger}
 e^{-\eta_{-,i}(\tau) \hat{b}} e^{\gamma(\tau)},
\end{equation}
where $W_i(\tau) = -\tilde{\mu}\tau + i \int\limits_0^\tau ds V_i(s)$, 
$\eta_{+,i} (\tau) = \int\limits_0^\tau ds {\cal F}_i(s) e^{W_i(s)}$, 
 $\eta_{-,i} (\tau) = \int\limits_0^\tau ds \bar{\cal F}_i(s) e^{-W_i(s)}$, and
 $\gamma_i(\tau) = \int\limits_0^\tau ds\int\limits_0^s ds' {\cal F}_i(s)\bar{\cal F}_i(s') \exp\left[ W_i(s') - W_i(s) \right]$.
 
To analyze the path integral (\ref{ZBH4}), we need to calculate the ``dynamic partition function,''
\begin{equation}
\label{rhoh4def}
{\rm Tr}\, \hat{{\rho}}_i(\beta) = e^{\gamma_i(\beta)} \sum\limits_{n=0}^{\infty} e^{- W_i(\beta) n} 
\left\langle n \left| e^{ - \eta_{+,i}(\beta) \hat{b}^\dagger} e^{ - \eta_{-,i}(\beta) \hat{b}} \right| n 
\right\rangle,
\end{equation}
where $|n\rangle = { \left(\hat{b^\dagger} \right)^n / \sqrt{n!} } |0\rangle$ are the usual harmonic oscillator eigenstates. 
As long as this series is convergent (the interaction $U$ is not too large), this trace can be evaluated exactly utilizing the 
Glauber coherent states and the result reads
\begin{equation}
\label{rhoh4exact}
{\rm Tr}\, \hat{{\rho}}_i(\beta) = {e^{\gamma_i(\beta)} \over 1 - e^{-W_i(\beta)} }\, 
\exp\left[ - {\eta_{+,i}(\beta)\eta_{-,i}(\beta) \over 1 - e^{-W_i(\beta)}} \right].
\end{equation}

The exact non-perturbative action is therefore given by the sum of these terms,
\begin{equation}
\label{S(f,a)}
S_{\rm eff} = \sum\limits_{i \in {\cal S}} 
\left\{ {\eta_{+,i}(\beta)\eta_{-,i}(\beta) \over 1 - e^{-W_i(\beta)}} - \gamma_i(\beta) + \ln \left[ 1 - e^{-W_i(\beta)} \right]\right\},
\end{equation}
where the functions involved in Eq.~(\ref{S(f,a)}) are defined below Eq.~(\ref{rhoh4}) and we recall that ${\cal F}_i(\tau) = {\rm curl}_i a(\tau)$, with $a_l(\tau)$ being a complex field defined on the links, $V_i(\tau)$ is a real field defined on the sites. {\em The exact partition function} is a path integral over both $a_l(\tau)$ and $V_i(\tau)$,
\begin{eqnarray}
\label{finalZh4}
\nonumber
{\cal Z}_{\rm BH} = \int  &&\!\!\!\!\!\! \prod\limits_l \left[{\cal D}^2 a_l(\tau)\right] \prod\limits_i \left[{\cal D}V_i(\tau)\right] e^{-S_{\rm eff}[a,V]}\\
&&\!\!\!\!\!\! \times  e^{ -  \int\limits_0^\beta d\tau \left[ {1 \over t} \sum\limits_l \left| a_l(\tau) \right|^2 - {1 \over 4U} \sum\limits_i V_i^2(\tau) \right]}.
\end{eqnarray}
We emphasize that the full action (\ref{S(f,a)})  contains simple Riemann integrals over imaginary time. Also, notice that had the field, $V_i(\tau)$, been absent, the partition function (\ref{rhoh4exact}) would have been manifestly real and positive definite and therefore it would not exhibit any sign problems. The presence of this field complicates matters in the quantum limit. However, since the main part of the action, $S_{\rm eff}[a,V]$, has been evaluated into an explicit form and the other two terms are comparatively trivial,  the numerical analysis of the action~(\ref{finalZh4}) may be an alternative to the conventional Monte-Carlo technique.

Also note that the fields $a_l$ and $V_i$ can be shown to be related to the superfluid and Mott order parameter correspondingly. Therefore, the exact non-linear action (\ref{ZBH4}) should contain a theory of the two competing orders, as well as a complete description of the superfluid-to-Mott-insulator quantum phase transition.  However, if the interaction becomes relatively strong, the series (\ref{rhoh4def}) becomes
manifestly divergent~\cite{Pollet_etal} and its careful resummation is required to get a meaningful result~\cite{VG_tbp}. These complications are not unexpected because we know that the $U \to \infty$ ``hardcore boson'' limit gives rise to the $XY$ quantum spin model~\cite{Assa_book} formulated in the complex extension of $\mathfrak{su}(2)$, and the corresponding $\mathfrak{sl}(2,\mathbb{C})$ Hubbard-Stratonovich dynamical system is not exactly solvable for an arbitrary quantum fluctuation field~\cite{VG_Qfl}. Note here that interestingly, the hardcore-boson limit can be viewed mathematically as a contraction of the infinite-dimensional single-site Bose-Hubbard algebra, 
\begin{equation}
\label{h^2}
\mathfrak{h}^2 = {\rm span}\, \left\{ \hat{n}^2,\hat{n},\hat{b}^\dagger,\hat{b},\hat{1}, \ldots \right\}
\end{equation}
 into $\mathfrak{su}(2)$ as $U \to \infty$. Clearly, the dual description of the corresponding $XY$-model should not include fluctuations of the site-field, $V_i(\tau)$, which will implement the no-double-occupancy constraint and reduce Eq.~(\ref{ZBH2}) to the dual partition function of the following type, (here by ``dual,'' we imply an unusual form of duality in lattice link-space, c.f., Ref.~[\onlinecite{Savit}], which is also different from duality in the preceding sections and Sec.~\ref{sec:DualHS}, which refers to the imaginary-time domain or equivalently quantum fluctuations):
\begin{eqnarray}
\label{ZXY3}
{\cal Z}_{XY} = \int  && \!\!\!\!\! \prod\limits_l  \left[ {\cal D}^2a_l(\tau)\right]  e^{-  {1 \over t} \sum\limits_l\int\limits_0^\beta d\tau  \left| a_l(\tau) \right|^2}\\
 && \!\!\!\!\!\!\!\!\!\! \times \prod\limits_i \left\{ 1 +  e^{-\mu_0\beta} \exp \left[ 2 {\rm Re}\, \int\limits_0^\beta d\tau\, {\rm curl}_i\, a(\tau) \right] \right\}.
 \nonumber
\end{eqnarray}
Here, $\mu_0$ is related to a uniform Zeeman magnetic field in the $z$-direction in the spin formulation and the curl of the link fields, $a_l(\tau)$ is defined in Eq.~(\ref{curl}).
  
 \subsubsection{Hubbard-Stratonovich dynamical system for the infinite-dimensional single-site Bose-Hubbard algebra, $\mathfrak{h}^2$}
 Finally, let us present yet another approach to analyze Bose-Hubbard model, which appears to be especially useful for  constructing various doping-dependent mean-field approximations of the theory~\cite{VG_tbp}. Notice, that action~(\ref{ZBH2}) was already local in site-fields and we introduced the additional on-site Hubbard-Stratonovich fields in order to reduce the problem to a tractable dynamical system for the solvable algebra, $\mathfrak{h}_4$. Alternatively, we can bypass this step and present the partition function in the following form instead
 \begin{eqnarray}
\label{ZBH21}
{\cal Z}_{\rm BH} = \int  \prod\limits_l  \left[ {\cal D}^2a_l(\tau)\right]
e^{-{1 \over t} \sum\limits_l \int\limits_0^\beta d\tau \left| a_l(\tau) \right|^2 } \! \prod\limits_i {\rm Tr}\, \hat{\cal P}_i(\beta),
\end{eqnarray} 
where now the ``density matrix'' is governed by the following Hubbard-Stratonovich dynamical system 
\begin{equation}
\label{nHSE3}
-\partial_\tau \hat{{\cal P}}_i(\tau) = \left\{  U \hat{n}_i^2 - \tilde{\mu} \hat{n}_i +
 \left[ \bar{\cal F}_i(\tau) \hat{b}_i + {\cal F}_i(\tau) \hat{b}_i^\dagger \right] \right\} \hat{{\cal P}}_i(\tau).
 \end{equation}
 Notice that the operators, $\left\{ \hat{n}_i^2,\hat{n}_i,\hat{b}_i,\hat{b}_i^\dagger,\hat{1} \right\}$, and their commutators do not close into a finite-dimensional algebra, but form an infinite-dimensional algebra instead. Eq.~(\ref{nHSE3}) does not appear to be solvable in the operator form for an arbitrary time-dependence of its coefficients. However, in the regime of strong interactions and on Mott insulating side, we can approximate the problematic interaction term as follows, $\hat{n}_i^2 \approx \hat{n}_i \langle n_i(\tau) \rangle$, which leads to a tractable self-consistent theory that retains quantum fluctuations. An analysis of the Bose-Hubbard model in the non-perturbative regime, including a careful analysis of its hardcore spin limit, will be presented elsewhere~\cite{VG_tbp}. 

\subsection{Dual action for a maximally-frustrated quantum spin model}
\label{sec:DualHS}
The single-particle part of this paper has focused on the dual approach to real-time dynamics for closed quantum systems  in which the trajectory in the algebra, i.e., the driving field, ${\bm b}(t)$, was related to the trajectory in the compact dynamic group, ${\bm \Phi}(t)$. It is interesting to see if  a similar dual approach can find its use in the many-particle context. The motivation for looking into this question is that a dual description gives the solution to quantum dynamics right away, i.e. it immediately determines ``the effective action,'' $S_{\rm eff} = -\sum\limits_i \ln \left[ {\rm Tr}\hat{\rho}_i(\beta) \right]$,  and circumvents the need to solve the corresponding differential equations. This approach is especially attractive in the cases where we integrate over all possible trajectories in the algebra and hence we may expect that such integration can be replaced by a path integral over the dual fields with readily available quantum dynamics.

As noted in the end of Sec.~\ref{sec:GenHS}, the direct mapping of the dual quantum-mechanical description based on the global exponentiation conjecture is not always possible for a generic Hubbard-Stratonovich dynamical system that arises in the many-particle context. This is because the corresponding dynamical system (\ref{nonunSE}), $i\partial_\tau \hat{\rho} \hat{\rho}^{-1} = {\bm \beta}(\tau) \cdot \hat{\bm J}$,  involves complex driving fields that double the dimensionality of the real algebra into its complex extension by adding non-compact generators. Due to the non-compactness of the corresponding dynamical group, there is a fundamental restriction that in this case we can not write down the non-unitary quantum evolution operator as a single exponential of algebra generators, i.e., generally, $\hat{\rho}(\tau) \ne \exp\left\{- \left[ {\bm R}(\tau) + i{\bm \Phi}(\tau) \right]\cdot \hat{\bm J} \right\}$ and we only know that a factorized version of the solution is available~\cite{WeiNorm1,WeiNorm2}. 

In some interacting many-particle models however the unitarity of quantum dynamics persists for Hubbard-Stratonovich dynamical systems as well.
A rule of thumb for the Hubbard-Stratonovich dynamical group to remain compact is that the interacting model should include only repulsion  (in the particle language) or antiferromagnetic terms (in the spin language). To provide an example, consider a quantum spin model with Heisenberg interactions
\begin{equation}
\label{Fspin}
\hat{\cal H} =  {J \over 2} \left( \sum\limits_{i=1}^N \hat{\bm \sigma}_i \right) \cdot \left( \sum\limits_{j=1}^N  \hat{\bm \sigma}_j \right).
\end{equation}
As evident from (\ref{Fspin}), the model assumes all spins interacting with each other, and therefore the geometry and dimensionality of the lattice play no role. Note that if the number of spins is three, $N=3$, the model (\ref{Fspin}) becomes a canonic toy model to illustrate the phenomenon of frustration. For a larger number of spins, it remains frustrated to a maximum due to the infinite-range antiferromagnetic interactions. Let us stress that the goal of this section is not to solve the spin model (\ref{Fspin}), but rather to illustrate the key features of the dual approach, which manifest themselves in more complicated models as well.

The thermodynamic partition function of the model (\ref{Fspin}) reads in the path-integral representation:
\begin{equation}
\label{ZFspin}
Z =  \int \prod \left[ {\cal D} \Omega_j(\tau) \right] e^{-\sum\limits_j S_{\rm WZ} [{\bm M}_j] - J [ \sum\limits_j {\bm M}_j(\tau) ]^2},
\end{equation}
where ${\bm M}_i(\tau)$ is the unit vector-field on the Bloch sphere for the $i$-th spin, $\Omega_i(\tau)$ is the corresponding solid angle, and $S_{\rm WZ}[{\bf M}]$ is the Wess-Zumino topological term in the action, whose explicit form is well-known~\cite{altland_condensed_2010}, but we do not need it (note that our general construction relies only on the existence of a path-integral). Let us now decouple the interaction term via a single Hubbard-Stratonovich field, ${\bm \Delta}(\tau)$, as follows (we omit below an insignificant overall constant):
\begin{eqnarray}
\label{ZFspinHS}
\nonumber
Z =  \int &&\!\!\!\!\!\!\!\! {\cal D}^3 {\bm \Delta}(\tau) \prod \left[ {\cal D} \Omega_j(\tau) \right] e^{-{1 \over 4J} \int\limits_0^\beta  d\tau {\bm \Delta}^2(\tau)}\\
&&\!\!\!\!\!\!\!\! \times  e^{-\sum\limits_j S_{\rm WZ} [{\bm M}_j]+ i \int\limits_0^\beta  d\tau {\bm \Delta}(\tau) \cdot {\bm M}(\tau)}.
\end{eqnarray}
Again, we can ``roll back'' the partition function into its Hamiltonian formulation to arrive to
\begin{eqnarray}
\label{ZFspin2}
Z =  \int {\cal D}^3 {\bm \Delta}(\tau) e^{-{1 \over 4J} \int\limits_0^\beta  d\tau {\bm \Delta}^2(\tau)} z^N[{\bm \Delta}(\tau)],
\end{eqnarray}
where $N$ is the number of spins and $z[{\bm \Delta}] = {\rm Tr}\, \hat{\rho}(\beta)$, where the matrix $\rho(\tau)$ follows from the Hubbard-Stratonovich  dynamical system
\begin{equation}
\label{SEFspin}
 \left\{
\begin{array}{ll}
i\partial_\tau \hat{\rho}(\tau) \hat{\rho}^{-1}(\tau)= {1\over 2} {\bm \Delta}(\tau) \cdot \hat{\bm \sigma}  \\
\hat{\rho}(0) = \hat{I}_2
\end{array}
\right.
\end{equation}
Due to the imaginary constant in Eq.~(\ref{SEFspin}), the dynamical group for this system is $SU(2)$ and therefore the problem maps onto unitary spin dynamics. Using the results of Sec.~\ref{sec:DSBEsu2},  we can write the solution in the dual form $\hat{\rho}(\tau) = \exp\left[ -{i \over 2} {\bm \Phi}(\tau) \cdot \hat{\bm \sigma} \right]$, where the dual field satisfies the dual Schr{\"o}dinger-Bloch equations (\ref{DSEsu(2)}) and thereby connects it to the Hubbard-Stratonovich field. Furthermore, using the elementary properties of the Pauli matrices, we find the ``dynamic partition function,'' $z[\Delta]$, in Eq.~(\ref{ZFspin2}) as follows:
\begin{equation}
\label{z1Fspin}
z[\Delta] = 2 \cos \left[\Phi(\beta)/2\right],
\end{equation} 
which takes on a very simple ``boundary form'' in the dual language. On the other hand, the Gaussian ${\bm \Delta}^2$-term in Eq.~(\ref{ZFspin2})
becomes non-trivial. 

If we now replace the integration over the original Hubbard-Stratonovich field with that over its dual counterpart, we can write symbolically the following ``dual'' expression for the partition function:
\begin{eqnarray}
\label{ZFspindual}
Z =  \int &&\!\!\!\!\!\!\!\!  {\cal D}^3 {\bm \Phi}(\tau)\,\, {\rm det} \left({\delta {\bm \Delta} \over \delta {\bm \Phi}}\right) 
e^{ N \ln \left\{ 2 \cos  \left[ \Phi(\beta) \over 2 \right] \right\} }\\
&&\!\!\!\!\!\!\!\!\!\!\!\!\!\!\!\!  \times \exp\left\{ -{4 \over J} \int\limits_0^\beta d\tau \left( {\dot{\bm n}}^2(\tau) \sin^2 \left[ {\Phi(\tau) \over 2} \right] + \left[ {\dot{\Phi}(\tau) \over 2} \right]^2 \right) \right\},
\nonumber
\end{eqnarray}
where ${\bm \Phi}(\tau) = \Phi(\tau) {\bm n}(\tau)$ with $|{\bm n}(\tau)| \equiv 1$ and ${\rm det} \left({\delta {\bm \Delta} \over \delta {\bm \Phi}}\right)$ denotes a Jacobian of the transform from the original Hubbard-Stratonovich
field covering all possible trajectories in the Euclidean space, $\mathbb{R}^3$ to the dual field covering all trajectories in the group, $SU(2) \sim S^3$. Let us reiterate that in doing this transformation, we ``trade off'' the  simplicity of the Gaussian term in the direct approach of Eq.~(\ref{ZFspin2}) for the simplicity of the quantum evolution term, ${\rm Tr}\, \hat{\rho}(\beta)$. The price we pay for this however is that the former, initially trivial Gaussian term that penalizes any fluctuations takes on the form of a non-linear functional, see the last term in Eq.~(\ref{ZFspindual}). Remarkably this factor in new variables describes a free particle moving in a three-dimensional sphere. Indeed, if we set ${\bf n} = \left( \sin\theta \cos\phi, \sin\theta \sin\phi, \cos\theta \right)$ and $w_1 = r \cos\theta$, $w_2 = r \sin\theta \cos\phi$, $w_3 = r \sin\theta \sin\phi \cos\left(\Phi/2\right)$, and $w_4  = r \sin\theta \sin\phi \sin\left(\Phi/2\right)$, then the Lagrangian in Eq.~(\ref{ZFspindual}) becomes ${\cal L} = {4 \over J}  \left( {\dot{\bm n}}^2(\tau) \sin^2 \left[ {\Phi(\tau) \over 2} \right] + \left[ {\dot{\Phi}(\tau) \over 2} \right]^2 \right)  = \sum_{i=1}^4 \dot{w}_i^2$ provided that $r = 2/\sqrt{J}$ is held constant.

Finally, let us mention a curious property of the dual representation (\ref{ZFspindual}): if the number of spins is even, all terms in the partition function (\ref{ZFspindual}) are manifestly real and positive definite. This is to be contrasted with the direct Hubbard-Stratonovich representation (\ref{ZFspin2}), which exhibits a ``sign problem'' because different trajectories of the field, ${\bm \Delta}(\tau)$, give rise to fast-oscillating terms in the path-integral. This observation suggests that the dual approach proposed here may be a promising avenue to cure the sign problem in certain theories.  

\section{Summary}
\label{sec:Summary}
In this paper, we have put together a general Lie-algebraic approach to analyze quantum dynamics. The punchline of the first part of the paper is that for a large class of non-equilibrium quantum systems, the notion of a Hilbert space, central to the conventional Schr{\"o}dinger formulation, is somewhat of a red herring, as the choice of a specific representation hides a more primitive Lie-algebraic structure of the Hamiltonian. We suggest and demonstrate on a number of explicit examples that the dual Schr{\"o}dinger-Bloch equation for the Hilbert-space-invariant generators is a conceptually viable and practically useful alternative to the conventional approach.  Let us emphasize that this and some other results and statements of Secs.~\ref{sec:InvariantForm} and \ref{sec:DualSBE} are not new and they appear to be scattered over the literature~\cite{Magnus,Wilcox,WeiNorm1,WeiNorm2,Alhassid1,Alhassid2,Vourdas,ChemPhysRev,MagnusRev} on quantum control theory~\cite{DAlessandro_book}, generalized coherent states~\cite{Gilmore_RMP}, as well as in the mathematical literature on the Lie theory and dynamical systems~\cite{Gilmore_book}. In fact, the equation identical to what we called here ``dual Schr{\"o}dinger-Bloch equation'' appears in the introductory chapters of many mathematical textbooks on Lie algebras~\cite{Sternberg} in the context of the derivation of the Baker-Campbell-Hausdorff formula. However, this particular form (\ref{dSE}) of the equation is not normally assigned major importance, perhaps because the mathematical researches often tend to study Lie groups in their utmost generality and  it has been long known since the late nineteenth century that for generic non-compact groups, a global exponential map generally does not exist. Due to this well-known fact, there has been no reason to study the equations of motion for ``global generators'' in the case of a generic dynamical system.  On the other hand, as emphasized in this paper, a large number of quantum-dynamical systems seem to give rise to dynamic groups with surjective exponential map, where quantum evolution operator  allows global exponentiation from the algebra. This observation seems to enhance the importance of the dual approach based on the Magnus representation~\cite{Magnus} and in fact one may argue that whether to take the Schr{\"o}dinger equation in a Hilbert space in its canonical form~(\ref{SE1}), or its dual Hilbert-space-invariant version (\ref{dSE}) as a starting point becomes a question of convention and convenience in this case.

We further argued in this paper, that the dual generators, ${\bm \Phi}(t)$, provide a transparent and simple way to establish a correspondence between
unitary quantum evolution and deterministic dynamics of the corresponding classical  system.  A concise summary of quantum-to-classical correspondence is as follows: Quantum dynamics maps an initial normalized state into a final normalized state in an $L$-dimensional Hilbert space, $|\Psi(t) \rangle = \exp \left[ - i {\bm \Phi}(t) \cdot \hat{\bm J}_L \right] | \Psi(0) \rangle \subset S^{2L -1} \subset  {\cal H}il(L)$, via action of a dynamic group in the $L$-dimensional unitary representation (i.e., $\hat{\bm J}_L$ above are $L \times L$ Hermitian matrices). Classical dynamics maps the initial coordinate-vector  onto a final coordinate-vector in such a way that both lie within a classical manifold, ${\cal M}$, embedded in a $d$-dimensional classical phase space,  ${\bm M}(t) = \exp \left[  -{\bm \Phi}(t) \cdot \hat{\bm f}_d \right] {\bm M}(0) \subset {\cal M} \subset \mathbb{R}^d$, and this dynamics occurs via action of a symmetry group in its $d$-dimensional adjoint representation (i.e., $\hat{\bm f}_d$ are $d \times d$ matrices with real entries). Quantum-to-classical correspondence is in that  the group action in both cases is governed by the same dual generators, ${\bm \Phi}(t)$, and thereby determines the solution to both the Schr{\"o}dinger equation and the corresponding generalized Bloch equations. The simplicity of this formulation raises the question of how general it is  and also concerns in relation to chaotic systems, where quantum-to-classical correspondence is far from clear-cut. Note  that our approach hinges on the existence of a global exponential map, which is guaranteed to exist for a certain (limited) class of finite-dimensional algebras. The general conditions to have such a map appears to remain an open mathematical question under debate and even less is known about infinite-dimensional algebras. Therefore, our construction does not directly apply to quantum models with infinite-dimensional algebras, nor to chaotic systems. 

Let us also point out that while the dual approach does not perform the miracle of solving differential equations that are otherwise unsolvable in the direct formulation, it does appear amazingly useful in constructing exactly-solvable non-linear models. Ref.~[\onlinecite{Our_TLS}] and Sec.~\ref{sec:DSBEsu2} have demonstrated that the method can be used to reverse-engineer a variety of exactly-solvable two-level-system dynamics and Sec.~\ref{sec:DSBEsu(N)} put forward a constructive five-step procedure of reverse-engineering exact solutions in more complicated higher-dimensional groups. In fact, the inspection of the latter five-step procedure indicates that it should be possible to start with a rather generic polynomial $P(\omega,t) = \prod\limits_k \left[ \omega - \omega_k(t) \right]^{\nu_k}$, which with some minor restrictions on the functions $\omega_k(t)$ and $\nu_k \in \mathbb{Z^+}$, can be viewed as a secular equation of an algebra, and then build upon it a quantum dynamical system, which would be exactly-solvable by construction. 

While quantum-to-classical correspondence and the general structure of the theory in the single-particle case (finite-dimensional algebra) appear to be understood, it is far from being so in the case of interacting many-particle models. In Sec.~\ref{sec:Lielattice} and Ref.~[\onlinecite{VG_Qfl}], we attempted to extend the Lie-algebraic view to a certain wide class of quantum lattice models (excluding various topological models, where interactions between different algebra generators, or equivalently generalized ``spin-orbit-couplings,'' would be allowed). The key new step proposed here is to use a generalized Hubbard-Stratonovich decomposition~\cite{VG_Qfl}, which maps a large, often infinite-dimensional, interacting algebra into an ensemble of finite-dimensional single-algebra (think single-particle) Hubbard-Stratonovich dynamical systems. One major difference between our approach and that of conventional Hubbard-Stratonovich decomposition is that we focus on algebra generators instead of counting the number of creation/annihilation operators in the conventional Fock representation. As demonstrated in Sec.~\ref{sec:BH}, this approach is quite useful especially for lattice models based on solvable algebras (such as the Bose-Hubbard model built of harmonic oscillators on lattice sites), where an arbitrary dynamical system, unitary or non-unitary, is explicitly exactly-solvable. 

In conclusion, let us note that an important wide-open mathematical question in addressing the many-particle lattice models of type~(\ref{Hint}) is the applicability of the path-integral approach itself~\cite{KlauderPI}. In particular, a recent Letter of Wilson and the author~\cite{JW_VG} has demonstrated that the standard normal-ordered path-integral seems to break down even for a model as simple as the single-site Bose-Hubbard model and while a way out was found to correct the theory into a sensible form, the latter non-normal-ordered construction is not based on any solid mathematical arguments, and therefore a careful investigation of this controversy is clearly called for. 
  
 \vspace*{0.1in} 
\begin{acknowledgments}
The author is grateful to Alex Dragt,  Anirban Gangopadhyay, Anatoli Polkovnikov, and Justin Wilson for discussions related to some aspects of  this work. This research was supported by DOE award DESC0001911.
\end{acknowledgments}

\appendix

\section{General Properties of Quantum Dynamics}
\label{sec:GenProp}
The explicit form of the Hilbert-space-invariant DSBE (\ref{dSE}) depends critically on the structure constants of the underlying Lie algebra. In this appendix, we make a few general remarks about unitary evolution and the structure of Eq.~(\ref{dSE}), which can be classified in a one-to-one correspondence with classification of Lie algebras. 

\subsection{Decomposition of the algebra into a standard form}
There exist powerful and effective methods of decomposing an arbitrary Lie  algebra into a standard form~\cite{Gilmore_book}. The first step involves factorization of ${\cal A}$ into a  solvable subalgebra,  ${\cal A}_0$, and a semisimple component, ${\cal A}_{\rm sem\,sim}$. The Levi decomposition~\cite{Levi} below exists for {\em any finite-dimensional algebra with no exceptions}
\begin{equation}
\label{Adec}
{\cal A} = {\cal A}_0 + {\cal A}_{\rm sem\,sim} = {\cal A}_0 + \sum\limits_{\nu = 1}^{N_{\rm s}} {\cal A}_{\nu},
\end{equation}
such that $\left[{\cal A}_{\rm sem\, sim},\, {\cal A}_{\rm sem\,sim} \right] \subset {\cal A}_{\rm sem\,sim}$, $\left[{\cal A}_{0},\, {\cal A}_{0} \right] \subset {\cal A}_{0}$, and $\left[{\cal A}_{\rm sem\,sim},\, {\cal A}_{0} \right] \subset {\cal A}_{0}$. By definition, the semisimple  component may  be decomposed into a combination of mutually-commuting simple  sub-albegras: ${\cal A}_{\rm sem\,sim} = \sum_{\nu = 1}^{N_{\rm s}} {\cal A}_{\nu}$, where we $N_{\rm s}$ is the number of such simple subalgebras. Note that our ability to actually decompose the algebra into this  form and also to extract the simple components is tied closely to our ability to solve a generic eigenvalue problem on the algebra, which corresponds in our case to finding the spectrum of a generic stationary Schr{\"o}dinger equation.

\subsection{Factorization of the evolution operator for solvable algebras}
\label{sec:Fact}
We recall that a solvable Lie algebra is spanned by the generators 
${\cal A}_{0} = {\rm span}\, \left\{ \check{J}_1^{(0)},\ldots,\check{J}_{d_0}^{(0)}\right\}$, such that $\forall \check{X} \in {\cal A}_{0},\, {\rm ad}_{\check{X}} \check{J}_a^{(0)} = \sum\limits_{ b \geq a} L^b_a \check{J}_b^{(0)}$ (nilpotent algebras include a strict inequality in the sum, i.e., $b > a$). In the language of the quantum dynamical problem defined by (\ref{HinA}), including its DSBE form (\ref{dSE}), it implies the following: The non-equilibrium dynamics in the solvable algebra reduces to a chain of ordinary differential equations that can be solved one-by-one using the graded structure. A solution can be obtained via factorization of the evolution operator as follows. Consider the Hamiltonian, which is a trajectory in a solvable algebra, ${\cal A}_0$, $\check{\cal H}_0(t) = b^{a}(t) \check{J}^{(0)}_a$. First, solve for dynamics of $\check{J}^{(0)}_1$,  as follows
$\check{U}_{1}(t) = \exp \left[ - i \check{J}^{(0)}_1 \int\limits_0^t dt' b^1(t') \right]$. Determine new ``driving fields'' for $1 < a \leq d_0$, $\sum_{a>1} b^{a}(t) \check{U}_{1}^\dagger (t)  \check{J}^{(0)}_a \check{U}_{1} (t) = \sum_{a>1} \tilde{b}^{a}(t)  \check{J}^{(0)}_a$. Note that no $\check{J}^{(0)}_1$-dependent terms are produced in the solvable case. Repeat the procedure by solving for dynamics of $\check{J}^{(0)}_2$ in the new field, $\check{U}_{2}(t) = \exp \left[ - i \check{J}^{(0)}_2 \int\limits_0^t dt' \tilde{b}^2(t') \right]$, {\em etc.}, until   the last generator is eliminated and the problem is solved in the following factorized form
\begin{equation}
\label{Factsolv}
\check{U}_{{\cal A}_0}(t) = \prod_{a = 1}^{d_0} \check{U}_{a}(t).
\end{equation}
The difference between this factorized solution~\cite{WeiNorm1,WeiNorm2} obtained  from the standard Schr{\"o}dinger equation (\ref{SE3}) and a solution to the  DSBE (\ref{dSE}) is that the latter would determine the evolution operator as a single exponential directly, effectively implementing the BCH identity in the product (\ref{Factsolv}), $\check{U}_{{\cal A}_0}(t) = \exp\left[ - i \sum\limits_{a=1}^{d_0} \Phi_0^a(t) \check{J}^{(0)}_a \right]$. However, the solvable structure remains in the DSBE as well, which should factorize into a chain of ordinary differential equations. An example of such a solution for the nilpotent algebra, $\mathfrak{h}_3$, is presented in Sec.~\ref{sec:DSBEh4}.

\subsection{Simple-algebra components}

The standard Lie-algebraic terminology  becomes a bit misleading, when we consider the simple algebra components, which in fact are the most complicated, because a simple algebra is an algebra whose generators  in a sense are all ``tied together'' via commutators (any element in a simple algebra is expressible as a commutator of two other elements). In this case, the factorization trick is not useful and the corresponding DSBE has the form of coupled matrix differential equations, as opposed to a chain of ordinary differential equations for solvable or nilpotent algebras.  On the other hand, the simple algebras have a very rigid structure. A rank-$r$ simple algebra can always be represented as an Abelian subalgebra of $r$-mutually commuting generators, $\check{h}_k$  [analogous to $\check{J}_z$ in $\mathfrak{su}(2)$] and ``raising/lowering'' operators $\check{E}_{\bm \alpha}$ and $\check{E}_{-{\bm \alpha}}$ [analogous to $\check{J}^\pm$ in $\mathfrak{su}(2)$] labelled by allowed root-vectors in the $r$-dimensional Euclidean root space and such that $\left[ \check{h}_i,\,\check{E}_{\pm {\bm \alpha}} \right] = \pm \alpha_i \check{E}_{\bm \alpha}$. The possible root vectors in the simple algebra are strictly constrained and so are the commutation relations, $\left[ \check{E}_{\bm \alpha},\,\check{E}_{\bm \beta}\right]$, which gives rise to a complete classification of the simple algebras based on the analysis of root spaces and Dynkin diagrams, as discussed in the vast literature on the subject (see, {\em e.g.}, Ref.~[\onlinecite{Gilmore_book}]).

If the dynamics of the simple component(s) of ${\cal A}$ have been determined, to treat the remaining non-universal solvable component, ${\cal A}_0$,  if any, becomes a straightforward exercise that can be accomplished via the same factorization trick. 
Indeed, assume that  the mutually-commuting evolution operators are known explicitly, $\check{U}_{{\cal A}_\nu}(t) = \exp \left[ - i {\bm \Phi}^{(\nu)} (t) \cdot \check{\bm J}^{(\nu)} \right]$. Then, the total evolution operator can be written in the following factorized form:
\begin{equation}
\label{Stot}
\check{U}_{\cal A}(t) = \left[ \prod_{\nu}  \check{U}_{{\cal A}_\nu}(t) \right] \check{U}_{{\cal A}_0}(t),
\end{equation}
where the last factor, $\check{U}_{{\cal A}_0}(t)= \exp \left[ - i {\bm \Phi}_0 (t) \cdot \check{\bm J}^{(0)} \right]$, 
satisfies the Schr{\"o}dinger equation in ``interaction representation,'' $i \partial_t \check{U}_{{\cal A}_0}(t) = \tilde{\bf b}_0(t) 
\cdot \check{\bm J}^{(0)} \check{U}_{{\cal A}_0}(t)$,
with the rotated dynamic fields, $\tilde{b}_0^a(t)$  determined via 
$$
b^a_0(t) \left[ \prod_{\nu}  \check{U}_{{\cal A}_\nu}(t) \right]^\dagger \check{J}^{(0)}_a \left[ \prod_{\nu}  \check{U}_{{\cal A}_\nu}(t) \right] = \tilde{b}^a_0(t) \check{J}^{(0)}_a.
$$
Note that since $\left[ {\cal A}_\nu,\, {\cal A}_0 \right] \subset {\cal A}_0$, the ``interaction representation form'' of the Schr{\"o}dinger equation closes within the solvable component and the remaining  exact solution in ${\cal A}_0$ can  be obtained from a chain of ordinary first-order differential equations by a variety of standard methods. 

 Therefore, classifying possible quantum dynamical systems and the corresponding DSBE reduces to an analysis of the corresponding equations within each of the classical algebras [i.e., $A_l \sim \mathfrak{su}(l+1)$, $B_l \sim \mathfrak{so}(2l+1)$, $D_l \sim \mathfrak{so}(2l)$, and $C_l \sim \mathfrak{sp}(l)$ ] and the five exceptional algebras, $G_2$, $F_4$, $E_6$, $E_7$, and $E_8$. These possibilities exhaust completely all  universal quantum dynamics that can possibly arise from an arbitrary non-equilibrium Hamiltonian matrix (modulo a non-universal solvable or nilpotent part, whose solution is comparatively trivial as discussed above). Apart from fundamental mathematical restrictions on the structure of the simple components, physics provides further constraints: if a problem can be formulated in a physical $L$-dimensional Hilbert space, then we must demand that the group, $G_{\cal A}$, be a subgroup of $SU(L)$. Eventhough, as we discussed above, quantum dynamics is completely decoupled from the Hilbert space, the very existence of it and the unitarity requirement provide certain constraints on types of allowed evolutions, $\check{U}(t) \subset G_{\cal A}\subset SU(L_{\rm min})$ (here, $L_{\rm min}$ can be taken to be the minimal dimension of a faithful unitary representation).

\subsection{Generalizing theory to certain infinite-dimensional representations}
\label{SSsec:Infinite_d}
Our original discussion has descended from the textbook Schr{\"o}dinger equation~(\ref{SE1}) with the assumption that the Hamiltonian be a finite-dimensional Hermitian matrix. This led to a finite-dimensional Lie algebra, which by construction had a finite-dimensional faithful representation. This in turn led to the conjecture about the global exponentiation of the algebra onto a compact Lie group. For our purposes motivated by applications to physics, it is the existence of a global exponential map that is crucial, while the compactness of the group is not germane and in fact there are many theories involving infinite-dimensional representations where it is manifestly not so ({\em  e.g.}, the harmonic oscillator). 

The global covering problem in its full generality is an old and very complicated mathematical question~\cite{EXPLie1} and we are in no position to comment on it here. However, we outline a class of models, where global exponentiation should hold irrespectively
of the compactness of the dynamic Lie group, $G_{\cal A}$. Consider, a Lie algebra, $\tilde{\cal A} = {\rm span} \left\{ \check{J}_1,\ldots,\check{J}_d \right\}$ with structure constants, $\tilde{f}_{ab}^{\phantom{ab}c}$, which admits global exponentiation. Form parameter-dependent  linear combinations of generators,  $\check{\cal J}_a = \sum\limits_{b=0}^d C_a^{\phantom{a}b}(\varepsilon) \check{J}_b$. This leads to parameter-dependent structure constants, $f_{ab}^{\phantom{ab}c}(\varepsilon)$. Taking the singular limit, $\varepsilon \to 0$, yields  in certain cases a new algebra ${\cal A} = \lim\limits_{\varepsilon \to 0} \tilde{\cal A}(\varepsilon)$ distinct from the original algebra. This procedure is called a Lie-algebraic contraction (a famous example of it is provided by the Galilean group viewed as a non-relativistic limit/contraction of the Lorentz group as the speed of light is taken to infinity, $\varepsilon =1/c \to 0$). Another example more relevant to the ongoing discussion is provided  in Sec.~\ref{sec:contra}, which briefly reviews how the harmonic oscillator algebra, $\mathfrak{h}_4$, can be obtained via contraction from the $\mathfrak{su}(2)$ spin algebra plus an Abelian $\mathfrak{u}(1)$-component. Both $\mathfrak{u}(2)$ and $\mathfrak{h}_4$ allow global exponentiation, eventhough the latter exponentiates in a non-compact group with no finite-dimensional unitary representation. We want to view this fact as a manifestation of the general conjecture that if an algebra, ${\cal A}$ can be contracted from a Lie algebra, ${\cal A} = \lim\limits_{\varepsilon \to 0}\tilde {\cal A}(\varepsilon)$ that allows global exponentiation onto a dynamic Lie group, then the former also does. In the context of models of relevance to condensed matter, this conjecture implies, in particular, that a variety of lattice boson models would allow global exponentiation even in the absence of a finite-dimensional matrix unitary representation. 

\section{Classical Manifold and Path Integral for Quantum Dynamics}
\label{app:clasM}
The interesting question reviewed in this appendix is how to generally obtain the classical manifold, ${\cal M}$, where classical motion is taking place. This is best elucidated in the context of the path integral formalism that can be generalized to describe dynamics descending from a generic Lie algebra and making use of generalized coherent states~\cite{Perelomov_book,Gilmore_RMP} introduced by Perelomov~\cite{Perelomov_CS} and Gilmore~\cite{Gilmore_CS} for an arbitrary Lie group. 

The usual path-integral construction~\cite{Kleinertbook} calculates averages or transition amplitudes directly and therefore must include information about the Hilbert space. This is despite the fact that the ``bulk'' of the path integral calculates the unitary 
evolution, which, as we just saw, is decoupled from both the initial conditions and the Hilbert space. The first step in building a path-integral representation is to express the evolution operator  as follows 
\begin{equation}
\label{Texp}
\check{U}(t) = T e^{-i \int\limits_0^t ds b^a(s) \check{J}_a} \approx \lim\limits_{N \to \infty}
\prod_{n=0}^N e^{-{it \over N}  b^a(t_n) \check{J}_a} \in G_{\cal A},
\end{equation}
where $t_n = tn/N$ represent time-slices. Then {\em a particular representation} is chosen, $T_L[{\cal A}]$, and a representation of unity in a proper basis, e.g., the overcomplete basis of  coherent states, is inserted between each pair of exponentials in the product (\ref{Texp}) for the given representation. This procedure is based on the existence of a set of states, $| {g} \rangle \in {\cal H}il(L)$, parametrized by the group, $g \in G_{\cal A}$,  such that 
\begin{equation}
\label{I1}
\hat{I}_L = \int\limits_{G_{\cal A}} d\mu({g}) | {g} \rangle \langle {g} |,
\end{equation}
where $d\mu({g})$ is a de Haar measure in the group, $G_{\cal A}$. Being parametrized by the continuous Lie group, the set of states, $| {g} \rangle$, is way overcomplete.  However, as emphasized by Perelemov~\cite{Perelomov_CS}, a part of the group integration is always redundant and in fact a more convenient parametrization exists, still overcomplete, but where the basis runs over a smaller manifold, ${\cal M}$, to be identified  with the manifold, where  classical dynamics  is taking place for the generalized Bloch equations (\ref{Bloch}).

The Perelomov construction is as follows: take an arbitrary normalized  state, $| \psi_0 \rangle$,  in the Hibert space. In the context of the dynamical quantum system at hand (\ref{SE1}), the initial condition provides the most natural choice of such a state, i.e., $|\psi_0 \rangle = |\Psi(0) \rangle$. Define a subgroup $H\left(| \psi_0 \rangle\right) \subset G_{\cal A}$ as a set of group elements, $\check{h} \in G_{\cal A}$, such that
\begin{equation}
\label{p}
\hat{h} | \psi_0 \rangle = e^{i \phi(h)}| \psi_0 \rangle,
\end{equation}
i.e., whose action on the initial condition reduces to a multiplication by a complex phase factor [in Eq.~(\ref{p}), $\hat{h} = \hat{T}_L(\check{h})$]. This maximum stability subgroup, $H\left(| \psi_0 \rangle\right)$, defines a coset ${\cal M} = G_{\cal A}/H$, which does not necessarily have a group structure, but retains a topological structure. The coherent states of Perelomov, $| {\bf M} \rangle$, are an overcomplete basis in the Hilbert space parametrized over the manifold, ${\bf M} \in {\cal M} = G_{\cal A}/H$.


As to the path integral itself, its precise form depends on what we wish to calculate using it. A rather generic structure is the transition amplitude between two coherent states, which we can call a Green function,
\begin{equation}
\label{PI}
{\cal G}({\bf M}_1,{\bf M}_2;t) = \int\limits_{{\bf M}(0)={\bf M}_1}^{{\bf M}(t) = {\bf M}_2} [{\cal D}{\bf M}(t)] \exp\left\{ -i S[{\bf M}(t)]\right\}.
\end{equation}
Here ${\bf M}(t) \subset {\cal M} \subset \mathbb{R}^d \sim {\cal A}^*$, {\em  i.e.}, the generalized coherent states  lie in the manifold ${\cal M}$ embedded without crossings into the $d$-dimensional Euclidean space, $\mathbb{R}^d$, associated with the underlying algebra. Since ${\rm dim}\left( G_{\cal A}\right) = {\rm dim} \left( {\cal A} \right) = d$ and ${\cal M} = G_{\cal A}/H$, the dimensionality of ${\cal M}$ is always smaller than $d$. Quite generally, we expect ${\rm dim} \left({\cal M}\right) \leq {\rm dim}({\cal A}) - 
\mathfrak{Ran}\mbox{\bf k} \left({\cal A}\right)$. E.g., the Bloch sphere, ${\cal M} = S^2 = SU(2)/U(1) \sim S^3/S^1$ is two dimensional and is embedded in the three-dimensional space associated with the $\mathfrak{su}(2)$ algebra. 

If we follow the usual logic in deriving the  path integral and postulate on physical grounds the continuity of the corresponding trajectories on ${\cal M}$, the explicit form of the action will take the form, $S = S_H[{\bf M}(t)] + S_{\rm Berry} [{\bf M}(t)]$, where the first Hamiltonian term i s$S_{\cal H}[{\bf M}(t)] \propto \int\limits_0^t ds {\bf b}(s) \cdot {\bf M}(s)$, and the second Berry phase term, $S_{\rm Berry} [{\bf M}(t)] \propto \int\limits_0^t ds \langle {\bf M}(s)| \partial_{s} |{\bf M}(s) \rangle$, incorporates commutation relations and/or topological effects such as the Wess-Zumino term in the spin path-integral. The exact form of the Berry phase action is determined by the structure constants of the algebra. E.g., for spin algebra with $f^{abc} = \varepsilon^{abc}$, it is the Wess-Zumino term, $S_{\rm Berry} [{\bf M}(t)] = i S \int_0^1 du \int\limits_0^t ds f^{abc}   \partial_u {M}_a(u,s) \partial_s {M}_b(u,s) M_c(u,s)$, where the dynamic magnetization, ${\bf M}(t)$, has been assigned an auxiliary argument $u$, in a non-unique way, but such that ${\bf M}(0,\tau) \equiv {\bf M}_0$ and ${\bf M}(1,s) = {\bf M}(s)$, therefore describing a string sweeping the sphere over time~\cite{altland_condensed_2010,Sachdevbook}. An interesting  question is to see if the topological terms arising from  dynamics in other compact groups may  be cast in such a form, being expressed through the structure constants of the algebra and measuring a volume covered on the manifold in the course of a particular realization of classical evolution. 

Path-integral lore due to Feynman suggests that if we now introduce a set of local coordinates, $(\theta_1,\ldots,\theta_D)$, in ${\cal M}$ [where $D = {\rm dim}\, {\cal M}$ is the dimensionality of the classical manifold], which is always possible since it is a topological space, then the minimization of the action with respect to $\theta_l$ will reproduce the classical equations of motion from Sec.~\ref{sec:QCC}, i.e., generalized Bloch equations~(\ref{Bloch}), but now manifestly defined on the classical manifold, ${\cal M}$.

Finally, let us make the following  speculative remarks: If we consider all possible Hilbert spaces associated with physical representations of a given algebra, ${\cal A}$, we shall obtain a set of classical manifolds. In complicated higher-rank algebras, such as $SU(3)$, they may in principle differ in topology and even dimensionality~\cite{Gilmore_RMP}. However, there exists  perhaps a reasonable  ``estimate'' on their possible dimensionality as follows, ${\rm  dim} \left[ G_{\cal A}/  U (r) \right] \leq {\rm  dim} \left(
{\cal M} \right) \leq {\rm  dim} \left[ G_{\cal A} / U^r(1 )\right] = d - r$, where $r =  \mathfrak{Ran}\mbox{\bf k} \left({\cal A}\right)$.  All possible classical manifolds, ${\cal M}$, can be embedded into the Euclidean space, $\mathbb{R}^d \sim {\cal A}^*$, which is the complete phase space of the classical problem (\ref{Bloch}) for the averages. If we parametrize all manifolds in some coordinate system in $\mathbb{R}^d$ and use certain units along the axes, which naturally should be related to the Planck constant, $\hbar$, we shall obtain a variety of possible classical manifolds that do not fill the entire phase space $\mathbb{R}^d$, but select the regions of dynamics for the averages, which are allowed by quantum mechanics. These manifolds also should respect and actually follow from the constraints imposed by $r$  independent Casimir invariants defined by the algebra. E.g., for $\mathfrak{su}(2)$, the set of allowed dynamic manifolds are the concentric Bloch spheres in $\mathbb{R}^3$ with the radii, $|{\bf M}| = {\hbar \over 2}, \hbar, {3 \hbar \over 2}, \ldots$. This is a way to visualize  the quantization constraints imposed by quantum mechanics. On the other hand the  Bloch equations (\ref{Bloch}) do not include the Planck constant and are formally not restricted by any quantization constraints. We can take arbitrary initial conditions for ${\bf M}(0)$ in $\mathbb{R}^d$, and even if they contradict the quantization constraints, the classical Bloch equations would remain solvable. The ``paradox'' may be resolved by taking the quasiclassical limit in the quantum problem, $\hbar \to 0$, which in the cartoon geometric picture proposed above would imply that the set of allowed manifolds fills up the entire phase space in this limit. E.g., the spacing between the Bloch spheres corresponding to different representations of the spin would disappear in the quasiclassical limit.

\bibliography{QCC-1}

\begin{thebibliography}{95}
\expandafter\ifx\csname natexlab\endcsname\relax\def\natexlab#1{#1}\fi
\expandafter\ifx\csname bibnamefont\endcsname\relax
  \def\bibnamefont#1{#1}\fi
\expandafter\ifx\csname bibfnamefont\endcsname\relax
  \def\bibfnamefont#1{#1}\fi
\expandafter\ifx\csname citenamefont\endcsname\relax
  \def\citenamefont#1{#1}\fi
\expandafter\ifx\csname url\endcsname\relax
  \def\url#1{\texttt{#1}}\fi
\expandafter\ifx\csname urlprefix\endcsname\relax\def\urlprefix{URL }\fi
\providecommand{\bibinfo}[2]{#2}
\providecommand{\eprint}[2][]{\url{#2}}

\bibitem[{\citenamefont{Feynman}()}]{Feynman_PI}
\bibinfo{author}{\bibfnamefont{R.~P.} \bibnamefont{Feynman}},
  \bibinfo{howpublished}{Rev. Mod. Phys. {\bf 20}, 367 (1948)}.

\bibitem[{\citenamefont{Glauber}()}]{Glauber}
\bibinfo{author}{\bibfnamefont{R.~J.} \bibnamefont{Glauber}},
  \bibinfo{howpublished}{Phys. Rev. {\bf 131}, 2766 (1963)}.

\bibitem[{\citenamefont{Klauder}({\natexlab{a}})}]{Klauder_CS}
\bibinfo{author}{\bibfnamefont{J.~R.} \bibnamefont{Klauder}},
  \bibinfo{howpublished}{J. Math. Phys. {\bf 4}, 1055 (1963); {\em ibid.}, {\bf
  4}, 1058 (1963)}.

\bibitem[{\citenamefont{Perelomov}({\natexlab{a}})}]{Perelomov_CS}
\bibinfo{author}{\bibfnamefont{A.}~\bibnamefont{Perelomov}},
  \bibinfo{howpublished}{Commun. Math. Phys. {\bf 26}, 222 (1972)}.

\bibitem[{\citenamefont{Gilmore}({\natexlab{a}})}]{Gilmore_CS}
\bibinfo{author}{\bibfnamefont{R.}~\bibnamefont{Gilmore}},
  \bibinfo{howpublished}{Ann. Phys. (NY) {\bf 74}, 391 (1972); Rev. Mev. de
  Fisica {\bf 23}, 142 (1974)}.

\bibitem[{\citenamefont{Thacker}()}]{Thacker_RMP}
\bibinfo{author}{\bibfnamefont{H.~B.} \bibnamefont{Thacker}},
  \bibinfo{howpublished}{Rev. Mod. Phys. {\bf 53}, 253 (1981)}.

\bibitem[{\citenamefont{Gutzwiller}()}]{Gutzwiller_book}
\bibinfo{author}{\bibfnamefont{M.~C.} \bibnamefont{Gutzwiller}},
  \bibinfo{howpublished}{``Chaos in Classical and Quantum Mechanics,'' Berlin
  etc., Springer-Verlag (1990)}.

\bibitem[{\citenamefont{Maldacena}()}]{AdSCFT}
\bibinfo{author}{\bibfnamefont{J.}~\bibnamefont{Maldacena}},
  \bibinfo{howpublished}{Adv. Theor. Math. Phys. {\bf 2}, 231 (1998); I. R.
  Klebanov and and E. Witten, Nucl. Phys. B {\bf 556}, 89 (1999); V. Kazakov,
  A. Marshakov, J. Minahan, and K. Zarembo, Jour. High Ener. Phys., Issue 5,
  024 (2004)}.

\bibitem[{\citenamefont{Kapustin and Witten}()}]{Langlands}
\bibinfo{author}{\bibfnamefont{A.}~\bibnamefont{Kapustin}} \bibnamefont{and}
  \bibinfo{author}{\bibfnamefont{E.}~\bibnamefont{Witten}},
  \bibinfo{howpublished}{Commun. Num. Theor. Phys. {\bf 1}, 1 (2007); E.
  Frenkel, Bull. Amer. Math. Soc. {\bf 41}, 151 (2004)}.

\bibitem[{\citenamefont{Altland and Simons}(2010)}]{altland_condensed_2010}
\bibinfo{author}{\bibfnamefont{A.}~\bibnamefont{Altland}} \bibnamefont{and}
  \bibinfo{author}{\bibfnamefont{B.~D.} \bibnamefont{Simons}},
  \emph{\bibinfo{title}{Condensed Matter Field Theory}}
  (\bibinfo{publisher}{Cambridge University Press}, \bibinfo{year}{2010}),
  \bibinfo{edition}{2nd} ed.

\bibitem[{\citenamefont{Ivanov et~al.}()\citenamefont{Ivanov, Corkum, and
  Dietrich}}]{TLS_laser}
\bibinfo{author}{\bibfnamefont{M.~Y.} \bibnamefont{Ivanov}},
  \bibinfo{author}{\bibfnamefont{P.~B.} \bibnamefont{Corkum}},
  \bibnamefont{and} \bibinfo{author}{\bibfnamefont{P.}~\bibnamefont{Dietrich}},
  \bibinfo{howpublished}{Laser Phys. {\bf 3}, 375 (1993)}.

\bibitem[{\citenamefont{Nakamura et~al.}()\citenamefont{Nakamura, Pashkin, and
  Tsai}}]{TLS_Nakamura}
\bibinfo{author}{\bibfnamefont{Y.}~\bibnamefont{Nakamura}},
  \bibinfo{author}{\bibfnamefont{Y.~A.} \bibnamefont{Pashkin}},
  \bibnamefont{and} \bibinfo{author}{\bibfnamefont{J.}~\bibnamefont{Tsai}},
  \bibinfo{howpublished}{Phys. Rev. Lett. {\bf 87}, 246601 (2001)}.

\bibitem[{\citenamefont{Zamith et~al.}()\citenamefont{Zamith, Degert, Stock,
  {de~Beauvoir}, Blanchet, Bouchene, and Girard}}]{TLS_Girard}
\bibinfo{author}{\bibfnamefont{S.}~\bibnamefont{Zamith}},
  \bibinfo{author}{\bibfnamefont{J.}~\bibnamefont{Degert}},
  \bibinfo{author}{\bibfnamefont{S.}~\bibnamefont{Stock}},
  \bibinfo{author}{\bibfnamefont{B.}~\bibnamefont{{de~Beauvoir}}},
  \bibinfo{author}{\bibfnamefont{V.}~\bibnamefont{Blanchet}},
  \bibinfo{author}{\bibfnamefont{M.~A.} \bibnamefont{Bouchene}},
  \bibnamefont{and} \bibinfo{author}{\bibfnamefont{B.}~\bibnamefont{Girard}},
  \bibinfo{howpublished}{Phys. Rev. Lett. {\bf 87} , 033001 (2001)}.

\bibitem[{\citenamefont{Deblock et~al.}()\citenamefont{Deblock, Onac, Gurevich,
  and Kouwenhoven}}]{TLS_Kou}
\bibinfo{author}{\bibfnamefont{R.}~\bibnamefont{Deblock}},
  \bibinfo{author}{\bibfnamefont{E.}~\bibnamefont{Onac}},
  \bibinfo{author}{\bibfnamefont{L.}~\bibnamefont{Gurevich}}, \bibnamefont{and}
  \bibinfo{author}{\bibfnamefont{L.~P.} \bibnamefont{Kouwenhoven}},
  \bibinfo{howpublished}{Science {\bf 301} , 203 (2003)}.

\bibitem[{\citenamefont{Schuler et~al.}()\citenamefont{Schuler, Speck, Tietz,
  Wrachtrup, and Seifert}}]{TLS_Seifert}
\bibinfo{author}{\bibfnamefont{S.}~\bibnamefont{Schuler}},
  \bibinfo{author}{\bibfnamefont{T.}~\bibnamefont{Speck}},
  \bibinfo{author}{\bibfnamefont{C.}~\bibnamefont{Tietz}},
  \bibinfo{author}{\bibfnamefont{J.}~\bibnamefont{Wrachtrup}},
  \bibnamefont{and} \bibinfo{author}{\bibfnamefont{U.}~\bibnamefont{Seifert}},
  \bibinfo{howpublished}{Phys. Rev. Lett. {\bf 94} , 180602 (2005)}.

\bibitem[{\citenamefont{Lupascu et~al.}()\citenamefont{Lupascu, Saito, Picot,
  {De~Groot}, Harmans, and Mooij}}]{TLS_Mooji}
\bibinfo{author}{\bibfnamefont{A.}~\bibnamefont{Lupascu}},
  \bibinfo{author}{\bibfnamefont{S.}~\bibnamefont{Saito}},
  \bibinfo{author}{\bibfnamefont{T.}~\bibnamefont{Picot}},
  \bibinfo{author}{\bibfnamefont{P.~C.} \bibnamefont{{De~Groot}}},
  \bibinfo{author}{\bibfnamefont{C.~J. P.~M.} \bibnamefont{Harmans}},
  \bibnamefont{and} \bibinfo{author}{\bibfnamefont{J.~E.} \bibnamefont{Mooij}},
  \bibinfo{howpublished}{Nature Phys. {\bf 3} , 119 (2007)}.

\bibitem[{\citenamefont{Fuchs et~al.}()\citenamefont{Fuchs, Dobrovitski, Toyli,
  Heremans, and Awschalom}}]{Awschalom_spin}
\bibinfo{author}{\bibfnamefont{G.~D.} \bibnamefont{Fuchs}},
  \bibinfo{author}{\bibfnamefont{V.~V.} \bibnamefont{Dobrovitski}},
  \bibinfo{author}{\bibfnamefont{D.~M.} \bibnamefont{Toyli}},
  \bibinfo{author}{\bibfnamefont{F.~J.} \bibnamefont{Heremans}},
  \bibnamefont{and} \bibinfo{author}{\bibfnamefont{D.~D.}
  \bibnamefont{Awschalom}}, \bibinfo{howpublished}{Science {\bf 326}, 1520
  (2009); G. Fuchs, V. Dobrovitski, D. Toyli, F. Heremans, C. Weis, T.
  Schenkel, and D. Awschalom, Nature Phys. {\bf 6}, 668 (2010)}.

\bibitem[{\citenamefont{Zhang et~al.}()\citenamefont{Zhang, Feng†, and
  Gilmore}}]{Gilmore_RMP}
\bibinfo{author}{\bibfnamefont{W.}~\bibnamefont{Zhang}},
  \bibinfo{author}{\bibfnamefont{D.}~\bibnamefont{Feng†}}, \bibnamefont{and}
  \bibinfo{author}{\bibfnamefont{R.}~\bibnamefont{Gilmore}},
  \bibinfo{howpublished}{Rev. Mod. Phys. {\bf 62}, 867–927 (1990)}.

\bibitem[{\citenamefont{Solari}(1987)}]{solari_semiclassical_1987}
\bibinfo{author}{\bibfnamefont{H.~G.} \bibnamefont{Solari}},
  \bibinfo{journal}{J. Math. Phys.} \textbf{\bibinfo{volume}{28}},
  \bibinfo{pages}{1097} (\bibinfo{year}{1987}).

\bibitem[{\citenamefont{Kochetov}(1998)}]{kochetov_quasiclassical_1998}
\bibinfo{author}{\bibfnamefont{E.~A.} \bibnamefont{Kochetov}},
  \bibinfo{journal}{J. Phys. A} \textbf{\bibinfo{volume}{31}},
  \bibinfo{pages}{4473} (\bibinfo{year}{1998}).

\bibitem[{\citenamefont{Stone et~al.}(2000)\citenamefont{Stone, Park, and
  Garg}}]{stone_semiclassical_2000}
\bibinfo{author}{\bibfnamefont{M.}~\bibnamefont{Stone}},
  \bibinfo{author}{\bibfnamefont{K.}~\bibnamefont{Park}}, \bibnamefont{and}
  \bibinfo{author}{\bibfnamefont{A.}~\bibnamefont{Garg}}, \bibinfo{journal}{J.
  Math. Phys.} \textbf{\bibinfo{volume}{41}}, \bibinfo{pages}{8025}
  (\bibinfo{year}{2000}).

\bibitem[{\citenamefont{Alscher and Grabert}()}]{Semiclas_JC}
\bibinfo{author}{\bibfnamefont{A.}~\bibnamefont{Alscher}} \bibnamefont{and}
  \bibinfo{author}{\bibfnamefont{H.}~\bibnamefont{Grabert}},
  \bibinfo{howpublished}{Eur. Phys. J. D {\bf 14}, 127 (2001)}.

\bibitem[{\citenamefont{Bloch et~al.}()\citenamefont{Bloch, Dalibard, and
  Zwerger}}]{noneq_RMP}
\bibinfo{author}{\bibfnamefont{I.}~\bibnamefont{Bloch}},
  \bibinfo{author}{\bibfnamefont{J.}~\bibnamefont{Dalibard}}, \bibnamefont{and}
  \bibinfo{author}{\bibfnamefont{W.}~\bibnamefont{Zwerger}},
  \bibinfo{howpublished}{Rev. Mod. Phys. {\bf 80}, 885 (2008)}.

\bibitem[{\citenamefont{Roberts et~al.}()\citenamefont{Roberts, Claussen,
  Cornish, Donley, Cornell, and Wieman}}]{noneq1}
\bibinfo{author}{\bibfnamefont{J.~L.} \bibnamefont{Roberts}},
  \bibinfo{author}{\bibfnamefont{N.~R.} \bibnamefont{Claussen}},
  \bibinfo{author}{\bibfnamefont{S.~L.} \bibnamefont{Cornish}},
  \bibinfo{author}{\bibfnamefont{E.~A.} \bibnamefont{Donley}},
  \bibinfo{author}{\bibfnamefont{E.~A.} \bibnamefont{Cornell}},
  \bibnamefont{and} \bibinfo{author}{\bibfnamefont{C.~E.}
  \bibnamefont{Wieman}}, \bibinfo{howpublished}{Phys. Rev. Lett. {\bf 86}, 4211
  (2001)}.

\bibitem[{\citenamefont{Greiner et~al.}({\natexlab{a}})\citenamefont{Greiner,
  Mandel, H{\"a}nsch, and Bloch}}]{noneq2}
\bibinfo{author}{\bibfnamefont{M.}~\bibnamefont{Greiner}},
  \bibinfo{author}{\bibfnamefont{O.}~\bibnamefont{Mandel}},
  \bibinfo{author}{\bibfnamefont{T.}~\bibnamefont{H{\"a}nsch}},
  \bibnamefont{and} \bibinfo{author}{\bibfnamefont{I.}~\bibnamefont{Bloch}},
  \bibinfo{howpublished}{Nature {\bf 419}, 51 (2002)}.

\bibitem[{\citenamefont{Sengupta et~al.}()\citenamefont{Sengupta, Powell, and
  Sachdev}}]{noneq3}
\bibinfo{author}{\bibfnamefont{K.}~\bibnamefont{Sengupta}},
  \bibinfo{author}{\bibfnamefont{S.}~\bibnamefont{Powell}}, \bibnamefont{and}
  \bibinfo{author}{\bibfnamefont{S.}~\bibnamefont{Sachdev}},
  \bibinfo{howpublished}{Phys. Rev. A. {\bf 69}, 053616 (2004)}.

\bibitem[{\citenamefont{Zurek et~al.}()\citenamefont{Zurek, Dorner, and
  Zoller}}]{noneq4}
\bibinfo{author}{\bibfnamefont{W.~H.} \bibnamefont{Zurek}},
  \bibinfo{author}{\bibfnamefont{U.}~\bibnamefont{Dorner}}, \bibnamefont{and}
  \bibinfo{author}{\bibfnamefont{P.}~\bibnamefont{Zoller}},
  \bibinfo{howpublished}{Phys. Rev. Lett. {\bf 95}, 105701 (2005)}.

\bibitem[{\citenamefont{Manmana et~al.}()\citenamefont{Manmana, Wessel, Noack,
  and Muramutsu}}]{noneq5}
\bibinfo{author}{\bibfnamefont{S.~R.} \bibnamefont{Manmana}},
  \bibinfo{author}{\bibfnamefont{S.}~\bibnamefont{Wessel}},
  \bibinfo{author}{\bibfnamefont{R.~M.} \bibnamefont{Noack}}, \bibnamefont{and}
  \bibinfo{author}{\bibfnamefont{A.}~\bibnamefont{Muramutsu}},
  \bibinfo{howpublished}{Phys. Rev. Lett. {\bf 98}, 210405 (2007)}.

\bibitem[{\citenamefont{Hofferberth et~al.}()\citenamefont{Hofferberth,
  Lesanovsky, Fischer, Schumm, and Schmiedmayer}}]{noneq6}
\bibinfo{author}{\bibfnamefont{S.}~\bibnamefont{Hofferberth}},
  \bibinfo{author}{\bibfnamefont{I.}~\bibnamefont{Lesanovsky}},
  \bibinfo{author}{\bibfnamefont{B.}~\bibnamefont{Fischer}},
  \bibinfo{author}{\bibfnamefont{T.}~\bibnamefont{Schumm}}, \bibnamefont{and}
  \bibinfo{author}{\bibfnamefont{J.}~\bibnamefont{Schmiedmayer}},
  \bibinfo{howpublished}{Nature {\bf 449}, 324 (2007)}.

\bibitem[{\citenamefont{Bistritzer and Altman}()}]{noneq7}
\bibinfo{author}{\bibfnamefont{R.}~\bibnamefont{Bistritzer}} \bibnamefont{and}
  \bibinfo{author}{\bibfnamefont{E.}~\bibnamefont{Altman}},
  \bibinfo{howpublished}{Proc. Natl. Acad. of Sc. {\bf 104}, 9955 (2007)}.

\bibitem[{\citenamefont{Polkovnikov and Gritsev}()}]{noneq8}
\bibinfo{author}{\bibfnamefont{A.}~\bibnamefont{Polkovnikov}} \bibnamefont{and}
  \bibinfo{author}{\bibfnamefont{V.}~\bibnamefont{Gritsev}},
  \bibinfo{howpublished}{Nature Phys. {\bf 4}, 477 (2008)}.

\bibitem[{\citenamefont{Dagnino et~al.}()\citenamefont{Dagnino, Barbern,
  Lewenstein, and Dalibard}}]{noneq9}
\bibinfo{author}{\bibfnamefont{D.}~\bibnamefont{Dagnino}},
  \bibinfo{author}{\bibfnamefont{N.}~\bibnamefont{Barbern}},
  \bibinfo{author}{\bibfnamefont{M.}~\bibnamefont{Lewenstein}},
  \bibnamefont{and} \bibinfo{author}{\bibfnamefont{J.}~\bibnamefont{Dalibard}},
  \bibinfo{howpublished}{Nature Phys. {\bf 5}, 431 (2009)}.

\bibitem[{\citenamefont{Babadi et~al.}()\citenamefont{Babadi, Pekker, Sensarma,
  Georges, and Demler}}]{noneq10}
\bibinfo{author}{\bibfnamefont{M.}~\bibnamefont{Babadi}},
  \bibinfo{author}{\bibfnamefont{D.}~\bibnamefont{Pekker}},
  \bibinfo{author}{\bibfnamefont{R.}~\bibnamefont{Sensarma}},
  \bibinfo{author}{\bibfnamefont{A.}~\bibnamefont{Georges}}, \bibnamefont{and}
  \bibinfo{author}{\bibfnamefont{E.}~\bibnamefont{Demler}},
  \bibinfo{howpublished}{arXiv:0908.3483v2 [cond-mat.quant-gas] (2009)}.

\bibitem[{\citenamefont{Muth et~al.}()\citenamefont{Muth, Unanyan, and
  Fleischhauer}}]{noneq11}
\bibinfo{author}{\bibfnamefont{D.}~\bibnamefont{Muth}},
  \bibinfo{author}{\bibfnamefont{R.}~\bibnamefont{Unanyan}}, \bibnamefont{and}
  \bibinfo{author}{\bibfnamefont{M.}~\bibnamefont{Fleischhauer}},
  \bibinfo{howpublished}{arXiv:1009.4646v2 [quant-ph] (2010)}.

\bibitem[{\citenamefont{Lindner et~al.}()\citenamefont{Lindner, Refael, and
  Galitski}}]{NLGRVG}
\bibinfo{author}{\bibfnamefont{N.~H.} \bibnamefont{Lindner}},
  \bibinfo{author}{\bibfnamefont{G.}~\bibnamefont{Refael}}, \bibnamefont{and}
  \bibinfo{author}{\bibfnamefont{V.~M.} \bibnamefont{Galitski}},
  \bibinfo{howpublished}{Nature Physics doi:10.1038/nphys1926 (published online
  on March 13, 2011)}.

\bibitem[{\citenamefont{Robertson et~al.}()\citenamefont{Robertson, Galitski,
  and Refael}}]{ARCGGR}
\bibinfo{author}{\bibfnamefont{A.}~\bibnamefont{Robertson}},
  \bibinfo{author}{\bibfnamefont{V.~M.} \bibnamefont{Galitski}},
  \bibnamefont{and} \bibinfo{author}{\bibfnamefont{G.}~\bibnamefont{Refael}},
  \bibinfo{howpublished}{Phys. Rev. Lett. {\bf 106}, 165701 (2011)}.

\bibitem[{\citenamefont{Bao et~al.}()\citenamefont{Bao, Dong, Silverstein, and
  Torroba}}]{AdSSC}
\bibinfo{author}{\bibfnamefont{N.}~\bibnamefont{Bao}},
  \bibinfo{author}{\bibfnamefont{X.}~\bibnamefont{Dong}},
  \bibinfo{author}{\bibfnamefont{E.}~\bibnamefont{Silverstein}},
  \bibnamefont{and} \bibinfo{author}{\bibfnamefont{G.}~\bibnamefont{Torroba}},
  \bibinfo{howpublished}{arXiv:1104.4098v2 [hep-th] (2011)}.

\bibitem[{\citenamefont{Paredes et~al.}()\citenamefont{Paredes, Widera, Murg,
  Mandel, F{\"o}lling, Cirac, Shlyapnikov, H{\"a}nsch, and Bloc}}]{TG_exp}
\bibinfo{author}{\bibfnamefont{B.}~\bibnamefont{Paredes}},
  \bibinfo{author}{\bibfnamefont{A.}~\bibnamefont{Widera}},
  \bibinfo{author}{\bibfnamefont{V.}~\bibnamefont{Murg}},
  \bibinfo{author}{\bibfnamefont{O.}~\bibnamefont{Mandel}},
  \bibinfo{author}{\bibfnamefont{S.}~\bibnamefont{F{\"o}lling}},
  \bibinfo{author}{\bibfnamefont{I.}~\bibnamefont{Cirac}},
  \bibinfo{author}{\bibfnamefont{G.}~\bibnamefont{Shlyapnikov}},
  \bibinfo{author}{\bibfnamefont{T.}~\bibnamefont{H{\"a}nsch}},
  \bibnamefont{and} \bibinfo{author}{\bibfnamefont{I.}~\bibnamefont{Bloc}},
  \bibinfo{howpublished}{Nature {\bf 429}, 277 (2004)}.

\bibitem[{\citenamefont{Kinoshita et~al.}()\citenamefont{Kinoshita, Wenger, and
  Weiss}}]{HO_exp}
\bibinfo{author}{\bibfnamefont{T.}~\bibnamefont{Kinoshita}},
  \bibinfo{author}{\bibfnamefont{T.}~\bibnamefont{Wenger}}, \bibnamefont{and}
  \bibinfo{author}{\bibfnamefont{D.}~\bibnamefont{Weiss}},
  \bibinfo{howpublished}{Nature (London) {\bf 440}, 900 (2006)}.

\bibitem[{\citenamefont{Rigol et~al.}({\natexlab{a}})\citenamefont{Rigol,
  Dunjko, and Olshanii}}]{Rigol_etal}
\bibinfo{author}{\bibfnamefont{M.}~\bibnamefont{Rigol}},
  \bibinfo{author}{\bibfnamefont{V.}~\bibnamefont{Dunjko}}, \bibnamefont{and}
  \bibinfo{author}{\bibfnamefont{M.}~\bibnamefont{Olshanii}},
  \bibinfo{howpublished}{Nature {\bf 452}, 854 (2008)}.

\bibitem[{\citenamefont{Rigol et~al.}({\natexlab{b}})\citenamefont{Rigol,
  Dunjko, Yurovsky, and Olshanii}}]{Integrable_relax}
\bibinfo{author}{\bibfnamefont{M.}~\bibnamefont{Rigol}},
  \bibinfo{author}{\bibfnamefont{V.}~\bibnamefont{Dunjko}},
  \bibinfo{author}{\bibfnamefont{V.}~\bibnamefont{Yurovsky}}, \bibnamefont{and}
  \bibinfo{author}{\bibfnamefont{M.}~\bibnamefont{Olshanii}},
  \bibinfo{howpublished}{Phys. Rev. Lett. {\bf 98}, 050405 (2007)}.

\bibitem[{\citenamefont{Brixner et~al.}()\citenamefont{Brixner, Damrauer,
  Niklaus, and Gerber}}]{Qcontr1}
\bibinfo{author}{\bibfnamefont{T.}~\bibnamefont{Brixner}},
  \bibinfo{author}{\bibfnamefont{N.~H.} \bibnamefont{Damrauer}},
  \bibinfo{author}{\bibfnamefont{P.}~\bibnamefont{Niklaus}}, \bibnamefont{and}
  \bibinfo{author}{\bibfnamefont{G.}~\bibnamefont{Gerber}},
  \bibinfo{howpublished}{Nature {\bf 414}, 57 (2001)}.

\bibitem[{\citenamefont{Shapiroa and Brume}()}]{Qcontr1.5}
\bibinfo{author}{\bibfnamefont{M.}~\bibnamefont{Shapiroa}} \bibnamefont{and}
  \bibinfo{author}{\bibfnamefont{P.}~\bibnamefont{Brume}},
  \bibinfo{howpublished}{Adv. AMO Phys. {\bf 42}, 287 (2000)}.

\bibitem[{\citenamefont{Geremia et~al.}()\citenamefont{Geremia, Stockton, and
  Mabuchi}}]{Qcontr2-}
\bibinfo{author}{\bibfnamefont{J.~M.} \bibnamefont{Geremia}},
  \bibinfo{author}{\bibfnamefont{J.}~\bibnamefont{Stockton}}, \bibnamefont{and}
  \bibinfo{author}{\bibfnamefont{H.}~\bibnamefont{Mabuchi}},
  \bibinfo{howpublished}{Science {\bf 304}, 270 (2004)}.

\bibitem[{\citenamefont{Taylor et~al.}()\citenamefont{Taylor, Petta, Johnson,
  Yacoby, Marcus, and Lukin}}]{Qcontr2}
\bibinfo{author}{\bibfnamefont{J.~M.} \bibnamefont{Taylor}},
  \bibinfo{author}{\bibfnamefont{J.~R.} \bibnamefont{Petta}},
  \bibinfo{author}{\bibfnamefont{A.~C.} \bibnamefont{Johnson}},
  \bibinfo{author}{\bibfnamefont{A.}~\bibnamefont{Yacoby}},
  \bibinfo{author}{\bibfnamefont{C.~M.} \bibnamefont{Marcus}},
  \bibnamefont{and} \bibinfo{author}{\bibfnamefont{M.~D.} \bibnamefont{Lukin}},
  \bibinfo{howpublished}{Phys. Rev. B {\bf 76}, 035315 (2007)}.

\bibitem[{\citenamefont{Press et~al.}()\citenamefont{Press, Ladd, Zhang, and
  Yamamoto}}]{Qcontr3}
\bibinfo{author}{\bibfnamefont{D.}~\bibnamefont{Press}},
  \bibinfo{author}{\bibfnamefont{T.~D.} \bibnamefont{Ladd}},
  \bibinfo{author}{\bibfnamefont{B.}~\bibnamefont{Zhang}}, \bibnamefont{and}
  \bibinfo{author}{\bibfnamefont{Y.}~\bibnamefont{Yamamoto}},
  \bibinfo{howpublished}{Nature {\bf 456}, 218 (2008)}.

\bibitem[{\citenamefont{Belhadj et~al.}()\citenamefont{Belhadj, Simon, Amand,
  Renucci, Chatel, Krebs, Lemaitre, Voisin, Marie, and Urbaszek}}]{Qcontr3.5}
\bibinfo{author}{\bibfnamefont{T.}~\bibnamefont{Belhadj}},
  \bibinfo{author}{\bibfnamefont{C.-M.} \bibnamefont{Simon}},
  \bibinfo{author}{\bibfnamefont{T.}~\bibnamefont{Amand}},
  \bibinfo{author}{\bibfnamefont{P.}~\bibnamefont{Renucci}},
  \bibinfo{author}{\bibfnamefont{B.}~\bibnamefont{Chatel}},
  \bibinfo{author}{\bibfnamefont{O.}~\bibnamefont{Krebs}},
  \bibinfo{author}{\bibfnamefont{A.}~\bibnamefont{Lemaitre}},
  \bibinfo{author}{\bibfnamefont{P.}~\bibnamefont{Voisin}},
  \bibinfo{author}{\bibfnamefont{X.}~\bibnamefont{Marie}}, \bibnamefont{and}
  \bibinfo{author}{\bibfnamefont{B.}~\bibnamefont{Urbaszek}},
  \bibinfo{howpublished}{Phys. Rev. Lett. {\bf 103}, 086601 (2009)}.

\bibitem[{\citenamefont{Foletti et~al.}()\citenamefont{Foletti, Bluhm, Mahalu,
  Umansky, and Yacob}}]{Qcontr4}
\bibinfo{author}{\bibfnamefont{S.}~\bibnamefont{Foletti}},
  \bibinfo{author}{\bibfnamefont{H.}~\bibnamefont{Bluhm}},
  \bibinfo{author}{\bibfnamefont{D.}~\bibnamefont{Mahalu}},
  \bibinfo{author}{\bibfnamefont{V.}~\bibnamefont{Umansky}}, \bibnamefont{and}
  \bibinfo{author}{\bibfnamefont{A.}~\bibnamefont{Yacob}},
  \bibinfo{howpublished}{Nature Phys. {\bf 5}, 903 (2009)}.

\bibitem[{\citenamefont{Polkovnikov}()}]{TP_Review}
\bibinfo{author}{\bibfnamefont{A.}~\bibnamefont{Polkovnikov}},
  \bibinfo{howpublished}{Annals of Phys. {\bf 325}, 1790 (2010)}.

\bibitem[{\citenamefont{Kamenev and Levchenko}()}]{Keldysh_Rev}
\bibinfo{author}{\bibfnamefont{A.}~\bibnamefont{Kamenev}} \bibnamefont{and}
  \bibinfo{author}{\bibfnamefont{A.}~\bibnamefont{Levchenko}},
  \bibinfo{howpublished}{Advances in Phys. {\bf 58}, 197 (2009)}.

\bibitem[{\citenamefont{{D'Alessandro}}()}]{DAlessandro_book}
\bibinfo{author}{\bibfnamefont{D.}~\bibnamefont{{D'Alessandro}}},
  \bibinfo{howpublished}{``Introduction to Quantum Control and Dynamics,''
  Chapman \& Hall/CRC (2007)}.

\bibitem[{\citenamefont{Gritsev et~al.}()\citenamefont{Gritsev, Barmettler, and
  Demler}}]{Grit1}
\bibinfo{author}{\bibfnamefont{V.}~\bibnamefont{Gritsev}},
  \bibinfo{author}{\bibfnamefont{P.}~\bibnamefont{Barmettler}},
  \bibnamefont{and} \bibinfo{author}{\bibfnamefont{E.}~\bibnamefont{Demler}},
  \bibinfo{howpublished}{New J. Phys. {\bf 12}, 113005 (2010)}.

\bibitem[{\citenamefont{Imambekov et~al.}()\citenamefont{Imambekov, Lukyanov,
  Glazman, and Gritsev}}]{Grit2}
\bibinfo{author}{\bibfnamefont{A.}~\bibnamefont{Imambekov}},
  \bibinfo{author}{\bibfnamefont{A.~A.} \bibnamefont{Lukyanov}},
  \bibinfo{author}{\bibfnamefont{L.~I.} \bibnamefont{Glazman}},
  \bibnamefont{and} \bibinfo{author}{\bibfnamefont{V.}~\bibnamefont{Gritsev}},
  \bibinfo{howpublished}{Phys. Rev. Lett. {\bf 104}, 040402 (2010)}.

\bibitem[{\citenamefont{Galitski}({\natexlab{a}})}]{VG_Qfl}
\bibinfo{author}{\bibfnamefont{V.}~\bibnamefont{Galitski}},
  \bibinfo{howpublished}{Phys. Rev. B {\bf 82}, 054511 (2010)}.

\bibitem[{\citenamefont{Galitski}({\natexlab{b}})}]{Fermionization}
\bibinfo{author}{\bibfnamefont{V.}~\bibnamefont{Galitski}},
  \bibinfo{howpublished}{Phys.Rev.B {\bf 82}, 060411(R) (2010)}.

\bibitem[{\citenamefont{Sedrakyan and Galitski}({\natexlab{a}})}]{BWZW}
\bibinfo{author}{\bibfnamefont{T.}~\bibnamefont{Sedrakyan}} \bibnamefont{and}
  \bibinfo{author}{\bibfnamefont{V.}~\bibnamefont{Galitski}},
  \bibinfo{howpublished}{Phys. Rev. B {\bf 82}, 214502 (2010)}.

\bibitem[{\citenamefont{Gangopadhyay et~al.}()\citenamefont{Gangopadhyay,
  Dzero, and Galitski}}]{Our_TLS}
\bibinfo{author}{\bibfnamefont{A.}~\bibnamefont{Gangopadhyay}},
  \bibinfo{author}{\bibfnamefont{M.}~\bibnamefont{Dzero}}, \bibnamefont{and}
  \bibinfo{author}{\bibfnamefont{V.}~\bibnamefont{Galitski}},
  \bibinfo{howpublished}{Phys. Rev. B {\bf 82}, 024303 (2010)}.

\bibitem[{\citenamefont{Magnus}()}]{Magnus}
\bibinfo{author}{\bibfnamefont{W.}~\bibnamefont{Magnus}},
  \bibinfo{howpublished}{Comm. Pure and Appl. Math. {\bf VII}, 649 (1954)}.

\bibitem[{\citenamefont{Wilcox}()}]{Wilcox}
\bibinfo{author}{\bibfnamefont{R.~M.} \bibnamefont{Wilcox}},
  \bibinfo{howpublished}{J. Math. Phys. {\bf 8}, 962 (1967)}.

\bibitem[{\citenamefont{Wei and Norman}({\natexlab{a}})}]{WeiNorm1}
\bibinfo{author}{\bibfnamefont{J.}~\bibnamefont{Wei}} \bibnamefont{and}
  \bibinfo{author}{\bibfnamefont{E.}~\bibnamefont{Norman}},
  \bibinfo{howpublished}{J. Math. Phys. {\bf 4}, 575 (1963)}.

\bibitem[{\citenamefont{Wei and Norman}({\natexlab{b}})}]{WeiNorm2}
\bibinfo{author}{\bibfnamefont{J.}~\bibnamefont{Wei}} \bibnamefont{and}
  \bibinfo{author}{\bibfnamefont{E.}~\bibnamefont{Norman}},
  \bibinfo{howpublished}{Proc. Am. Math. Soc. {\bf 15}, 327 (1964)}.

\bibitem[{\citenamefont{Alhassid and Levine}({\natexlab{a}})}]{Alhassid1}
\bibinfo{author}{\bibfnamefont{Y.}~\bibnamefont{Alhassid}} \bibnamefont{and}
  \bibinfo{author}{\bibfnamefont{R.~D.} \bibnamefont{Levine}},
  \bibinfo{howpublished}{J. Chem. Phys. {\bf 67}, 4321 (1977)}.

\bibitem[{\citenamefont{Alhassid and Levine}({\natexlab{b}})}]{Alhassid2}
\bibinfo{author}{\bibfnamefont{Y.}~\bibnamefont{Alhassid}} \bibnamefont{and}
  \bibinfo{author}{\bibfnamefont{R.~D.} \bibnamefont{Levine}},
  \bibinfo{howpublished}{Phys. Rev. A {\bf 18}, 89 (1978)}.

\bibitem[{\citenamefont{Vourdas}()}]{Vourdas}
\bibinfo{author}{\bibfnamefont{A.}~\bibnamefont{Vourdas}},
  \bibinfo{howpublished}{J. Phys. A: Math. Gen. {\bf 39}, R65 (2006)}.

\bibitem[{\citenamefont{Stock and Thoss}()}]{ChemPhysRev}
\bibinfo{author}{\bibfnamefont{G.}~\bibnamefont{Stock}} \bibnamefont{and}
  \bibinfo{author}{\bibfnamefont{M.}~\bibnamefont{Thoss}},
  \bibinfo{howpublished}{Adv. Chem. Phys. {\bf 131}, 243 (2005)}.

\bibitem[{\citenamefont{Blanes et~al.}()\citenamefont{Blanes, Casas, Oteo, and
  Ros}}]{MagnusRev}
\bibinfo{author}{\bibfnamefont{S.}~\bibnamefont{Blanes}},
  \bibinfo{author}{\bibfnamefont{F.}~\bibnamefont{Casas}},
  \bibinfo{author}{\bibfnamefont{J.}~\bibnamefont{Oteo}}, \bibnamefont{and}
  \bibinfo{author}{\bibfnamefont{J.}~\bibnamefont{Ros}},
  \bibinfo{howpublished}{Phys. Rep. {\bf 470}, 151 (2009)}.

\bibitem[{\citenamefont{Suzuki}()}]{Suzuki}
\bibinfo{author}{\bibfnamefont{M.}~\bibnamefont{Suzuki}},
  \bibinfo{howpublished}{Commun. Math. Phys. {\bf 57}, 193 (1977)}.

\bibitem[{\citenamefont{Gilmore}({\natexlab{b}})}]{Gilmore_book}
\bibinfo{author}{\bibfnamefont{R.}~\bibnamefont{Gilmore}},
  \bibinfo{howpublished}{``Lie Groups, Lie Algebras and Some of Their
  Applications,'' Wiley, New York (1974)}.

\bibitem[{\citenamefont{Vinjanampathy and Rau}()}]{BlochSU(3)}
\bibinfo{author}{\bibfnamefont{S.}~\bibnamefont{Vinjanampathy}}
  \bibnamefont{and} \bibinfo{author}{\bibfnamefont{A.~R.~P.}
  \bibnamefont{Rau}}, \bibinfo{howpublished}{J. Phys. A: Math. Theor. {\bf 42},
  425303 (2009)}.

\bibitem[{\citenamefont{Bulgac and Kusnezov}()}]{BulKuz}
\bibinfo{author}{\bibfnamefont{A.}~\bibnamefont{Bulgac}} \bibnamefont{and}
  \bibinfo{author}{\bibfnamefont{D.}~\bibnamefont{Kusnezov}},
  \bibinfo{howpublished}{Annals of Physics {\bf 19}, 187-224 (1990)}.

\bibitem[{\citenamefont{Mosseri and Dandoloff}()}]{2Hopf_entangl}
\bibinfo{author}{\bibfnamefont{R.}~\bibnamefont{Mosseri}} \bibnamefont{and}
  \bibinfo{author}{\bibfnamefont{R.}~\bibnamefont{Dandoloff}},
  \bibinfo{howpublished}{J. Phys. A: Math. Gen. {\bf 34}, 10243 (2001)}.

\bibitem[{\citenamefont{Bernevig and Chen}()}]{3Hopf_entangl}
\bibinfo{author}{\bibfnamefont{B.}~\bibnamefont{Bernevig}} \bibnamefont{and}
  \bibinfo{author}{\bibfnamefont{H.}~\bibnamefont{Chen}},
  \bibinfo{howpublished}{J. Phys. A: Math. Gen. {\bf 36}, 8325 (2003)}.

\bibitem[{com()}]{comment}
\bibinfo{howpublished}{A possible construction is to use Hadamard identity,
  $\exp\left[{-i {\rm ad}_{{\bm \Phi} \cdot \check{\bm J}}} \right]{\bm \Theta}
  \cdot \check{\bm J} = {\bm \Gamma} \cdot \check{\bm J}$, and define ${\rm
  Had}\,({\bm \Phi},{\bm \Theta}) = {\bm \Gamma}$. Then, call two elements
  equivalent ${\bm \Phi}_1 \sim {\bm \Phi}_2$, iff, ${\rm Had}\,({\bm
  \Phi}_1,{\bm \Theta}) \equiv {\rm Had}\,({\bm \Phi}_2,{\bm \Theta})$, for all
  ${\bm \Theta} \in {\cal A}^*$. The quotient set ${\cal A}^*/\sim$ by this
  equivalence relation gives rise to a new structure, which appears related to
  the corresponding group.}

\bibitem[{\citenamefont{Landau}()}]{LZ}
\bibinfo{author}{\bibfnamefont{L.}~\bibnamefont{Landau}},
  \bibinfo{howpublished}{Phys. Z. Sowjetunion {\bf 2}, 46 (1932); C. Zener,
  Proc. R. Soc. London, Ser.~A {\bf 137}, 696 (1932)}.

\bibitem[{\citenamefont{Lim and Berry}()}]{Lim_Berry}
\bibinfo{author}{\bibfnamefont{R.}~\bibnamefont{Lim}} \bibnamefont{and}
  \bibinfo{author}{\bibfnamefont{M.~V.} \bibnamefont{Berry}},
  \bibinfo{howpublished}{J. Phys. A:~Math. Gen. {\bf 24}, 3255 (1991)}.

\bibitem[{\citenamefont{Altland et~al.}()\citenamefont{Altland, Gurarie,
  Kriecherbauer, and Polkovnikov}}]{Altland_etal}
\bibinfo{author}{\bibfnamefont{A.}~\bibnamefont{Altland}},
  \bibinfo{author}{\bibfnamefont{V.}~\bibnamefont{Gurarie}},
  \bibinfo{author}{\bibfnamefont{T.}~\bibnamefont{Kriecherbauer}},
  \bibnamefont{and}
  \bibinfo{author}{\bibfnamefont{A.}~\bibnamefont{Polkovnikov}},
  \bibinfo{howpublished}{Phys. Rev. A {\bf 79}, 042703 (2009)}.

\bibitem[{\citenamefont{Sedrakyan and Galitski}({\natexlab{b}})}]{TSVGTLS}
\bibinfo{author}{\bibfnamefont{T.}~\bibnamefont{Sedrakyan}} \bibnamefont{and}
  \bibinfo{author}{\bibfnamefont{V.}~\bibnamefont{Galitski}},
  \bibinfo{howpublished}{arXiv:1012.2005v1 [quant-ph] (2010)}.

\bibitem[{\citenamefont{Fisher et~al.}()\citenamefont{Fisher, Weichman,
  Grinstein, and Fisher}}]{MPABH}
\bibinfo{author}{\bibfnamefont{M.~P.~A.} \bibnamefont{Fisher}},
  \bibinfo{author}{\bibfnamefont{P.~B.} \bibnamefont{Weichman}},
  \bibinfo{author}{\bibfnamefont{G.}~\bibnamefont{Grinstein}},
  \bibnamefont{and} \bibinfo{author}{\bibfnamefont{D.~S.}
  \bibnamefont{Fisher}}, \bibinfo{howpublished}{Phys. Rev. B {\bf 40}, 546
  (1989)}.

\bibitem[{\citenamefont{Jaksch et~al.}()\citenamefont{Jaksch, Bruder, Cirac,
  Gardiner, and Zoller}}]{BH_Zoller}
\bibinfo{author}{\bibfnamefont{D.}~\bibnamefont{Jaksch}},
  \bibinfo{author}{\bibfnamefont{C.}~\bibnamefont{Bruder}},
  \bibinfo{author}{\bibfnamefont{J.}~\bibnamefont{Cirac}},
  \bibinfo{author}{\bibfnamefont{C.}~\bibnamefont{Gardiner}}, \bibnamefont{and}
  \bibinfo{author}{\bibfnamefont{P.}~\bibnamefont{Zoller}},
  \bibinfo{howpublished}{Phys. Rev. Lett. {\bf 81}, 3108 (1998)}.

\bibitem[{\citenamefont{Greiner et~al.}({\natexlab{b}})\citenamefont{Greiner,
  Mandel, Esslinger, H{\"a}nsch, and Bloch}}]{BHexp}
\bibinfo{author}{\bibfnamefont{M.}~\bibnamefont{Greiner}},
  \bibinfo{author}{\bibfnamefont{O.}~\bibnamefont{Mandel}},
  \bibinfo{author}{\bibfnamefont{T.}~\bibnamefont{Esslinger}},
  \bibinfo{author}{\bibfnamefont{T.~W.} \bibnamefont{H{\"a}nsch}},
  \bibnamefont{and} \bibinfo{author}{\bibfnamefont{I.}~\bibnamefont{Bloch}},
  \bibinfo{howpublished}{Nature {\bf 415}, 39 (2002)}.

\bibitem[{\citenamefont{Wilson and Galitski}()}]{JW_VG}
\bibinfo{author}{\bibfnamefont{J.}~\bibnamefont{Wilson}} \bibnamefont{and}
  \bibinfo{author}{\bibfnamefont{V.}~\bibnamefont{Galitski}},
  \bibinfo{howpublished}{Phys. Rev. Lett. {\bf 106}, 110401 (2011)}.

\bibitem[{\citenamefont{Kleinert}({\natexlab{a}})}]{Kleinert1}
\bibinfo{author}{\bibfnamefont{H.}~\bibnamefont{Kleinert}},
  \bibinfo{howpublished}{``Collective Quantum Fields,'' Lectures presented at
  the Erice Summer School on Low-Temperature Physics (1977) in Fortschr. Physik
  {\bf 26}, 565-671 (1978)}.

\bibitem[{\citenamefont{Kleinert}({\natexlab{b}})}]{Kleinert2}
\bibinfo{author}{\bibfnamefont{H.}~\bibnamefont{Kleinert}},
  \bibinfo{howpublished}{``On the Hadronization of Quark Theories,'' Lectures
  presented at the Erice Summer Institute (1976) in ``Understanding the
  Fundamental Constituents of Matter,'' Plenum Press, New York, A.~Zichichi ed.
  (1978)}.

\bibitem[{\citenamefont{Kleinert}({\natexlab{c}})}]{Kleinert3}
\bibinfo{author}{\bibfnamefont{H.}~\bibnamefont{Kleinert}},
  \bibinfo{howpublished}{Annals of Physics {\bf 266}, 135 (1998)}.

\bibitem[{\citenamefont{Pollet et~al.}()\citenamefont{Pollet, Prokof'ev, and
  Svistunov}}]{Pollet_etal}
\bibinfo{author}{\bibfnamefont{L.}~\bibnamefont{Pollet}},
  \bibinfo{author}{\bibfnamefont{N.~V.} \bibnamefont{Prokof'ev}},
  \bibnamefont{and} \bibinfo{author}{\bibfnamefont{B.~V.}
  \bibnamefont{Svistunov}}, \bibinfo{howpublished}{Phys.Rev.Lett. {\bf 105},
  210601 (2010)}.

\bibitem[{\citenamefont{Galitski}({\natexlab{c}})}]{VG_tbp}
\bibinfo{author}{\bibfnamefont{V.}~\bibnamefont{Galitski}},
  \bibinfo{howpublished}{to be published}.

\bibitem[{\citenamefont{Auerbach}()}]{Assa_book}
\bibinfo{author}{\bibfnamefont{A.}~\bibnamefont{Auerbach}},
  \bibinfo{howpublished}{``Interacting Electrons and Quantum Magnetism,''
  Springer-Verlag, N.Y. (1998)}.

\bibitem[{\citenamefont{Savit}()}]{Savit}
\bibinfo{author}{\bibfnamefont{R.}~\bibnamefont{Savit}},
  \bibinfo{howpublished}{Rev. Mod. Phys. {\bf 52}, 453 (1980)}.

\bibitem[{\citenamefont{Sternberg}()}]{Sternberg}
\bibinfo{author}{\bibfnamefont{S.}~\bibnamefont{Sternberg}},
  \bibinfo{howpublished}{``Lie algebras,'' online book:
  http://www.math.harvard.edu/~shlomo/ (2004)}.

\bibitem[{\citenamefont{Klauder}({\natexlab{b}})}]{KlauderPI}
\bibinfo{author}{\bibfnamefont{J.~R.} \bibnamefont{Klauder}},
  \bibinfo{howpublished}{arXiv:quant-ph/0303034v1 (2003)}.

\bibitem[{\citenamefont{Levi}()}]{Levi}
\bibinfo{author}{\bibfnamefont{E.}~\bibnamefont{Levi}},
  \bibinfo{howpublished}{Atti. Accad. Sci. Torino Cl. Sci. Fis. Mat. Natur.
  {\bf 40}, 3 (1906)}.

\bibitem[{\citenamefont{{W{\"u}stner}}()}]{EXPLie1}
\bibinfo{author}{\bibfnamefont{M.}~\bibnamefont{{W{\"u}stner}}},
  \bibinfo{howpublished}{J. Lie Theor. {\bf15}, 269 (2005)}.

\bibitem[{\citenamefont{Perelomov}({\natexlab{b}})}]{Perelomov_book}
\bibinfo{author}{\bibfnamefont{A.}~\bibnamefont{Perelomov}},
  \bibinfo{howpublished}{``Generalized Coherent States and Their
  Applications,'' Springer, Berlin (1986)}.

\bibitem[{\citenamefont{Kleinert}({\natexlab{d}})}]{Kleinertbook}
\bibinfo{author}{\bibfnamefont{H.}~\bibnamefont{Kleinert}},
  \bibinfo{howpublished}{``Path Integrals in Quantum Mechanics, Statistics,
  Polymer Physics, and Financial Markets,'' 4th edition, World Scientific
  (Singapore, 2009)}.

\bibitem[{\citenamefont{Sachdev}()}]{Sachdevbook}
\bibinfo{author}{\bibfnamefont{S.}~\bibnamefont{Sachdev}},
  \bibinfo{howpublished}{``Quantum phase transitions,'' Cambridge University
  Press (2001)}.

\end{thebibliography}

\end{document}